\documentclass[11pt,reqno]{amsart}
\usepackage{mathrsfs,amsmath,amsxtra,amssymb,amsthm,amsfonts,graphicx,float,hyperref,braket,subcaption,textalpha}
\newcommand{\kla}{\left ( }
\newcommand{\mer}{\right ) }
\renewcommand{\for}{\begin{eqnarray*}}
\newcommand{\mel}{\end{eqnarray*}}
\def\fr{\begin{align*}}

\newcommand{\kl}{\pl \le \pl}
\newcommand{\gl}{\pl \ge \pl}

\newcommand{\lel}{\pl = \pl}

\newcommand{\nz}{{\mathbb N}}
\newcommand{\nen}{n \in \nz}

\newcommand{\rz}{{\mathbb R}}
\newcommand{\zz}{{\mathbb Z}}

\newcommand{\cz}{{\mathbb C}}

\newcommand{\ten}{\otimes}

\newcommand{\pl}{\hspace{.1cm}}

\newcommand{\pla}{\hspace{1.5cm}}

\newcommand{\ran}{\rangle}
\newcommand{\lan}{\langle}

\newcommand{\om}{\omega}

\newcommand{\al}{\alpha}
\newcommand{\si}{\sigma}

\newcommand{\La}{\Lambda}
\newcommand{\la}{\lambda}

\renewcommand{\phi}{\varphi}

\renewcommand{\L}{{\mathcal L}}
\newcommand{\F}{{\mathcal F}}
\newcommand{\E}{{\mathcal E}}

\newcommand{\A}{{\mathcal A}}
\newcommand{\B}{{\mathcal B}}

\newcommand{\M}{{\mathcal M}}
\newcommand{\K}{{\mathcal K}}

\renewcommand{\H}{{\mathcal H}}

\newcommand{\N}{{\mathcal N}}

\newcommand{\U}{{\mathcal U}}

\newcommand{\ez}{{\mathbb E}\vspace{0.1cm}}

\newcommand{\norm}[2]{\parallel \! #1 \! \parallel_{#2}}

\newtheorem{lemma}{Lemma}[section]
\newtheorem{prop}[lemma]{Proposition}
\newtheorem{theorem}[lemma]{Theorem}
\newtheorem{cor}[lemma]{Corollary}

\newtheorem{rem}[lemma]{Remark}

\newcommand{\re}{\begin{rem}\rm}
\newcommand{\mar}{\end{rem}}

\newcommand{\qd}{\end{proof}\vspace{0.5ex}}
\newcommand{\prf}{\begin{proof}[\bf Proof:]}

\newcommand{\xspace}{\hbox{\kern-2.5pt}}

\newtheorem*{theorem*}{Theorem}

\allowdisplaybreaks

\usepackage{color}

\definecolor{LightGray}{rgb}{0.94,0.94,0.94}
\definecolor{VeryLightBlue}{rgb}{0.9,0.9,1}
\definecolor{LightBlue}{rgb}{0.8,0.8,1}
\definecolor{DarkBlue}{rgb}{0,0,0.6}
\definecolor{LightGreen}{rgb}{0.88,1,0.88}
\definecolor{MidGreen}{rgb}{0.6,1,0.6}
\definecolor{DarkGreen}{rgb}{0,0.6,0}
\definecolor{DarkGrreen}{rgb}{0,0.8,0}

\definecolor{VeryLightYellow}{rgb}{1,1,0.9}
\definecolor{LightYellow}{rgb}{1,1,0.6}
\definecolor{MidYellow}{rgb}{1,1,0.5}
\definecolor{DarkYellow}{rgb}{0.8,1,0.3}
\definecolor{VeryLightRed}{rgb}{1,0.9,0.9}
\definecolor{LightRed}{rgb}{1,0.8,0.8}
\definecolor{DarkRed}{rgb}{0.8,0.2,0}
\definecolor{DarkRedb}{rgb}{0.6,0.2,0}
\definecolor{DarkLila}{rgb}{0.8,0,1}
\definecolor{Beige}{rgb}{0.96,0.96,0.86}
\definecolor{Gold}{rgb}{1.,0.84,0.}
\definecolor{Goldb}{rgb}{0.7,0.3,0.5}
\definecolor{MyYellow}{rgb}{1.,0.84,0.8}

\usepackage{verbatim}
\usepackage{amsmath}
\usepackage{amsthm}
\usepackage{graphicx}
\usepackage{amsfonts}
\usepackage{latexsym}
\usepackage{color}
\usepackage{braket}

\title[Multivariate Trace Inequalities \& Universal Recovery Beyond Tracial Settings]{Multivariate Trace Inequalities, p-Fidelity, and Universal Recovery Beyond Tracial Settings}
\author[M. Junge]{Marius Junge}
\address{Department of Mathematics\\
University of Illinois, Urbana, IL 61801, USA} \email[Marius
Junge]{mjunge@illinois.edu}

\author[N. LaRacuente]{Nicholas LaRacuente}
\address{University of Chicago, Chicago, IL 60637, USA} \email[Nicholas LaRacuente]{nlaracuente@uchicago.edu}

\thanks{NL is supported by IBM as a Postdoctoral Scholar at the University of Chicago and The Chicago Quantum Exchange. NL was previously supported by the Department of Physics at the Univerity of Illinois at Urbana-Champaign. MJ is partially supported by NSF grants DMS 1800872 and Raise-TAQS 1839177.}

\DeclareMathOperator{\tr}{tr}

	\newcommand{\RR}{\mathbb{R}}
	\newcommand{\CC}{\mathbb{C}}
	\newcommand{\BB}{\mathbb{B}}
	\newcommand{\NN}{\mathbb{N}}
	\newcommand{\Hil}{\mathcal{H}}

	\newcommand{\id}{\mathbf{1}}
	
	\renewcommand{\L}{\mathcal{L}}

\DeclareMathOperator{\Lin}{Lin}

\DeclareMathOperator{\mdom}{mdom}

\begin{document}

\begin{abstract}
Trace inequalities are general techniques with many applications in quantum information theory, often replacing classical functional calculus in noncommutative settings. The physics of quantum field theory and holography, however, motivate entropy inequalities in type III von Neumann algebras that lack a semifinite trace. The Haagerup and Kosaki $L_p$ spaces enable re-expressing trace inequalities in non-tracial von Neumann algebras. In particular, we show this for the generalized Araki-Lieb-Thirring and Golden-Thompson inequalities from (Sutter, Berta \& Tomamichel 2017 \cite{sutter_multivariate_2017}). Then, using the Haagerup approximation method, we prove a general von Neumann algebra version of univeral recovery map corrections to the data processing inequality for relative entropy. We also show subharmonicity of a logarithmic p-fidelity of recovery. Furthermore, we prove that non-decrease of relative entropy is equivalent to existence of an $L_1$-isometry implementing the channel on both input states.
\end{abstract}

\maketitle

\section{Introduction}
Trace inequalities are extremely powerful in studying quantum information and probabilities. Often a classical inequality that follows from functional calculus will yield a quantum generalization from an inequality on traces of matrix products. A well-known example is the Golden-Thompson inequality, stating that for a pair of Hermitian matrices $x,y$,
\begin{equation}
\tr \exp(x + y) \leq \tr(\exp(x) \exp(y)) \pl.
\end{equation}
For classical vectors or simultaneously diagonalizable matrices, the equality holds almost trivially. In \cite{sutter_multivariate_2017}, Sutter, Berta, and Tomamichel generalize the Golden-Thompson inequality to show that for Hermitian matrices $\{H_k\}_{k=1}^n$ and $p \geq 1$,
\begin{equation} \label{eq:gtsutter}
\ln \Big \| \exp \sum_{k=1}^n H_k  \Big \|_{p}
	\leq \int_{\RR} dt \beta_0(t) \ln \Big \| \prod_{k=1}^n \exp((1 + it) H_k) \Big \|_{p} \pl,
\end{equation}
where $\| \cdot \|_p$ is the Schatten $p$-norm on matrices, and
\begin{equation} \label{eq:beta0}
\beta_0(t) = \frac{\pi}{2} \big ( \cosh(\pi t) + 1\big )^{-1} \pl.
\end{equation}
This generalized Golden-Thompson inequality follows from a generalization of the Araki-Lieb-Thirring inequality.

The 4-input version of equation \eqref{eq:gtsutter} implies a key result in quantum information theory. The quantum channel is a general model of how the state of an open quantum system changes when interacting with an initially uncoupled environment. Due to this environmental interaction, the effect of a channel is generally not invertible - it may lose information about the system. In some special cases, it is nonetheless possible to recover the original input state. For example, quantum error correction defines a `code space' within a larger system, such that perturbations of states in the code space are effectively invertible \cite{terhal_quantum_2015, barnum_reversing_2000}. In the theory of quantum communication \cite{wilde_quantum_2013, beigi_decoding_2016}, one asks how many bits of information one may recover from the output of a quantum channel with arbitrarily powerful encoding and decoding. Holography in high energy physics relies on a reversible map between bulk and boundary theories \cite{ryu_holographic_2006, cotler_entanglement_2019, penington_replica_2020}.

A key quantity in quantum information is the relative entropy between quantum density matrices, denoted $D(\rho \| \eta)$ for densities $\rho$ and $\eta$. One of the most fundamental inequalities in quantum information theory is the data processing inequality for relative entropy, which states that for any quantum channel $\Phi$,
\[ D(\rho \| \eta) \geq D(\Phi(\rho) \| \Phi(\eta)) \pl. \]
We recall and denote by $R_{\eta, \Phi}$ the Petz recovery map, given by a normalized and re-weighted adjoint of $\Phi$ \cite{petz_sufficient_1986, petz_sufficiency_1988}. It is always the case that $R_{\eta, \Phi} \circ \Phi(\eta) = \eta$. The Petz map for $\eta, \Phi$ sometimes acts as an inverse on $\rho$ as well. In particular,
\begin{equation} \label{eq:petzeq}
D(\rho \| \eta) = D(\Phi(\rho) \| \Phi(\eta)) \iff R_{\eta, \Phi} \circ \Phi(\rho) = \Phi(\rho) \pl.
\end{equation}
The intuition for data processing is that no stochastic or quantum process may increase the distinction between two probability distributions or densities. Equality of relative entropy faithfully indicates that $\Phi$ also doesn't destroy any information in $\rho$ relative to $\eta$.

A natural question is whether a small difference in relative entropy implies approximate recovery. Holographic theories, for instance, consider approximately invertible maps between subsystems of a bulk spacetime and corresponding quantum boundary \cite{cotler_entanglement_2019}. Quantum information applications such as error correction and communication may work with only approximately preserved code spaces, formally outside the strict criteria for perfect recovery via Petz map. A number of recent works have begun to quantitatively link relative entropy difference to fidelity of recovered states.

A resurgence of activity on approximate recovery started with Fawzi and Renner's approximate Markov chain result \cite{fawzi_quantum_2015}. A special form of relative entropy is the conditional mutual information on a tripartite system $A \otimes B \otimes C$, given by
\[ I(A:B|C)_\rho = D \Big (\rho^{ABC} \Big \| \frac{\id}{|A|} \otimes \rho^{BC} \Big ) - D \Big (\rho^{AC} \Big \| \frac{\id}{|A|} \otimes \rho^C \Big ) \pl, \]
where $\rho^{BC}, \rho^{AC}$, and $\rho^C$ refer to respective marginals of $\rho$. Fawzi and Renner show that
\[ I(A:B|C)_\rho \geq - 2 \ln f_1(\rho, R^{FW}(\rho^{AC})) \pl, \]
where $f_1(\rho, \eta) = tr(|\sqrt{\rho} \sqrt{\eta}|)$ is the usual fidelity, and for some channel $R^{FW}$ (not necessarily the Petz map). If one can perfectly recover $\rho^{ABC}$ from $\rho^{AC}$ by acting only on $C$, then the system is called a quantum Markov chain \cite{sutter_approximate_2018}. In \cite{sutter_universal_2016}, the same inequality is shown for a universal recovery map, which depends only on $\rho^{AC}$ rather than on $\rho^{ABC}$. Li and Winter use this form of recovery in \cite{li_squashed_2018} to show a monogamy of entanglement.

Wilde extends approximate recovery to general relative entropy differences in \cite{wilde_recoverability_2015}, showing
\begin{equation} \label{rt}
D(\rho \| \eta) - D(\Phi(\rho) \| \Phi(\eta)) \geq - 2 \ln \Big ( \sup_{t \in \RR} f_1(\rho, R^{t}_{\eta, \Phi} (\rho^{AC})) \Big )
\end{equation}
for a twirled recovery map $R^{t}_{\eta, \Phi}$ parameterized by $t$. In \cite{junge_universal_2018}, Junge, Renner, Sutter, Wilde, and Winter show that
\begin{equation} \label{jungerecov}
D(\rho \| \eta) - D(\Phi(\rho) \| \Phi(\eta)) \geq - 2 \int_{\rz} \ln f_1(\rho, R_{\eta, \Phi}^t(\Phi(\rho))) d \beta_0(t) \pl,
\end{equation}
where $d \beta_0(t) = (\pi/2) (\cosh(\pi t) + 1)^{-1} dt$. Using convexity, one may move the integral inside the logarithm and fidelity to construct the explicit, universal recovery map given by
\begin{equation} \label{eq:univ}
\tilde{R}_{\eta, \Phi}(\rho) = - 2 \int_{\rz} R_{\eta, \Phi}^t(\rho) d \beta_0(t) \pl.
\end{equation}
 Another result by Sutter, Tomamichel and Harrow \cite[Corollary 4.2]{sutter_strengthened_2016, sutter_multivariate_2017} strengthens the inequality as a corollary of equation \eqref{eq:gtsutter}. Let $D_M(\rho \| \eta) := \sup_{\M \in POVMS} D(\M(\rho) \| \M(\eta))$ denote the measured relative entropy. Then
\begin{equation} \label{eq:recmeas}
D(\rho \| \eta) - D(\Phi(\rho) \| \Phi(\eta)) \geq D_M(\rho \| \tilde{R}_{\eta, \Phi} \circ \Phi(\eta))
\end{equation}
for a recovery map $\tilde{R}_{\eta, \Phi}$ as defined in Theorem \ref{entropybound}.

More recently, Carlen and Vershynina show (corollary 1.7 in \cite{carlen_recovery_2020}) that
\begin{equation} \label{cveq}
D(\rho \| \eta) - D(\E(\rho) \| \E(\eta)) \geq \Big ( \frac{\pi}{8} \Big )^4 \| \Delta_{\rho, \eta} \|^{-2} \| R_{\rho, \E}(\E(\eta)) - \eta \|^4_1 \pl,
\end{equation}
where $\Delta_{\rho, \sigma}$ is the relative modular operator, and $\E$ is a conditional expectation that restricts a density to a matrix subalgebra. A recent work by Gily\'en, Lloyd, Marvian, Quek, and Wilde suggests a quantum algorithm that implements the Petz recovery map in special cases \cite{gilyen_quantum_2020}.

For recovery's applications to quantum field theory \cite{witten_aps_2018}, it is desirable to extend finite-dimensional results to infinite-dimensional von Neumann algebras, including type III factors that lack a finite or even semifinite trace. Applications of recovery appear in finite-dimensional analogs of the the Ads/CFT correspondence \cite{cotler_entanglement_2019}. Recovery may underpin eventual proofs of ideas related the the Ryu-Takayanagi conjecture and analogies to error correction, but field theories are widely believed to be type III, non-tracial algebras, in which much of the finite-dimensional quantum information machinery remains conjecture. Two very recent works address the type III extension of recovery maps. One, by Gao and Wilde, extends equations \eqref{rt} and \eqref{cveq} to the von Neumann algebra setting, also addressing generalizations to optimized $f$-divergences \cite{gao_recoverability_2020}. Faulkner, Hollands, Swingle, and Wang prove an equation in the form of \eqref{jungerecov} for subalgebraic restriction/inclusion, with applications in high energy physics \cite{faulkner_approximate_2020}. In a later work, Faulkner and Hollands extend these results to 2-positive channels \cite{faulkner_approximate_2020-1}, and in a follow up, Hollands \cite{hollands_trace-_2021}. derives a result in the form of equation \eqref{eq:recmeas}.

\subsection{Primary Contributions}
In this work, we show how the multivariate trace inequalities of \cite{sutter_multivariate_2017} still hold and apply in arbitrary von Neumann algebras, surprisingly including the non-tracial types. This set of results consists of two inequalities, given as Theorems \ref{alt} and \ref{gt}. These Theorems are similar in form to those of \cite{hollands_trace-_2021} but were derived independently. First, we show a generalized Araki-Lieb-Thirring inequality extending (\cite[theorem 3.2]{sutter_multivariate_2017}) to von Neumann algebras, and slightly generalizing the form of \cite[Corollary 1]{hollands_trace-_2021} to a range of Kosaki norms:
\begin{theorem}[Araki-Lieb-Thirring] \label{alt}
Let $\rho, \eta$ be normal, faithful states on von Neumann algebra $M$, $p \geq 1$, $n \in \NN$, $w \in [0,1]$, and $\{x_k\}_{k=1}^n \subseteq M$ be positive semidefinite, bounded operators,
\begin{equation}
\ln \bigg \|\prod_{k=1}^n x_k^r \bigg \|_{L_{p/r}^w(M, \rho, \eta)}
	\leq r \int_{-\infty}^{\infty} dt \beta_r(t) \ln \bigg \| \prod_{k=1}^n x_k^{1+it} \bigg \|_{L_{p}^w(M, \rho, \eta)} \pl.
\end{equation}
\end{theorem}
The technical version of this Theorem appears as Theorem \ref{alt}. Here the norms are Kosaki $L_p$ norms, given for an operator $x \in M$ by
\begin{equation} \label{eq:kosakintro}
\| x \|_{L^w(M, \rho)} = \| \rho^{(1 - w)/p} x \rho^{w/p}\|_{L_p(M)} \pl.
\end{equation}
The norm $\| \cdot \|_{L_p(M)}$ is the Haagerup $L_p$ space norm, bypassing the potentially traceless nature of the original algebra and reducing to the usual $L_p$ norm for tracial algebras (see Section \ref{sec:haaglp}). The weight $\beta_r$, generalizing $\beta_0$ as in equation \eqref{eq:beta0}, is given by
\begin{equation} \label{eq:betar}
\beta_\theta(t) := \frac{\sin(\pi \theta)}{2 \theta(\cosh(\pi t) + \cos(\pi t))} \pl.
\end{equation}
We also derive an analog of the generalized Golden-Thompson inequality (\cite[corollary 3.2]{sutter_multivariate_2017}, equation \eqref{eq:gtsutter} in this introduction), with a slightly different dependence on $p$. This inequality has a similar but not identical form to that of \cite[Corollary 3]{hollands_trace-_2021}.
\begin{theorem}[Golden-Thompson] \label{gt}
Let $\{H_k\}_{k=1}^n \subseteq M$ be bounded Hermitian operators and $\rho = \exp(H_0) \in L_1(M)$ have full support. Then
\begin{equation}
\ln \bigg \| \exp \Big ( \frac{H_0}{p} + \sum_{k=1}^n H_k \Big ) \bigg \|_{L_p(M)}
	\leq \int_{\RR} dt \beta_0(t) \ln \bigg \| \prod_{k=1}^n \exp((1 + it) H_k) \bigg \|_{L_{p}^1(M,\rho)} \pl.
\end{equation}
\end{theorem}
Almost immediately from the same argument, we obtain a generalization of Lieb's theorem:
\begin{rem} \label{liebthm}
Let $\rho$ be Hermitian such that $\exp(\rho) \in L_1(M)$. Then the function $f : M \rightarrow M$ given  by
\[ f(X) = \| \exp(\rho/p + \ln X) \|_{L_p(M)} \]
is concave on the positive definite cone.
\end{rem}
We then rederive equation \eqref{eq:recmeas} for arbitrary von Neumann algebras. This result is identical to but derived independently from \cite[Theorem 1]{hollands_trace-_2021}.
\begin{theorem} \label{sutterrecov}
Let $D(\rho \| \eta)$ denote the quantum relative entropy between normal states $\rho$ and $\eta$, and $\Phi$ denote a quantum channel (a completely positive, normal map). Let $D_M(\rho \| \eta) := \sup_{\M \in POVMS} D(\M(\rho) \| \M(\eta))$ denote the measured relative entropy. Then
\begin{equation}
D(\rho \| \eta) - D(\Phi(\rho) \| \Phi(\eta)) \geq D_M(\rho \| \tilde{R}_{\eta, \Phi} \circ \Phi(\eta)) \pl,
\end{equation}
where $\tilde{R}_{\eta, \Phi}$ is as in Equation \eqref{eq:univ}.
\end{theorem}

Furthermore, we generalize the universal recovery map in the style of \eqref{jungerecov} to channels on all von Neumann algebras, and for a $p$-generalization of the fidelity similar to that of Liang et al's in equation (2.14) \cite{liang_quantum_2019}, given by
\begin{equation}
f_p(\rho, \eta) = \|\sqrt{\rho} \sqrt{\eta} \|_p \pl.
\end{equation}
We denote a \textit{twirled recovery map} in equivalent form to Wilde's \cite{wilde_recoverability_2015}, but parameterized by complex $z$,
\begin{equation}
R_{\eta, \Phi}^z (\hat{\rho}) \lel \eta^{\bar{z}/2}\Phi^{\dag}(\hat{\eta}^{-\bar{z}/2}\hat{\rho}\hat{\eta}^{-z/2})\eta^{z/2}  \pl,
\end{equation}
and a \textit{logarithmic, twirled $p$-fidelity of recovery} given by
\begin{equation}
FR_{\eta, \Phi}^{z}(\hat{\rho}) = -\ln f_{1/Re(z)}(\rho^{Re(z)}, R_{\eta, \Phi}^{z}(\hat{\rho}^{Re(z)})) \pl.
\end{equation}
For convenience of notation, we may denote $R_z = R^{z}_{\eta, \Phi}$ when $\eta$ and $\Phi$ are clear from context. Our notion of fidelity of recovery is closely related to that considered earlier in the field \cite{berta_fidelity_2016}, though we have included the logarithm in the quantity for convenience. Then we show that:
\begin{theorem} \label{thm:main}
Let $\Phi : \M \rightarrow \N$ be a normal, completely positive map from von Neumann algebra $\M$ to algebra $\N$. Let $\rho, \sigma$ be densities on $\M$. Then
\[ D(\rho \| \eta) \geq D(\Phi(\rho) \| \Phi(\eta)) + 2 p \int_{\rz} FR^{(1+it)/p}_{\eta, \Phi}(\rho) \pl  \beta_0(t) dt \]
for $p \geq 1$.
\end{theorem}
As with equation \eqref{jungerecov}, we can use convexity of the $p$-fidelity and negative logarithm to move the integral inside, constructing an explicit, universal recovery map (see Theorem \ref{cor:univrecov}). Equation \eqref{jungerecov} follows as the $p=1$ case. Theorem \ref{thm:main} follows a more general result for $p$-fidelity of recovery:
\begin{theorem} \label{thm:fidhar}
$FR_{\eta, \Phi}^{z}$ is subharmonic.
\end{theorem}
Theorem \ref{thm:fidhar} is justified by Remark \ref{inH} in section \ref{sec:fidelity}. Theorem \ref{thm:fidhar} converts a mathematical comparison from complex interpolation theory into a direct bound on physical quantities.

For $p=2$ and $M\subset \mathbb{B}(L_2(M))$ represented in so-called standard form \cite{haagerup_standard_1975} we may always assume that $\rho(x)=(\sqrt{d_{\rho}},x\sqrt{d_{\rho}})$ is implemented by its natural `purification.' Then we deduce (see Remark \ref{rem:nonlin}) that
\begin{equation} \label{eqnonlin1}
\|d_{\rho}^{1/2}-R^{1/2}_{\eta, \Phi}(d_{\hat{\rho}}^{1/2})\|_2^2
  \kl D(\rho \|\eta)-D(\Phi(\rho) \|\Phi(\eta)) \pl .
\end{equation}
This implies
\begin{equation} \label{eqnonlin2}
\| d_{\rho}-R^{1/2}_{\eta, \Phi}(d_{\hat{\rho}}^{1/2})^2\|_1^2
 \kl 4  (D(\rho \|\eta)-D(\Phi(\rho) \|\Phi(\eta)))
 \pl .
\end{equation}
Thus using non-linear recovery maps enables us to obtain a quadratic error formula, which qualitatively resembles equation \eqref{cveq} and the results in \cite{gao_recoverability_2020}.

Using the same techniques, we prove a data processing inequality for $p$-fidelity, that for any quantum channel $\Phi$ and pair of states $\rho, \eta$,
\begin{equation}
f_p(\Phi(\rho), \Phi(\eta)) \geq f_p(\rho, \eta) \pl.
\end{equation}

Finally, we derive a new condition for equality in data processing for states with shared support:
\begin{theorem}[Introduction version of \ref{l1iso}] \label{l1intro} Let $\rho, \eta$ be states such that $\rho\kl \la \eta$, and $\Phi:L_1(M)\to L_1(\hat{M})$ be a quantum channel for von Neumann algebras $M,\hat{M}$. Then the following are equivalent
\begin{enumerate}
\item[i)] $D(\Phi(\rho) \|\Phi(\eta))=D(\rho \|\eta)$;
 \item[ii)] There exists a $\eta$-conditioned subalgebra $M_0\subset M$ and an completely positive $L_1$-isometry $u : \hat{M} \rightarrow M_0$ such that
  \[ u(\eta)\lel \Phi(\eta)\pl,\pl u(\rho)\lel \Phi(\rho) \pl .\]
\end{enumerate}
\end{theorem}
Theorem \ref{l1iso} is intuitive for finite-dimensional channels with equivalent input and output spaces, for which perfect recoverability for all states implies unitarity. In the infinite dimensional situation and with different input and output spaces Petz's map gives a precise recovery. However, Theorem \ref{l1iso} improves on Petz's recovery map by providing a local lift from the states space of the output to back to the input, motivated by Kirchberg's work. Assuming equality in an AdS/CFT correspondence, this amount to a an exact lift from boundary to bulk states.

A first, key realization in our method is that the Haagerup $L_p$ spaces as detailed in \cite{haagerup_reduction_2008} can often serve as a substitute for the usual trace. A second is that the interpolation spaces defined by Kosaki \cite{kosaki_applications_1984} coincide with these Haagerup spaces. The trace inequalities in \cite{sutter_multivariate_2017} actually follow two proof strategies, one using traditional information-theoretic techniques that mirror those of \cite{sutter_strengthened_2016}, and another using the complex interpolation methods roots of \cite{junge_universal_2018}. Kosaki's interpolation results let us rederive the main trace inequalities of \cite{sutter_multivariate_2017} with minor adjustments, based on the Kosaki analog of the basic interpolation theorem underlying them (stated as \cite[theorem 3.1]{sutter_multivariate_2017} and in our case as theorem \ref{hir}). These do not lead as quickly to Corollary \ref{sutterrecov}, because the analyticity and definitions of functions such as the operator logarithm are more subtle. Instead, we return to settings with finite trace, and then apply the Haagerup approximation method of \cite{haagerup_reduction_2008} via the continuity results we derived previously in \cite{junge_universal_2020}. This approach suggests the Haagerup approximation as a general method for entropy inequalities beyond tracial settings.

Section \ref{sec:background} reviews the mathematical background of the rest of the text. In Section \ref{sec:traceineq}, we prove the generalized Araki-Lieb-Thirring (Theorem \ref{alt}) and Golden-Thompson (Theorem \ref{gt}) inequalities. In Section \ref{sec:lp}, we re-introduce the rotated recovery maps and show some necessary $L_p$ inequalities for the recovery results. In Section \ref{sec:fidelity}, we introduce the form of $p$-fidelity that will underlie one form of recovery inequality and prove results on differentiation of quantities that will yield the desired relative entropy comparisons. In Section \ref{sec:finiterecov}, we show the finite von Neumann algebra cases of the recovery Theorems \ref{sutterrecov} and \ref{thm:main}. In Section \ref{contsec}, we show continuity bounds on relative entropy, and in Section \ref{sec:approx}, we prove the needed results to approximate relative entropy in type III by entropy in lower-type algebras and to remove assumptions of states sharing support. In Section \ref{sec:genrecov}, we present the technical versions and proofs of the recovery Theorems \ref{sutterrecov} and \ref{thm:main}. In Section \ref{sec:posvec}, we show an analogous recovery bound for Hilbert space vectors. In Section \ref{pfiddp}, we show a data processing inequality for $p$-fidelity. In Section \ref{l1}, we prove the $L_1$-isometry equivalence to saturation of data processing (Theorem \ref{l1intro}). We conclude with Section \ref{sec:conclusion}.

\section{Background} \label{sec:background}

By $\BB(\Hil)$ we denote the bounded operators on Hilbert space $\Hil$, and we will consider general von Neumann algebras of the form $M \subseteq \BB(\Hil)$, including infinite-dimensional and non-separable Hilbert spaces. By $\rho, \eta$ we commonly denote normal, positive semidefinite states in the predual $M_*$, which in finite dimension would be density matrices. By $\id$ we denote the identity operator. By a factor, we refer to a von Neumann algebra with trivial center, the subalgebra of operators that commute with the whole algebra. Physically, we may think of a center as a classical probability space attached to a potentially quantum system.

Von Neumann algebra factors may have type $I_d, I_\infty, II_1; II_\infty; III_0, III_\lambda$, or $III_1$. Type $I_d$ factors are subalgebras of the bounded operators (matrices) on finite-dimensional Hilbert spaces, and type $I_\infty$ arises from the straightforward $d \rightarrow \infty$ limit. We denote the trace in type $I$ by $tr$. In $I_\infty$, $tr(\id) = \infty$ - here the trace is semifinite in the sense of not being infinite on all elements of the algebra, but it is not finite. Type $II_1$ factors are infinite dimensional with a finite, normalized trace $tr$ such that $tr(\id) = 1$. Algebras of type $II_\infty$ have the form $M \otimes \BB(\Hil)$ for $M$ of type $II_1$ and infinite-dimensional $\Hil$. In type $II_\infty$ the trace $tr$ is semifinite, and $\tr(\id) = \infty$.

Algebras of type $III$ are non-tracial, in that there is not even a semifinite trace. For a physically-motivated review of how type III arises, see the hyperfinite construction of $II_1, III_\lambda$, and $III_1$ factors in \cite{witten_aps_2018}. Type III is nonetheless a relevant model of quantum field theory, matching observed divergences of the trace and other features, such as divergent entanglement between spatial subregions.

A von Neumann algebra with non-trivial center is a direct sum of factors. While the full algebra may have mixed type, each factor will have a type as lifted above. Hence to show the results on this paper for general von Neumann algebras, it is sufficient to show that our constructions and results hold consistently on factors of all types. For a thorough treatment of operator algebra theory, see \cite{takesaki_theory_2002}.

\subsection{Basic Modular Theory}
Starting from a von Neumann algebra $M$ and state $\omega$, the GNS construction allows one to define an inner product given by
\begin{equation}
\braket{x|y}_\omega = \omega(x^* y)
\end{equation}
and via completion construct a corresponding Hilbert space and representation of operators in $M$. See \cite{strocchi_introduction_2008} for an introduction with emphasis on physical relevance.

In full generality, a von Neumann algebra $M$ may contain bounded operators from Hilbert space $\Hil$ to Hilbert space $\Hil'$. Let $\ket{\eta} \in \Hil$ and $\ket{\rho} \in \Hil'$ be a pair of normalized vectors for which $\ket{\eta}$ is
\begin{enumerate}
\item Cyclic, in that $\{ a \ket{\eta} : a \in M\}$ is dense in $\Hil$.
\item Separating, in that if $a \in M$ and $a \ket{\eta} = 0$, then $a = 0$.
\end{enumerate}
The Tomita-Takesaki operator $S_{\eta, \rho}$ is given by $S_{\eta, \rho} a \ket{\eta} = a^\dagger \ket{\rho}$. $S_{\eta, \rho}$ has polar decomposition
\[ S_{\eta, \rho} = J_{\eta, \rho} \Delta^{1/2}_{\eta, \rho} \pl, \]
where we call $J_{\eta,\rho}$ the relative modular conjugation. $\Delta_{\eta, \rho}$ is Hermitian and is called the relative modular operator. We define $\Delta_{\eta, \rho}$ for pairs of states (in the algebra sense) $\rho, \eta \in M_*$, letting $\ket{\rho} = \rho^{1/2}$ and $\ket{\eta} = \eta^{1/2}$ as canonical purifications. In finite dimension, $\Delta_{\eta, \rho}(x) = \rho^{-1} x \eta$ for any $x \in \M$. $\Delta^{it}_{\eta, \rho}$ is analogous to unitary time-evolution, leading to the interpretation of $\ln \Delta_{\rho, \eta}$ as a modular Hamiltonian in quantum field theory. For more information on modular theory in physics, see \cite{summers_tomita-takesaki_2005, zhang_tomita-takesaki_2013, witten_aps_2018}.

\subsection{Haagerup Spaces} \label{sec:haaglp}
For a von Neumann algebra $M$ on Hilbert space $\Hil$, faithful state $\rho$, and group $G$, we denote by $M \rtimes G = M \rtimes_{\sigma^\rho} G$ the crossed product of $M$ by $G$ with respect to the modular automorphism group $\sigma = \sigma^{\rho}$. Details of this construction appear in \cite[Section 1.2]{haagerup_reduction_2008}, from which we take all subsequent constructions in this Subsection. $M \rtimes G$ is the von Neumann algebra on $L_2(G, \Hil)$ generated by $\pi_\sigma(x)$ for $x \in M$ and $\lambda(g)$ for $g \in G$, defined by
\begin{equation}
(\pi_\sigma(x) \xi)(h) = \sigma_h^{-1}(x) \xi(h), \pl (\lambda(g) \xi)(h) = \xi(h - g) \pla
	\text{ for } \xi \in L_2(G, \Hil), h \in G \pl.
\end{equation}
$M \rtimes \RR$ is of type $II_\infty$, so there exists a semifinite trace $\tau$ on the crossed product. For the rest of this subsection, we will assume that $G = \RR$. Let $L_0(M \rtimes \RR, \tau)$ denote the topological involutive algebra of all operators on $L_2(\RR, \Hil)$ that are measurable with respect to $(M \rtimes \RR, \tau)$. Let $\hat{\sigma}_t$ be be dual automorphism of $\sigma$ given by
\begin{equation}
\hat{\sigma}_s(\lambda(t)) = e^{i t s} \lambda(t) \text{ for } t \in \RR,
	\pla \hat{\sigma}_s(\pi(x)) = \pi(x) \text{ for } x \in M \pl.
\end{equation}
We then have the Haagerup $L_p$ spaces, given as
\begin{equation}
L_p(M) = \{ x \in L_0(M \rtimes \RR, \tau) : \hat{\sigma}_s(x) = e^{-s/p} x \text{ } \forall s \in \RR\} \pl.
\end{equation}
In particular, $L_\infty(M)$ coincides with $M$. As we will recall in section \ref{sec:koslp}, Haagerup $L_p$ spaces defined for the same $M$ but different $\rho$ are isometric, so we will not explicitly refer to $\rho$ in denoting them. $L_p(M)$ is a 
linear subspace of $M$ and an $M$-bimodule.

The map $\omega \mapsto d_\omega$, which maps a state $\omega \in M_*^+$ to its unique, implementing density in $L_1(M)$, extends to a linear homomorphism from $M_*$ to $L_1(M)$. Here $d_\omega$ is fixed by the relation that $\omega(x) = Tr(x d_\omega)$ for any $x \in M$. Hence one may transfer the norm of $M_*$ to a norm on $L_1(M)$, denoted $\| \cdot \|_{L_1(M)}$. Furthermore, $L_1(M)$ is equipped with a distinguished, contractive linear functional $Tr$, the Haagerup trace, defined by
\begin{equation}
Tr(d_\omega) : = \omega(\id) \text{ for } \omega \in M_* \pl.
\end{equation}
Consequently, $\| \eta \|_1 = Tr(|\eta|)$ for every $\eta \in L_1(M)$. It then holds, as expected, that
\begin{equation} \label{eq:lphagtr}
\|a\|_{L_p(M)} = Tr(|a|^p)^{1/p}, \text{ and } Tr(a b) = Tr(b a)
\end{equation}
for $a \in L_p(M)$, $b \in L_{p'}(M)$, and $1 = 1/p + 1/p'$ as H\"older conjugates. The H\"older inequality holds for Haagerup $L_p$ norms, and $L_p(M)^* = L_{p'}(M)$ for $1 \leq p < \infty$. Finally, for any $a \in L_p$, there is a unique polar decomposition
\begin{equation}
a = u |d_\psi|^{1/p} \pl,
\end{equation}
where $u \in M$, $\psi \in M_*^+$, and $d_\psi$ implements $\psi$ in $L_1(M)$.

If we start with a tracial von Neumann algebra $M$ and construct Haagerup $L_p$ spaces from $(M, tr)$, then we will find that $Tr = tr$. Hence as seen via Equation \eqref{eq:lphagtr}, this $L_p$ space coincides with the expected $L_p$ space or Schatten class on a tracial algebra, with norm given as $\| x \|_p = tr(|x|^{p})^{1/p}$. With respect to the trace in $M \rtimes \RR$, every normalized density in $L_p(M)$ has the same singular numbers, and hence the same distribution, as shown in \cite{fack_generalized_1986}. Nonetheless, for quantities that depend on the $L_p$ norms rather than directly on the detailed spectrum of densities, we are free to use the Haagerup construction everywhere.

Formally, one should distinguish between a state $\rho \in M_*^+$ and its implementing density $d_\rho \in L_1(M)$. We will however often denote $d_\rho$ by $\rho$, such as in Equation \eqref{eq:kosakintro}. As shorthand, we may denote $\| \cdot \|_{L_p(M)}$ by $\| \cdot\|_p$ when the relevant von Neumann algebra is clear from context.

\begin{rem}\label{cc}{\rm   Let $\delta>0$, let $\eta \in M_*^+$ be a normal, faithful state, and assume $\delta \eta \le \rho \le \delta^{-1}\eta$. The operator $d_\eta^{it}$ is a unitary in $M \rtimes G$, not necessarily in $M$. However, the function
 \[ g_{\eta,\rho}(it) := d_{\eta}^{it}d_{\rho}^{-it} \]
satisfies $\hat{\sigma}_s(g_{\eta,\rho}(it))=g_{\eta,\rho}(it)$ and hence does belong to $\pi(M)\cong M$. In fact for $z=\theta+it$, $\theta\le 1/2$ we deduce form
 \[  d_{\eta}^{2\theta}\le \delta^{-2\theta} \pl  d_{\rho}^{-2\theta} \]
that
 \[ \|d_{\eta}^{\theta}d_{\rho}^{-\theta}\|^2 \lel
  \| d_{\rho}^{-\theta}d_{\eta}^{2\theta}d_{\rho}^{-\theta}\| \kl \delta^{-2\theta}  \]
is bounded. This implies that on $\{z|0<\Re(z)<\frac{1}{2}\}$ the function
 \[ g_{\eta,\rho}(z) := d_{\eta}^{z}d_{\rho}^{-z} \]
is well-defined and analytic and, thanks to
 \[ \hat{\sigma}_s( g_{\eta,\rho}(z))
 \lel (e^{zs}d_{\eta}^z)(e^{-zs}d_{\rho}^z) \lel g_{\eta,\rho}(z)  \]
having values in $M$. As noted in \cite{araki_relative_1973}, $g_{\eta,\rho}(it)$ intertwines the modular automorphisms of $\eta$ and $\rho$. Forms of $g_{\eta,\rho}(z)$ appear naturally and usefully in modular theory.

The same argument applies to the modular semigroup
\[ \si_t^{\eta,\rho}(\pi(x)) \lel d_{\eta}^{it}\pi(x)d_{\rho}^{-it} \pl, \]
which satisfies $\theta_s(\si_t^{\eta,\rho}(\pi(x))\lel \si_t^{\eta,\rho}(\pi(x))$ and
 \[ g_{\eta,\rho}(it) \lel \si_t^{\eta,\rho}(\pi(1)) \in \pi(N) \pl .\]
Moreover, let $\si_z^{\eta,\rho}$ be the  unique linear extension of the modular group. Then
 \[ g_{\eta,\rho}(z) \lel \si_z^{\eta,\rho}(1)  \in N \]
at least for $0\le \Re(z)\le 1/2$.}
\end{rem}

\subsection{The Haagerup Reduction}
Like the Haagerup $L_p$ spaces, the reduction method starts with a crossed product. Instead of working with $\rz$, we use the discrete group $G=\bigcup_{n} 2^{-n}\zz\subset \rz$, constructing $\tilde{M} = M \rtimes_{\sigma^\eta} G$ for some normal, faithful state $\eta \in M_*^+$. The advantage of using a discrete group is that we have conditional expectation $\E:\tilde{M}\to M$ given by
\begin{equation} \label{x0}
\E(\sum_g x_{g}\la(g))=x_0 \pl.
\end{equation}
$\E$ is norm-preserving, and a well-known result by Marie Choda \cite{choda_remark_1968, choda_remark_1973} implies that such a conditional expectation may not go from a von Neumann algebra of lower type to one of higher type. Hence $\tilde{M}$ remains of type $III$ and will not allow us to construct Haagerup spaces. Instead, we rely on the following properties (see \cite{haagerup_reduction_2008}):
\begin{enumerate}
\item[Hi)] $\E$ and $\tilde{\eta}=\eta\circ \E$ are faithful.
\item[Hii)] There exists an increasing family of subalgebras $\tilde{M}_k$ and normal conditional expectation $F_k:\tilde{M}\to \tilde{M}_k$ such that $\tilde{\eta}F_k=\tilde{\eta}$;
\item[Hiii)] $\lim_k \|F_k(\psi)-\psi\|_{\tilde{M}_*}=0$ for every normal state $\psi\in \tilde{M}_*$;
\item[Hiv)] For every $k$ there exists a normal faithful trace trace $\tau_k(x)=\tilde{\eta}(d_k(x))$ such that $d_k\in (\tilde{M}_k)_*^+$, and $a_k\le d_k\le a_k^{-1}$ for some scalars $a_k \in \RR^+$. Hence $\tilde{M}_k$ is of type $II_1$.
\end{enumerate}
The Haagerup approximation then yields a method for proving results in type $III$: first, prove the result in type $II_1$, and then show convergence in the limit as $k \rightarrow \infty$.

\subsection{Complex Interpolation} \label{sec:interp}
Within finite-dimensional matrix algebras, many of the desired entropy \cite{sutter_strengthened_2016} and trace \cite{sutter_multivariate_2017} inequalities follow from identifying typical sets of eigenvalues. One can easily imagine that these techniques encounter challenges for infinite-dimensional operators. As noted in \cite{sutter_multivariate_2017}, however, the mathematical technique known as complex interpolation presents an alternate route to many of the same conclusions. Long-studied in operator theory, complex interpolation has strong results that hold without finite-dimensional assumptions. In this chapter, we review the basic tools of complex interpolation that power main results of this paper. For an in-depth treatment of the topic, the reader may consult \cite{bergh_interpolation_1976}.

Two Banach spaces $A_0$ and $A_1$ are compatible if both are subspaces of a Hausdorff topological space $A$. The sum space
\[ A_0 + A_1 := \{x = x_0 + x_1 | x_0 \in A_0, x_1 \in A_1 \} \]
is then a Banach space, equipped with norm
\[ \| x \|_{A_0 + A_1} = \inf_{x_0 \in A_0, x_1 \in A_1} \{\|x_0\|_{A_0} + \|x_1\|_{A_1}\} \pl. \]
Let $S := \{z \in \CC : 0 \leq Re(z) \leq 1\}$ be the vertical strip on the complex plane. By $\F(A_0, A_1)$ we denote the space of functions $f : S \rightarrow A_0 + A_1$ that are bounded on continuous on $S$ and holomorphic on its interior such that
\[ \{f(it) | t \in \RR \} \subset A_0, \{f(1+it) | t \in \RR \} \subset A_1 \pl. \]
$\F(A_0, A_1)$ is again a Banach space with norm
\[ \|f\|_\F = \max \{\sup_{t \in \RR} \|f(it)\|_{A_0}, \sup_{t \in \RR} \|f(1+it)\|_{A_1} \} \pl. \]
For $\theta \in [0,1]$, we define the complex interpolation space
\begin{equation}
[A_0, A_1]_\theta := \{x \in A_0 + A_1 | x = f(\theta), f \in \F(A_0, A_1)\}
\end{equation}
with norm
\begin{equation}
\|x\|_{[A_0, A_1]_\theta } := \inf\{ \|x\|_\F | f(\theta) = x \} \pl.
\end{equation}
For interpolation spaces, we recall Stein's interpolation theorem on norms of maps:
\begin{theorem}[Stein's Interpolation, \cite{bergh_interpolation_1976}]\label{stein}
Let $(A_0,A_1)$ and $(B_0,B_1)$ be two couples of Banach spaces that are each compatible. Let $\{T_z| z\in S\}\subset \BB(A_0+A_1, B_0+B_1)$ be a bounded analytic family of maps such that
\[\{T_{it}|\pl t\in \mathbb{R}\}\subset \BB(A_0,B_0)\pl ,\pl \{T_{1+it}|\pl t\in \mathbb{R}\}\subset  \BB(A_1,B_1)\pl.\]
Suppose $\Lambda_0=\sup_t{\norm{T_{it}}{\B(A_0,B_0)}}$ and  $\Lambda_1=\sup_t{\norm{T_{1+it}}{\B(A_1,B_1)}}$ are both finite, then for $0< \theta < 1$, $T_\theta$ is a bounded linear map from $(A_0,A_1)_\theta$ to $(B_0,B_1)_\theta$ and
\[\norm{T_\theta}{\BB((A_0,A_1)_\theta ,(B_0,B_1)_\theta)}\le \Lambda_0^{1-\theta}\Lambda_1^{\theta}\pl .\]
\end{theorem}
To derive most of the results of this paper, we will rely on a different form of complex interpolation, known Hirschman's strengthening of Hadamard's three line theorem:
\begin{lemma}[Generalized Hirshman / Hadamard, \cite{hirschman_convexity_1952, hiai_generalized_2017}] \label{lem:hir1}
Let $g(z) : S \rightarrow \CC$ be bounded and continuous on $S$ and holomorphic on its interior. Then for $\theta \in [0,1]$,
\begin{equation*}
\ln |g(\theta)| \leq \int_{-\infty}^\infty \ln |g(it)|^{1-\theta} d \beta_{1-\theta}(t)
	+ \int_{-\infty}^\infty \ln |g(1+it)|^\theta d \beta_\theta(t) \pl.
\end{equation*}
\end{lemma}
\noindent Here  $\beta_0(t) dt$ as in equation \eqref{eq:beta0} is obtained as the pointwise limit of the measures $\beta_{\theta}(t) dt$, given in equation \eqref{eq:betar}. For interpolation spaces, the same idea appears as:
\begin{lemma}\label{Hi} Let $A_0, A_1$ be a pair of compatible Banach spaces, and $w = \theta + is$.
 \begin{enumerate}
 \item Let $F : S \rightarrow M$ be an analytic function vanishing at infinity. Then
  \[ \ln \|F(w)\|_{[A_0,A_1]_{\theta}} \kl \int_{\partial S} \ln \|F(z)\|_{[A_0, A_1]_{Re(z)}} d\mu_{w}(z) \pl .\]
 \item $\mu_w(i\rz)=1-\theta$, $\mu_w(1+i\rz)=\theta$;
 \item $\mu_{w|i \RR}=f_w^0H^1$ and $\mu_{w|1+i\rz}=f_w^1H^1$ are absolutely continuous with respect to the $1$-dimensional Hausdorff measure $H_1$, and moreover
  \[ f_w^0(it) \lel \frac{ e^{\pi(s-t)}\sin \pi \theta}{\sin^2(\pi\theta)+(\cos(\pi \theta)-e^{-\pi(s-t)})^2} \pl, \]
and
  \[ f_w^1(1+it) \lel \frac{ e^{\pi(s-t)}\sin \pi \theta}{\sin^2(\pi\theta)+(\cos(\pi \theta)+e^{-\pi(s-t)})^2} \pl . \]
\end{enumerate}
\end{lemma}
\subsection{Kosaki Spaces and Norms} \label{sec:koslp}
In \cite{kosaki_applications_1984}, Kosaki constructs a family of spaces via complex interpolation, which coincide with the Haagerup spaces. In general, we will denote the Kosaki $L_p$ norm of $x \in M$ for a pair of states $\rho, \eta$ with implementing densities $d_\rho, d_\eta \in L_1(M)$ by
\begin{equation} \label{eq:kosnorm}
\| x \|_{L^w_p(M, \rho, \eta)} = \| d_\rho^{(1 - w)/p} x d_\eta^{w/p}\|_{L_p(M)}
\end{equation}
for $w \in [0,1]$. Multiplication by powers of densities in $L_1(M)$ enforces that the normed element is in $L_p(M)$, even if $x \in M$ may not be. When $\eta = \rho$, we denote $L^w(M, \rho) := L^w_p(M, \rho, \rho)$. At $p=\infty$, one can see that the Kosaki norm reduces to $\| \cdot \|_{L_\infty(M)}$. By the left and right Kosaki norms, we refer respectively to the $w = 0$ and $w = 1$ cases.
\begin{prop} \label{kos1} Let $2\le p\le \infty$ and $d_{\omega}$ be the density of a normal faithful state and $\xi_{\omega}$ the GNS vector representing $\omega$. Then $i_p(x \xi_{\omega}) \lel x d_{\omega}^{1/2p}$
extends to isometric isomorphism between $[N,L_2(N)]_{2/p}$ and $L_p(N)$. Moreover, the map
$\iota_\omega^{w,p}(x) \lel d_{\phi}^{(1-w)/p} x d_{\phi}^{w/p}$
extends to an isometry between $ [\iota_{w,1}(N),L_1(N)]_{1/p} $ and $L_p(N)$.
\end{prop}
Kosaki's original construction starts with a von Neumann algebra $M$ and faithful state $\rho$, constructs the space $L_2(M, \rho)$ as the closure of $M$ with respect to the inner product $(a,b)_\rho = \rho(a b^*)$ for $a,b \in M$ and resulting norm, and finally uses the first isomorphism of proposition \ref{kos1} to define $L^w_{p, \rho, \eta}$. Kosaki proves a key isomorphism to the Haagerup spaces:
\begin{theorem}[Kosaki] The map $\iota_p(x)\lel x d_{\rho}^{\frac{1}{p}}$ extends to a completely isometric isomorphism between $L_p^1(\N,\rho)$ and the complemented subspace $L_p(\N e)$ of the Haagerup $L_p$ spaces $L_p(\N)$. Here $e$ is the support of $\rho$.
\end{theorem}
It follows almost immediately that while the choice of reference state in constructing the crossed product may change the map $\omega \mapsto d_\omega$, the Haagerup spaces defined for different such choices are isometric.

For our purpose we need a slight extension of Kosaki's $L_p$ spaces for non-faithful states $\phi$ with support projection $e$. This can easily be obtained by approximation. Let us assume that $N$ is $\si$-finite and $\psi$ is a normal faithful state. Then
 \[ D \lel d_{\phi}+(1-e)d_{\psi}(1-e) \]
is a faithful normal density in $L_1(N)$. Note $D$ commutes with $e$.
\begin{cor}\label{intfam} The norms $\| x \|_{L^1_p(N, \eta)}$ form an interpolation family on $N e$ for $1\le p \le \infty$, as do $\| x \|_{L^0_p(M, \eta)}$ on $e N$.
\end{cor}
\begin{proof} Recall that
 \[ \|x\|_{L_p^1(N,\eta)} \lel \|x \eta^{1/p}\|_{L_p(N)} \]
form an interpolation family and the space $L_p(N)e$ is complemented in the Haagerup  $L_p$ space. Then we observe that
 \[ \iota_{\eta,p}(x)e \lel x \eta^{1/p}e \lel x e \eta^{1/p} \lel x e d_\eta^{1/p}
 \lel \iota_{\eta,p}(x e) \pl .\]
This shows that $R_e(x)=x e$ extends to a contraction from $L_p^1(N,\eta)$ to $L_p^1(Ne,\eta)$.
\qd

We then have a statement of Hirschmann's Lemma for Kosaki $L_p$ norms. Via the re-iteration theorem (see \cite{bergh_interpolation_1976}) and lemma \ref{Hi}:
\begin{lemma} \label{hir}
Let $G : S \rightarrow M$ be analytic, $2 \leq q_0, q_1$, and $1/q(\theta) = (1-\theta)/q_0 + \theta/q_1$. Then for all $\theta$ in the complex strip,
\begin{equation*}
\begin{split}
\ln \|G(\theta)\|_{L_{q(\theta)}^w(M,\rho, \eta)}
  \kl & (1-\theta)\int \ln \|G(it)\|_{L_{q_0}^w(M,\rho, \eta)} \beta_{1 - \theta}(t) dt \\+ & \theta \int \ln \|G(1 + it)\|_{L_{q_1}^w(M,\rho, \eta)} \beta_\theta(t) dt \pl .
\end{split}
\end{equation*}
\end{lemma}
For a finite von Neumann algebra $M$ with identity $\id$, $\| x \|_{L^w_p(M, \id, \id)} = \|x \|_p$ as the usual $p$-norm for any $w \in [0,1]$ and $x \in M$. When $M$ lacks finite trace, $\id \not \in L_1(M)$ by definition. As shown in the following section, it may at times be useful to take Kosaki spaces from finite von Neumann algebras, such as for proving monotonicity of relative entropies. More broadly, rewriting norm inequalities with Kosaki spaces both gives weighted generalizations and helps bypass the distinction between different algebraic types.

\subsection{Quantum Channels}
A quantum channel is a general model of an open quantum process with an initially uncorrelated environment. In tracial settings, a channel is a completely positive, trace preserving (in general, normal) map $\Phi:L_1(M)\to L_1(N)$. Recall that the anti-linear duality bracket
 \[ (x,\rho) \lel Tr(x \rho^*) \]
allows us to identify $\bar{M}_*$ with $L_1(M)$ and hence
 \[ (\Phi^{\dag}(x), \rho)\lel Tr(x \Phi(\rho)^*) \]
defines a normal, unital, completely positive map $\Phi^{\dag} : N \to M$. As denoted, this construction may use the Haagerup trace.

The usual, finite-dimensional Stinespring dilation is one of the core techniques of quantum information theory, rewriting any quantum channel as an isometry followed by a partial trace. Even in semifinite von Neumann algebras, this Stinespring dilation may fail. We replace it by a more general form. The following fact is well-known. Since it is crucial for all our arguments we indicate a short proof.
\begin{lemma}\label{adjoint} Let $\Psi:N\to M$ be a normal completely positive unital map (the dual of a channel). Then there exists a Hilbert space $H$  normal $^*$-homomorphism $\pi:N\to \BB(H)\bar{\ten}M$, and a projection $e\in \BB(H)$ such that
 \[ \Psi(x) \lel (e\ten 1)\pi(x)(e\ten 1) \pl .\]
\end{lemma}

\begin{proof} We will use the standard GNS construction, see \cite{lance_hilbert_1995, paulsen_completely_2003}. Let $\K=N\ten_{\Phi}M$ be the Hilbert $C^*$-module over $M$ with inner product
 \[ (a\ten x,b\ten y) \lel x^*\Psi(a^*b)y \pl .\]
Let $\tilde{\K}$ be the closure of $\K$ in the strong operator topology of the module (see \cite{junge_noncommutative_2005}). Then $\tilde{\K}$ admits a module basis and hence is of the form $\tilde{K} \lel f(H\bar{\ten}M)$ for some projection $f\in \BB(H)\bar{\ten}M$. The subspace $1\ten M\subset \K$ is an $M$ right module and hence the orthogonal projection $q$ onto $(1\ten M)$ is in in $(M^{op})'=\mathbb{B}(H)\bar{\ten}M$. We may define $^*$-representation, see \cite{lance_hilbert_1995}
 \[ \pi(\al)(a\ten x) \lel \al a \ten x \]
Then we deduce that for $e=qf$ we have
 \[ \Psi(x) \lel  e\pi(x)e \pl .\]
It remains to show that $\pi$ extends to the strong closure of $\tilde{K}$, and that $\pi$ is normal. For simplicity we assume that $\eta$ is a normal faithful state and define the Hilbert space $L_2(\K,\eta)$ via the inner product
 \[ (\xi,\phi)_{\eta} \lel \eta((\xi,\phi)) \pl .\]
Note that $L_2(\tilde{K},\eta)=L_2(\K,\eta)$ and the inclusion $\tilde{K}^{\eta}\subset L_2(\tilde{K},\eta)$ is dense, faithful because $\eta$ is faithful. Then we see that for all $a,b,x,y$ the function
 \[ \om_{a,b,x,y}(\al)\lel \eta(x^*\Psi(a^*\al b)y) \]
is normal, thanks to $\Psi$ being normal. By norm approximation, we deduce that $\pi$ extends to a normal representation on $\L_2(\K,\eta)=L_2(\tilde{K},\eta)$. Since this is true for all $\eta$, we see that $\pi$ extends to a representation on the closure $\tilde{K}$. Finally, we observe that weak$^*$ closure of the adjointable maps on $\tilde{K}$ satisfies
 \[ \L_w(\tilde{K}) \lel e(\mathbb{B}\bar{\ten}M)e \pl .\]
Since our map $\pi:N\to \L_w(\tilde{K})$ is normal, we see that, after identification, that  $\pi:N\to (\mathbb{B}\bar{\ten}M)$ is a normal, not necessarily unital $^*$-homomorphism.
\qd

\subsection{Relative Entropy}
In any von Neumann algebra $M$, we define the relative entropy
\begin{equation}
D(\rho \| \eta) : = \braket{\rho^{1/2}, \ln \Delta_{\eta, \rho} \rho^{1/2}} \pl,
\end{equation}
where the inner product is given by the GNS construction for an (algebra, state) pair $(M, \omega)$ if needed. With a semifinite trace, there is an equivalent form
\[ D(\rho \| \eta) = tr(\rho(\log \rho - \log \eta)) \]
that is more familiar in quantum information.

We may write a wide variety of generalized R\'enyi entropies in terms of the Kosaki norms of $g^{1/p'}_{\rho, \eta}$, were $p'$ is the H\"older conjugate of $p$. In particular, we recall the $\alpha-z$ R\'enyi entropies defined and analyzed in \cite{audenaert_-z-renyi_2015, zhang_equality_2020, zhang_wigner-yanase-dyson_2020, chehade_saturating_2020} for real $\alpha, z \geq 0$, given (up to a constant from taking the natural rather than base 2 logarithm) by
\begin{equation} \label{eq:azorig}
D_{\alpha, z}(\rho \| \eta) =
	\frac{1}{\alpha - 1} \ln tr \Big ( \big ( \rho^{\alpha / z} \sigma^{(1-\alpha)/z} \big )^z \Big ) \pl.
\end{equation}
We recall that when $\alpha = z$, this form recovers the sandwiched R\'enyi relative $\alpha$-entropy \cite{muller-lennert_quantum_2013, wilde_strong_2014, jencova_renyi_2018}, and when $z = 1$, the Petz-R\'enyi relative $\alpha$-entropy \cite{petz_quasi-entropies_1986}. When $\alpha = z = 1$, these forms coincide as the usual relative entropy. Via the Kosaki norm, we re-express the $\alpha-z$ R\'enyi entropy for $z \geq \alpha \geq 1$ as
\begin{equation}
D_{\alpha, z}(\rho \| \eta) = \frac{z}{\alpha - 1} \ln \| g_{\rho, \eta}^{1/z'} \|_{L_z^{z-\alpha}(M, \rho, \eta)} \pl,
\end{equation}
where $z'$ is the H\"older conjugate of $z$. When $M$ is a general von Neumann algebra, the Kosaki form is nonetheless a sensible expression. The range of $\alpha$ and $z$ may extend to $\alpha, z \geq 0$ in finite dimension by formally interpreting equation \eqref{eq:kosnorm} for $w \not \in [0,1]$, though it might not always be valid to construct the Kosaki space on arbitrary von Neumann algebras. The Petz-R\'enyi relative $\alpha$-entropy corresponds to the left Kosaki norm, and the sandwiched relative entropy to the right Kosaki norm.

Kosaki $L_p$ spaces provide an extremely convenient tool to prove data processing inequalities for the sandwiched relative entropy. Data processing for $p>1$ was originally shown using other methods in \cite{beigi_sandwiched_2013}. Here we briefly sketch the Kosaki space argument. Let $\Phi:L_1(M)\to L_1(\hat{M})$ be a completely positive trace preserving map and $\eta$ a normal faithful sate, which we call the \emph{reference} state. Let $\hat{\eta}=\Phi(\eta)$ be the image with support $\hat{e}$. By continuity $\Phi(L_1(M))\subset \L_1(\hat{e}\hat{M}\hat{e})\hat{e}$ and hence we and will assume $\hat{e}=1$. We obtain an
induced map $\Phi_{\infty}:M\to \hat{M}$ given by
 \[ \hat{\eta}^{1/2}\Phi_{\infty}(x)\hat{\eta}^{1/2}
 \lel \Phi(\eta^{1/2}x\eta^{1/2}) \]
More generally, it is easy to show by interpolation that the map
 \[ \Phi_p(\eta^{1/2p}x\eta^{1/2p})\lel \hat{\eta}^{1/2p}\Phi_{\infty}(x)\hat{\eta}^{1/2p} \]
is a contraction. Of course interpolation applies exactly because  $\La_p(\eta)=\eta^{1/2p}M\eta^{1/2p}$ is dense in the image of the symmetric Kosaki map $\iota_p^{1/2}:[\iota^{1/2}(M),L_1(M)]_{1/p} \to L_p(M)$.

We refer to \cite{junge_decomposable_2004} for the fact that $\Phi_{\infty}$ is indeed a normal completely positive unital map. Therefore $\Phi_{\infty}$ admits a Stinespring dilation
 \[ \Phi_{\infty}(x) \lel e\pi(x)e \pl , \]
where $\pi:M\to \L(H_{\hat{M}})$ is obtained from the $W^*$-module $M\ten_{\Phi_{\infty}}\hat{M}$.

\begin{lemma} Let $2\le p\le \infty$ and $y\in M$. Then
 \[ \|\pi(y)e\|_{L_{2p}^1(\L,\hat{\eta})}
 \kl \|y\|_{L_{2p}^1(M,\eta)} \pl .\]
\end{lemma}
Indeed, for $p=\infty$ this is obvious and for $p=2$ we have
\begin{align*}
  \|\pi(y)e\|_{2}^2
 &=  \hat{\eta}(e\pi(y^*y)e)
 \lel \hat{\eta}(\Phi_{\infty}(y^*y))
 \lel Tr(\eta^{1/2}\Phi_{\infty}(y^*y)\hat{\eta}^{1/2}) \\
 &= Tr(\Phi(\eta^{1/2}y^*y\eta^{1/2})
 \kl  Tr(\eta^{1/2}y^*y\eta^{1/2}) \pl .
 \end{align*}
Here we only had to use the trace-reducing property of the original map $\Phi$. In combination with Kosaki's embedding result we deduce that
 \begin{align*}
  \|\hat{\eta}^{-1/2p'}\Phi(\eta^{1/2}y^*y\eta^{1/2})\hat{\eta}^{-1/2p'}\|_p
  &= \| \hat{\eta}^{-1/2p'}\hat{\eta}^{1/2}\Phi_{\infty}(y^*y)\hat{\eta}^{1/2}\hat{\eta}^{-1/2p'}\|_p \\
  &= \|\hat{\eta}^{1/2p}\Phi_{\infty}(y^*y)\hat{\eta}^{1/2p}\|_p \\
 &= \|\pi(y)e\|_{L_{2p}^1(\L,\hat{\eta})}^2 \\
 &\le \|y\|_{L_{2p}^1(M,\eta)}^2 \\
 &= \|\eta^{1/2p}y^*y\eta^{1/2p}\|_p \lel
  \|\eta^{-1/2p'}\eta^{1/2}y^*y\eta^{1/2}\eta^{-1/2p'}\|_p \pl .
  \end{align*}
Thus by density we deduce the sandwiched $p$-Renyi data processing inequality:

\begin{theorem}\label{slp} Let $\eta$ be faithful and $1\le p\le \infty$. Then
 \[ \|\Phi(\eta)^{-1/2p'}\Phi(\rho)\Phi(\eta)^{-1/2p'}\|_p
 \kl \|\eta^{-1/2p'} \rho \eta^{-1/2p'}\|_p \]
for all $\rho \in L_1(M)$. Here $\|\cdot\|_p$ may refer to Haagerup $L_p$ norms and $^{-1/2p'}$ to the pseudo inverse on the support. In terms of sandwiched R\'enyi entropy, the inequality is equivalent to
\[ D_p(\Phi(\rho)\|\Phi(\eta)) \kl D_p(\rho\|\eta) \pl. \]
\end{theorem}

\section{Trace Inequalities} \label{sec:traceineq}
From the Kosaki $L_p$ version of Hirschmann's Lemma (Lemma \ref{hir}) follows the Kosaki $L_p$ version of the two main results of \cite{sutter_multivariate_2017}, the extended Araki-Lieb-Thirring (ALT) and Golden-Thompson (GT) inequalities. First, a generalizing reproof of the former from \cite[theorem 3.2]{sutter_multivariate_2017}:
\begin{proof}[(Proof of Theorem \ref{alt})]
Assume for now that $x_k$ are positive definite for all $k$, and that $\rho, \eta$ are faithful. When $r=1$, $\beta_r(t)$ acts like a delta distribution at $0$, and the inequality follows trivially, so suppose $r \in (0,1)$. Let $G(z) := \prod_{k=1}^n x_k^z$. Positive definiteness and boundedness of $x_k$ for all $k \in 1...n$ ensures analyticity of $G$. We apply Lemma \ref{hir} with $\theta = r, q_0 = \infty, q_1 = p$. Then $q_\theta = p/r$,
\[ \theta \ln \| G(1+it) \|_{L_{q_1}^w(M, \rho, \eta)} =
	 r \ln \bigg \| \prod_{k=1}^n x_k^{1+it} \bigg \|_{L_{p}^w(M, \rho, \eta)} \pl, \]
\[ (1-\theta) \ln \| G(it) \|_{L_{q_0}^w(M, \rho, \eta)} =
	(1-r) \ln \bigg \| \prod_{k=1}^n x_k^{it} \bigg \|_{L_{\infty}^w(M, \rho, \eta)} \pl, \]
and
\[ \ln \|G(\theta)\|_{L_{p/r}^w(M, \rho, \eta)} =
	\ln \bigg \| \prod_{k=1}^n x_k^{r} \bigg \|_{L_{p/r}^w(M, \rho, \eta)} \pl. \]
As $\prod_k x_k^{it}$ is unitary, and because the ${L_{\infty}^w(M, \rho, \eta)}$ norm is essentially just the operator norm on $M$,
\[ \ln \bigg \| \prod_{k=1}^n x_k^{it} \bigg \|_{L_{\infty}^w(M, \rho, \eta)} = 0 \pl , \]
completing the Theorem.

If $x_k$ is merely positive semidefinite, we interpret
\[ \bigg \| \prod_{k=1}^n x_k^{1+it} \bigg \|_{L_{p}^w(M, \rho, \eta)} 
	= \lim_{\epsilon \rightarrow 0} \bigg \| \prod_{k=1}^n (x_k + \epsilon \id)^{1+it} \bigg \|_{L_{p}^w(M, \rho, \eta)} \]
for some positive definite $\id$. Then the inequality holds.

If $\rho, \eta$ are not faithful, we interpret $\rho = \rho + \epsilon (1 - e_\rho) \omega, \eta = \eta + \epsilon (1 - e_\eta) \omega$ for a faithful state $\omega \in M_*^+$, and take the limit as $\epsilon \rightarrow 0$, where $e_\rho$ and $e_\eta$ are the respective support projections of $\rho$ and $\omega$.

\end{proof}
The generalized Golden-Thompson inequality \cite[corollary 3.2]{sutter_multivariate_2017} requires a generalized Kato-Lie-Suzuki-Trotter formula. Unfortunately, this result this result is not so simple when we combine elements of a type III von Neumann algebra $M$ with an unbounded element of $L_1(M)$. Instead, we use the Trotter formula in finite algebras with the Haagerup approximation method to extend to the desired result.
\begin{lemma} \label{lkts}
Let $\{H_k\}_{k=1}^n \subseteq M$ be a collection of bounded operators in $M$, $\rho = \exp(H_0)$ be such that $\rho \in L_p(M)$ (equivalently, $\rho^p \in L_1(M)$), and $x_k = \exp(H_k)$ for each $k \in 1...n$. Then:
\begin{enumerate}
	\item $ \al_r \lel  (\rho^{r/2} x_1^{r/2} ... x_{n-1}^{r/2} x_n^r x_{n-1}^{r/2} ... x_1^{r/2} \rho^{r/2} )^{1/r} \in L_p(M)$ and is bounded in $L_p$ norm.
	\item Let $M$ be a finite von Neumann algebra and $x_k$ be bounded. Then
     \[ \lim_r \al_r \lel \exp(H_0+\sum_k H_k) \pl .\] 
\end{enumerate}
\end{lemma}
\begin{proof}
By H\"older's inequality we deduce that 
 \[ \|\al_r\|_{L_p(M)} \kl \prod \|x_k\|_{L_\infty(M)} \|\rho\|_{L_p(M)} \] 
is uniformly bounded. We use the embedding of $L_p(M)$ into $L_{p,\infty}(M\rtimes \rz,tr)$ so that all the $\al_r$ are indeed affiliated to $M\rtimes \rz$. Let $e$ be a spectral projection of $\rho$ so that $\rho e$ is bounded. Using $a\le b$ implies $a^r\le b^r$ we deduce that $\al_re$ is also bounded. By the Trotter formula \cite{suzuki_convergence_1996, suzuki_compact_1997} we deduce for the $\infty$ norm that
 \[ \lim_{r \rightarrow 0} \si(\al_r) e \lel \exp(H_0+\sum_{k=1}^n H_k)e \pl. \] 
This may not hold in the general $L_p$ spaces, where $\rho \in L_1(M)$ is unbounded. By extracting the exponential of a positive multiple of the identity, we can make all $H_0 ... H_k$ effectively negative operators, thereby satisfying the conditions of the Trotter formula. Hence $\al_r$ converges in the measure topology to $\exp(H_0+\sum_{k=1}^n H_k)$. On the image of $L_p(M)$ the norm and the measure topology coincide, so $\al_r$ converges in $L_p$, and definitely weakly to $\exp(H_0+\sum_k H_k)$. Note that \[ \si_s(\exp(H_0+\sum_k H_k))
  \lel \exp(\si_s(H_0)+\sum_k H_k) \pl .\]
Since
 \[ \si_s(\exp(H_0))\lel e^{-s/p}\exp(H_0) \] we deduce that 
  \[ \si_s(H_0) \lel -\frac{s}{p}+H_0 \pl .\]
 This implies 
  \[  \exp(\si_s(H_0)+\sum_k H_k)
  \lel e^{-s/p}\exp(H_0+\sum_k H_k) \pl .\] 
In other words the limit is in $L_p$. Then weak converge already implies
 \[ \|\exp(H_0+\sum_k H_k)\|_{L_p(M)}
 \kl \limsup_{r} \|\al_r\|_{L_p(M)} \pl .\]
This concludes the proof for the Haagerup $L_p$ space.
\end{proof}
\begin{rem} $\exp(H_0+H)$ has to be interpreted very carefully. This can be done using the embedding of $L_1(N)$ into $L_{1,\infty}(N\rtimes \rz)$. Using this formalism, the density for $\exp(H_0+H)$ is the unique positive functional $\psi$ such that 
 \[ g_{\psi, \phi}(it) = (D\psi:D\phi)_t \lel \exp(it(H_0+H))\exp(-it H_0)\]
in the sense of Connes' cocycle. (The actual densities are then obtained by analytic continuation, or by a power series.)
 In \cite{petz_sufficient_1986} this object is defined as $\om^h$ provided $\om(x)=tr(\exp(H_0)x)$. Since the density $\exp(H_0)$ is $L_0$ measurable the logarithm $H_0$ is actually well-defined by functional calculus. This construction is used in the description of relative entropy. \end{rem} 
Due to the subtleties therein, the generalized Golden-Thompson inequality is stated as a Theorem rather than a Corollary.
\begin{proof}[(Proof of Theorem \ref{gt})]
First, we handle the finite case, in which the proof follows simply from that of the original \cite[corollary 3.2]{sutter_multivariate_2017}. Let $x_k = \exp(2 H_k)$ for $k=1...n$. Theorem \ref{alt} implies that
\begin{equation} \label{eq:trot1}
\begin{split}
\ln \bigg \| \prod_{k=1}^n x_k^r \bigg \|_{L_{2 p/ r}^1(M,\rho)} \leq \int dt \beta_r(t) \ln tr (\rho^{1/2 p} A_1^{\frac{1+it}{2}} ... A_{n}^{\frac{1+it}{2}} \Big )^{p} \pl .
\end{split}
\end{equation}
For an operator $y \in L_q(M)$, it will hold generally that $\| y \|_q = \|y^* y\|_{q/2}^{1/2}$. For the Kosaki norms, 
\begin{equation} \label{eq:kosdouble}
 \| y \|_{L^1_{q}(\tilde{M}, \rho)} = \|  y \rho^{1/q} \|_{L_q(M)} = \| \rho^{1/q} y^* y \rho^{1/q} \|_{L_{q/2}(M)}^{1/2} \pl.
\end{equation}
Hence
\begin{equation}
\begin{split}
\ln \bigg \| \prod_{k=1}^n x_k^r \bigg \|_{L_{2 p/ r}^1(\tilde{M},\rho)} & 
= \ln tr \Big (
		\big | x_1^{r/2} ... x_{n-1}^{r/2} x_n^{r/2} \rho^{r/2 p} \big |^{2 p/r}
		\Big )
	\\ & = \ln tr \Big (
		\big (\rho^{r/2 p} x_1^{r/2} ... x_{n-1}^{r/2} x_n^r x_{n-1}^{r/2} ... x_1^{r/2} \rho^{r/2 p} \big )^{p/r}
		\Big ) \pl.
\end{split}
\end{equation}
Compared with \cite[corollary 3.2]{sutter_multivariate_2017}, we must be more careful to show that the limit as $r \rightarrow 0$ exists and converges to something that is still in the correct Haagerup $L_p$ space. Now we consider the family of operators
 \[   \al_r \lel  (\rho^{r/2 p} x_1^{r/2} ... x_{n-1}^{r/2} x_n^r x_{n-1}^{r/2} ... x_1^{r/2} \rho^{r/2 p})^{1/r}      \pl .\]
We apply Lemma \ref{lkts} to complete the finite case, substituting $\rho^{1/p}$ for $\rho$.

Now we consider the general Theorem in arbitrary von Neumann algebras. Let us first indicate the proof for $p=2$. We apply the Haagerup construction for $\phi(x)=tr(dx)$ and assume $tr(d)=1$, i.e. $\phi$ is a normal faithful state. Then $N\rtimes G\lel \bigcup_k M_k$ and there exists conditional expectation $E_k:N\rtimes G$ such that $E_k(x)$ converges strongly to $x$ and $E_k(\psi)$ converges in $L_1$ norm. The good news is that $M_k$ is a finite von Neumann algebra with trace $\tau_k$ and the new extended state $\hat{\phi}$ satisfies
 \begin{enumerate}
 \item[i)] $E_k(\hat{\phi})=\hat{\phi}$;
 \item[ii)] The density $d_k=\exp(H_0(k))$ of $\hat{\phi}$ with respect to $\tau_k$ is bounded from above and below.
 \end{enumerate} 
This allows us to define the new bounded elements $H_j(k)=E_k(H_j)$. In this context Lemma 3.2 2) applies and we can use the Lie-Trotter-Kato formula, and deduce
 \[ \|\exp(\frac{H_0(k)}{2}+\sum_{j=1}^n
 E_k(H_j))\|
 \kl \int d\beta_0(t) \|\prod \exp(1+it E_k(H_j)\|_{L_2^1} \pl .\]
Since $E_k(H_j)$ converges to $H_j$ strongly,
and hence $\exp(itE_k(H_j)$ converges strongly (this series is uniformly absolutely convergence because the elements are uniformly bounded)  the dominated convergence implies convergence to the correct right hand side in $L_p(N\rtimes G)$. Applying the conditional expectation yields the correct upper bound.

Taking the limit for $k\to \infty$ on the left hand side is more problematic, but well known thanks to the work of Araki \cite{araki_relative_1973}. 

Let us denote by $b_k=\sum_{j=1}^n E_k(H_j)$. Then
 \[ \tilde{d}_k \exp(H_0(k)+b_k)^{1/2}\exp(H_0(k)/2)d_k^{1/2}
 \lel \exp(H_0(k)+b_k)^{1/2} \]
is exactly the GNS vector implementing the functional $\phi(k)(x)=\tau_k(\exp(H_0(k)+b_k)x)$ and the relative modular group is given by 
 \[ (D\phi(k):D\hat{\phi})_t
 \lel \tilde{d_k}^{it}d_k^{-it} \pl .\]
This particularly simple formula here is die to the trace. However, the corresponding cocycle also makes the sense in the not necessarily finite von Neumann algebra $N\rtimes G$. Moreover, thanks to the work of Araki, there is a clear interpretation of the density obtained from a bounded perturbation $\om^h$ by a bounded element $h\in N\rtimes G$. More precisely, the implementing vector is given by (see in particular \cite{araki_relative_1973}{Prop 4.12} )
 \[ \xi_{\exp(\log(\hat{d}+h))} = \exp((\log \Delta + h)/2)(\xi_{\hat{\phi}}) \pl. \]
Araki writes down the explicit Feynamann-Katz for this power series and the new density $\Psi(h)$. In the semifinite case, there is no need to use the modular operator $\Delta=L_d^{1/2}R_{d}^{-1/2}$, because the exponential function is additive for commuting operators. Now we may apply \cite[Prop 4.1]{araki_relative_1973} which includes the strong converges of $\Psi(h_k)$ to $\Psi(h)$. This shows that 
 \[  \xi_{\exp(\hat{H}_0+\sum_j H_j)} \lel 
 \Psi(\sum_j H_j/2) \lel \lim_k 
 \Psi(\sum_j E_k(H_j)) \lel \lim_k \xi_{\phi(k)} \pl .\]
Here we use the conditional expectation of $N\rtimes G\to M_k$ to defined the unique embedding on the $L_2$ space level. Thus passing to the limit for $k\to \infty$, the norm estimate remains true, thank to the dominated convergence theorem.

Finally, for other values of $p$, we may use Ricard's estimate of the Mazur map to the strong convergence on the $L_p$ level from rescaling the bounded hamiltonian and the density. This means the estimate is only true for $p\gl 1$.
\end{proof}
\begin{proof}[Proof of Remark \ref{liebthm}]
This inequality is immediate in the finite case, following the arguments of \cite{tropp_joint_2012}. We then apply the continuity argument from the proof of \ref{gt} for $\exp(\rho/p + Y)$, where in this case $Y = \ln X \in \M$. 
\end{proof}
%\begin{cor}
%We obtain an inequality in the form of usual Golden-Thompson for $n=2$,
%\begin{equation*}
%\| \exp(H_0 / p + H) \|_{L_p(M)} \leq \|\exp(H) \|_{L_p^{1/2}(M, \rho)} \pl.
%\end{equation*}
%\end{cor}
%We here see a possible interleaved Kosaki-like norm, which merges $\rho$ with $X$ via the exponential. A norm of the form $\| %\exp ((1/p) \ln \rho + \ln X) \|_{L_p(M)}$ would be finite for all $\rho$ and $X$ that would a sandwiched ($w = 1/2$) Kosaki %norm wherever the Golden-Thompson inequality holds.
\begin{rem}
The generalization of the ALT and GT inequalities to unitarily invariant norms in \cite{hiai_generalized_2017} holds automatically in type I and with small modifications in type II, where there is a semifinite trace. In non-tracial algebras, there may not exist unitarily invariant norms in this sense.
\end{rem}
\begin{rem}
Taking a Kosaki norm on a finite von Neumann algebra $M$, such as of finite dimension or type $II_1$, we have that $\id \in L_p(M)$. In this case, the Haagerup trace $Tr$ coincides with the finite trace $tr$, and we may take the Kosaki norm $\| \cdot \|_{L_p^w(p, \id, \id)}$. Doing so recovers the original ALT and GT inequalities from \cite{sutter_multivariate_2017}.
\end{rem}

\section{$L_p$ Estimates and Recovery Maps for Quantum Channels} \label{sec:lp}
In this section we present a priori estimates on $L_p$ spaces which are required to formulate the recovery Theorem in the von Neumann algebra setting.  The arguments are very closely related to the first author's lecture notes for proving the data processing inequality for the sandwiched entropy.

In the following, we will fix $\Phi:L_1(M)\to L_1(\hat{M})$, $\Psi=\Phi^{\dag}:\hat{\M}\to M$, $e\in \BB(B(H))\bar{\ten}M=\tilde{M}$ and the normal $^*$-homomorphism  $\pi:\hat{M}\to \tilde{M}$.

\begin{lemma}\label{easint} Let $\Phi(\eta)\lel \hat{\eta}$ with support $s(\eta)$, $s(\hat{\eta})$ respectively. Then for all $1\le p\le \infty$.
 \[ \|\pi(y)e s(\eta) \|_{L_{2p}^1(\tilde{M},\eta)}\kl \|ys(\hat{\eta})\|_{L_{2p}(\hat{M},\hat{\eta})} \pl .\]
\end{lemma}

\begin{proof} Since $\Phi$ is trace preserving we note that
 \begin{align*}
 \|\pi(y)e\|_{L_2(\eta))}^2
 &= Tr(d_{\eta}e\pi(y^*y)e) \lel Tr(d_{\eta}\Psi(y^*y))\\
 &= Tr(\Phi(d_{\eta})y^*y) \lel \|y\|_{L_2(\hat{\eta})}^2 \pl .
 \end{align*}
Thus interpolation according to Lemma \ref{intfam} implies the assertion.\qd

\begin{prop}\label{pest} Let $d\in L_1(N)$ be the density of a state $\eta$ and $\hat{d}=\Phi(d)$, with support $s=s(d)$ and $\hat{s}=s(\hat{d})$. Let $1\le p\le \infty$. Then
 \[ R_p(x) \lel d^{1/2p}\Phi^{\dag}(\hat{d}^{-1/2p}x\hat{d}^{-1/2p}) d^{1/2p} \]
extends to contraction from $L_p(\hat{M}))$ to $L_p(M)$.
\end{prop}
\begin{proof} Let us recall the abstract (Markinciewicz) interpolation theorem: Let $(A_0,A_1)\subset V$, $(\hat{A}_0,\hat{A}_1)\subset \hat{V}$ be interpolation couples and $T:A_0+A_1\to \hat{A}_0+\hat{A}_1$ be a linear map such that $T(A_0)\subset \hat{A}_0$ and $T(A_1)\subset \hat{A}_1$. Then
 \[ \|T:A_\theta\to \hat{A}_{\theta}\|\kl \|T:A_0\to \hat{A}_0\|^{1-\theta}
 \|T:A_1\to \hat{A}_1\|^{\theta} \pl .\]
For the proof one considers the analytic function  $G(z)=T(F(z))$, and then takes the infimum over $F$ such that $F(\theta)=x$. In our situation $A_0=\hat{s}\hat{M}\hat{s}$ and $A_1=\hat{s}L_1(\hat{M})\hat{s}$, $\hat{A}_0=sMs$, $\hat{A}_1=sL_1(M)s$. The map is given by
$T(\hat{d}^{1/2}x\hat{d}^{1/2})=
d^{1/2}\Phi^{\dag}(x)d^{1/2}$.
We also use the map $T_{\infty}(x)=s\Phi^{\dag}(x)s$, and observe the following  commuting diagram
 \[ \begin{array}{ccc} \hat{s}\hat{M}\hat{s}&\stackrel{T_{\infty}}{\to} & M \\
 \downarrow_{\iota_{p,\hat{d}}}& &\downarrow_{\iota_{p,d}} \\
  \hat{s}L_p(\hat{M})\hat{s}&\stackrel{R_p}{\to}&L_p(M)\\ \downarrow_{\gamma_{p',\hat{d}}}& &\downarrow_{\gamma_{p',d}} \\
 \hat{s}L_1(\hat{M})\hat{s}&\stackrel{T}{\to}& L_1(M)
 \end{array} \]
Here $\gamma_{p,d}(x)=d^{1/2p'}xd^{1/2p'}$ is chosen such that  $\gamma_{p,d}\iota_{p,d}=\iota_{1,d}$ is the symmetric Kosaki embedding. We may think of $T_{\infty}$ as a densely defined map on $\iota_{1}(\hat{s}\hat{M}\hat{s})$. Thus it remains to show that $\iota_1$ is indeed a contraction. By H\"older's inequality the map $q:L_2(\hat{M})\ten L_2(\hat{M})\to L_1(M)$, $q(x\ten y)=xy$ is a contraction, and indeed a metric surjection, because the adjoint $q^*:\hat{M}\to \BB(L_2(\hat{M}))$ is isometric. The same is true for $\hat{q}(x\ten y)=\hat{s}xy\hat{s}$ as a map $\hat{q}:\hat{s}L_2(\hat{M})\ten L_2(\hat{M})\hat{s}\to \hat{s}L_1(\hat{M})\hat{s}$. Note that $\hat{M}\hat{d}^{1/2}$ is dense in $L_2(M)$. This shows that the set $D_1$ of elements
  \[ \hat{x} \hat{d}^{1/2}xy\hat{d}^{1/2} \quad, \quad \|\hat{d}^{1/2}x\|_2 <1\pl,\pl \|y\hat{d}^{1/2}\|<1  \]
is dense in the unit ball of $\hat{s}L_1(\hat{M})\hat{s}$. Then we recall that
 \begin{align*}
  \|\pi(y)ed^{1/2}\|_2^2 \lel Tr(d\Phi^{\dag}(y^*y)) \lel Tr(\hat{d}y^*y) \lel \|y\hat{d}^{1/2}\|_2^2 \pl .
  \end{align*}
Taking $^*$'s we see that similarly $\|d^{1/2}e\pi(x)\|_2=\|\hat{d}^{1/2}x\|_2$. Let $u\in M$ be contraction. Then we deduce (where $Tr$ is the Haagerup trace) that
 \begin{align*}
 Tr(uT(\hat{d}^{1/2}xy\hat{d}^{1/2}))
 &= Tr(ud^{1/2}\Phi^{\dag}(xy)d^{1/2})\\
 &= Tr(ud^{1/2}e\pi(xy)ed^{1/2}) \\
 &=   (\pi(x)ed^{1/2} ,\pi(y)ed^{1/2}u) \pl .
 \end{align*}
Thanks to the right module property of $L_2(\M)$ we deduce
 \begin{align*}
  |Tr(uT(\hat{d}^{1/2}xy\hat{d}^{1/2})|&\le \|\pi(x)ed^{1/2}\|_{L_2(\tilde{M})}
  \|\pi(y)ed^{1/2}u\|_{L_2(\tilde{M})} \\
 &\le  \|\pi(x^*)ed^{1/2}\|_{L_2(\tilde{M})}
  \|\pi(y^*)ed^{1/2}\|_{L_2(\tilde{M})} \|u\|\\
 &\lel \|u\|\pl  \|\hat{d}^{1/2}x\|_2  \pl \|y\hat{d}^{1/2}\|_2     \pl .
  \end{align*}
Taking the supremum over $\|u\|\le 1$, we    deduce that $T(D_1)$ belongs to the unit ball of $L_1(M)$, and hence $T$ extends to a contraction on $\hat{s}L_1(M)\hat{s}$. By the abstract Markinkiewic theorem we deduce $R_p$ is also a contraction, and the continuous extension of the map $R_p(\hat{d}^{1/2p}x\hat{d}^{1/2p})=d^{1/2p}\Phi^{\dag}(x)d^{1/2p}$. \qd

As an application, we deduce the contraction property of the (twirled) Petz recovery maps, on $L_p$:

\begin{lemma} \label{norme} Let $\eta$ be a state and $\hat{\eta}=\Phi(\eta)$ the image under $\eta$ with support $\hat{e}$. Then
 \[ R_z(\hat{x}) \lel \eta^{\bar{z}/2}\Phi^{\dag}(\hat{\eta}^{-\bar{z}/2}\hat{x}\hat{\eta}^{-z/2})\eta^{z/2} \]
extends to a (completely) bounded  operator on $L_{p(z)}(\hat{M})$ with values in $L_{p(z)}(M)$ for
 \[ \frac{1}{p(z)} \lel Re(z) \pl .\]
\end{lemma}

\begin{proof}
First, we handle the semifinite case. Let $\La_{\hat{\eta},p(z)}=\hat{\eta}^{1/2p(z)}\hat{M}\hat{\eta}^{1/2p(z)}$ be the image of the symmetric Kosaki map in $L_{p(z)}(\hat{e}\hat{M}\hat{e})$. We consider Kosaki's right-sided interpolation space
 \[ L_{2p(z)} \lel [\hat{M}, L_2(\hat{M},\hat{\eta})]_{1/p(z)} \pl .\]
For an element $\hat{x} \in L_{2p(z)}$ of norm $<1$. we can find an analytic function $g(z)\in \hat{M}\hat{e}$ such that
 \[ \|g(it)\|_{\infty} \le 1 \pl ,\pl  \hat{\eta}(g(1+it)^*g(1+it))\le 1 \]
for all $t$. This allows us to consider
 \[ G(z) \lel \pi(g(z))e_N \in \L(H_\M) \]
and deduce that
 \[ \|G(z)\|_{L_{2p(z)}(\L(H_\M),\eta)} \kl 1 \pl .\]
Indeed, this is obvious for $z=it$. For $z=1+it$ we note that
 \begin{align*}
  \| G(1+it)\|_{L_2(\L(\H_\M),\eta)}^2
  &= \|\eta^{1/2} G(1+it)^*G(1+it)\eta^{1/2}\|_1 \\
  &=  Tr(\eta^{1/2}\Phi^{\dag}(g(z+it)^*g(1+it))\eta^{1/2}) \\
  &= Tr(\Phi(\eta)g(1+it)^*g(1+it)) \lel \|g(1+it)\|_{L_2(\hat{M},\hat{\eta})} \le 1 \pl.
  \end{align*}
There  we have shown that $V_z:L_{2p(z)}(\hat{M}\hat{e})\to L_{2p(z)}(\L(H_\M))$,
 \[ V_z(\hat{x}\hat{\eta}^{z/2}) \lel \pi(\hat{x})e\eta^{z/2} \]
extends to a contraction on $L_{2p(z)}(\hat{M}\hat{e})$ with values in $L_{2p(z)}(\L(H_\M))$. Now, we consider an element $\hat{x}\in \La_{p(z),\hat{\si}}(\hat{M})$. Note that
$L_p(\hat{M})=L_{2p}(\hat{M})L_{2p}(\hat{M})$, i.e. we can write $\hat{x}\lel \hat{x}_1\hat{x}_2$ such that $\hat{e}\hat{x}_1=\hat{x}_1$ and $\hat{x}_2\hat{e}=\hat{x}_2$. By the argument above we know that
 \[ \|R_z(\hat{x}_j^*\hat{x}_j)\|_{p(z)}
 \lel \|(V_z(\hat{x}_j)^*V_z\hat{x_j})\|_{p(z)}
  \kl \|V_z(\hat{x}_j)\|_{2p(z)}^2 \kl \|x_j\|_{2p(z)}^2 \]
 holds for $j=1,2$. Therefore
  \[ \|R_z(\hat{x}_2^*\hat{x}_1)\|_{p(z)}
  \kl \|(V_z\hat{x}_2)^*\|_{2p(z)} \|V_z(\hat{x}_1)\|_{2p(z)}
   \kl \|\hat{x}_2\|_{2p(z)} \|\hat{x}_1\|_{2p(z)} \pl .\]
Taking the infimum over all such decompositions, implies the assertion.

In Haagerup spaces, let $z=\theta+it$ and $p=\theta^{-1}$. Then we have a factorization \[ R_z \lel   \si_{-t}^{d}  R_p \si_{t}^{\hat{d}} \pl .\]
Here we use the $L_p$ version of the modular group
 \[ \si_t^d(x) \lel e^{-itd}xe^{itd} \pl.\]
Note that
 \[ \theta_s(\si_t^d(x)) \lel e^{-itd}e^{itd} e^{-s/p}x \lel e^{-s/p}x \pl .\]
Thus, by the definition of the Haagerup $L_p$ space, $\si_t^{d}$ is a contraction with inverse $\si_{-t}^d$.\qd

\section{$p$-fidelities and interpolation} \label{sec:fidelity}

A main tool in our analysis of recovery maps will be given by a new definition of $p$-fidelity from \cite{liang_quantum_2019}
 \[ F_p(x,y) \lel  \frac{\|\sqrt{y}\sqrt{x}\|_p}{\max\|x\|_p,\|y\|_p} \pl \]
and for $x,y\in L_p$
 \[ f_p(x,y) \lel \|\sqrt{x} \sqrt{y}\|_p \pl .\]

\begin{lemma}\label{fup} Let $1\le p\le\infty$ and $\eta$ be faithful. Let $E:\tilde{M}\to M$ be a conditional expectation and
  \[ \tilde{\rho}\lel \rho\circ E\pl ,\pl \tilde{\eta}\lel \eta\circ E \]
such that $\tilde{\eta}$ is also faithful.   Then 
 \[ f_p(\tilde{\rho}^{1/p},\tilde{\eta}^{1/p})
 \lel f_p(\rho^{1/p},\eta^{1/p}) \pl .\] 
 \end{lemma}

\begin{proof} We have to rewrite fidelity by duality as follows
 \begin{align*}
 f_p(x,y) &= \sup_{\|z\|_{p'}\le 1} Tr(z^*x^{1/2p}y^{1/2p})\\
 &=\sup_{\|ay^{1/p'}\|_{p'}\le 1} Tr(y^{1/2}a^*x^{1/2p}y^{1/2}y^{-1/2p}) \\
 &=\sup_{\|ay^{1/p'}\|_{p'}\le 1} Tr(ay^{1/2},\Delta_{x, y}^{1/2p}(y^{1/2}))
  \pl. \end{align*}
According to our assumption $M\subset \tilde{M}$ and also $M_2(M)\subset M_2(\tilde{M})$. According to Connes' 2x2 matrix trick (see \cite{junge_noncommutative_2003}) we know that $L_2(M_2(M))\subset L_2(M_2(\tilde{M}))$. By approximation we may assume that $\rho$ and hence $\tilde{\rho}$ are also faithful. Then $\psi(x)=\frac{\rho(x_{11})+\eta(x_{22})}{2}$ is a faithful state on $M_2(M)$ and $\tilde{\psi}=\psi\circ E$ is the corresponding extension. We also have a canonical embedding $\iota_2:L_2(M_2(M))\to L_2(M_2(\tilde{M}))$ given by $\iota_2(xd_{\psi}^{1/2})=x\tilde{d}_{\psi}^{1/2}$ (see \cite{junge_noncommutative_2003}). Moreover,  we have the following commutation relation \[ \iota_2 \circ \si_t^{\psi} \lel \si_{t}^{\tilde{\psi}}\iota_2 \pl, \] 
which implies
 \[ \iota_2 \Delta_{\psi}^z\lel \Delta_{\tilde{\psi}}^z\iota_2 \pl .\] 
Let us also recall that for the matrix unit $e_{12}=|1\ran\lan 2|$ we have 
 \[ e_{12} \ten \Delta_{\rho,\eta}(\xi) \lel \Delta_{\psi}(e_{12}\ten \xi) \pl .\] 
In particular, $\iota_2(d_{\eta}^{1/2})=d_{\tilde{\eta}}^{1/2}$ and 
 \[  \Delta_{\tilde{\rho},\tilde{\eta}}^{1/2p}(d_{\tilde{\eta}}^{1/2})
  \lel \Delta_{\tilde{\rho},\tilde{\eta}}^{1/2p}(\iota_2(d_{\eta}^{1/2}))
  \lel \iota_2(\Delta_{\rho,\eta}^{1/2p}(d_{\eta}^{1/2}) \pl. \] 
Now, it is easy to conclude. The map $\iota_{p'}(ad_{\eta}^{1/2})=ad_{\tilde{\eta}}^{1/2}$ extends to an isometric embedding of $L_{p'}(M)\subset L_{p'}(\tilde{M})$ and hence
  \[ f_p(\rho^{1/p},\eta^{1/p})\kl f_p(\tilde{\rho}^{1/p},\tilde{\eta}^{1/p}) \pl .\] 
On the other hand for $a \in \tilde{M}$, we see that for $x\in M$ we have 
 \[ (ad_{\tilde{\eta}}^{1/2},xd_{\tilde{\eta}}^{1/2})
 \lel (E(a)d_{\eta}^{1/2},xd_{\eta}^{1/2}) \pl .\]
Since the conditional expectation is extends to a  contraction $E_{p'}(ad_{\tilde{\eta}^{1/p'}})=E(a)d_{\eta}^{1/p'}$, we also find the reverse inequality $f_p(\tilde{\rho}^{1/p},\tilde{\eta}^{1/p})\le f_p(\rho^{1/p},\eta^{1/p})$. 
\qd

\subsection{Interpolation formula for comparable states}

In the following we will assume that $\eta$ and $\rho$ are densities in $L_1(M)$ such that
 \[ \delta \eta \kl \rho \kl \delta^{-1}\eta \pl .\]
Formally we should probably write $d_{\eta}$ for the density such that $\eta(x)=tr(xd_{\eta})$ holds for all $x$, but we decided to follow Takesaki's convention. Let $\Phi:L_1(M)\to L_1(\hat{M})$ be a completely positive and (sub-)trace preserving map, i.e. the dual map $\Phi^{\dag}:\hat{M}\to M$ defined by
 \[ Tr(\Phi^{\dag}(x^*)\eta) \lel Tr(x\Phi(\eta)) \]
is completely positive and (sub-)unital. Let us recall the Stinesping factorization
 \[ \Phi^{\dag}(x)\lel e\pi(x) e \]
for some normal $^*$-homomorphism  $\pi:\hat{M}\to \BB(H)\bar{\ten}M$ and some projection $e\in M'$. We will use the notation $\tilde{M}=e(\BB(H)\bar{\ten}M)e$ and $f$ for the support of $\eta$ and $\hat{f}$ for the support of $\hat{\eta}=\Phi(\eta)$ or $\hat{\rho}=\eta(\rho)$. Indeed, by positivity,
 \[ \delta \Phi(\eta)\le \Phi(\rho)\le \delta^{-1}\eta(\eta)\]
shows that the support projections (both in $\hat{M}$) coincide.

\begin{lemma}\label{fidint} Let $2\le q_0,q_1$ and $\frac{1}{q(\theta)}=\frac{1-\theta}{q_0}+\frac{\theta}{q}$. Let $\beta_{\theta}$ as given in \eqref{eq:betar} represent $\theta$ on the boundary of the strip $\{0\le \Re(z)\le 1\}$. Then
 \[ G(z) \lel \pi(\hat{\rho}^{z/2}\hat{\eta}^{-z/2}\hat{f})ef\eta^{z/2}\rho^{-z/2} \]
is analytic in $\tilde{M}$ and
 \begin{enumerate}
 \item[i)] For all $\theta$ in the complex strip, \begin{align*}
\ln \|G(\theta)\|_{L_{q(\theta)}^1(\tilde{M},\rho)}
  \kl & (1-\theta)\int \ln \|G(it)\|_{L_{q_0}^1(\tilde{M},\rho)} \beta_{1 - \theta}(t) dt \\+ & \theta \int \ln \|G(1 + it)\|_{L_{q_0}^1(\tilde{M},\rho)} \beta_\theta(t) dt;
\end{align*}
 \item[ii)] $\int -\ln \|G(1+it)\|_{q_1} \beta_{\theta}(t) dt \kl \frac{-\ln \|G(\theta)\|_{q(\theta)}}{\theta}$;
 \item[iii)] $\int -\ln \|G(1+it)\|_{q_1} \beta_0(t) dt \kl \liminf_{\theta \to 0} \frac{-\ln \|G(\theta)\|_{q(\theta)}}{\theta} \pl .$
\end{enumerate}
\end{lemma}
\begin{proof} Let us recall that $\mu_{\theta}$ is the unique measure such that
\begin{equation} \label{eq:hir}
f(\theta) \lel (1-\theta)\int f(it) d\mu_{1-\theta}(t)+ \theta \int f(1+it) d\mu_{\theta}(t) \pl .
\end{equation}
Therefore i) is a reformulation of Lemmas \ref{Hi} and \ref{hir} so that % and the fact that $\mu_{\theta}(i\rz)=(1-\theta)$ and  $\mu_{\theta}(1+i\rz)=\theta$, i.e.
\[ d \mu_\theta(1 + i t) = \frac{1}{\theta} \beta_\theta(t) dt, d \mu_{1 - \theta}(t) = \frac{1}{1 - \theta} \beta_{1 - \theta}(t) dt \pl. \]
The analyticity of $G$ follows from Remark \ref{cc} and $\Re(z)\le 1$. For $z=it$ the element $\hat{\rho}^{it}\hat{\eta}^{-it}$ is in $\hat{M}$ and a partial isometry, the same applies to $\eta^{it}\hat{\rho}^{-it}$ and hence
 \[ \|G(it)\|_{L_{q_0}^1(\tilde{M}),\rho)}\kl Tr(\rho) \le 1 \pl .\]
Thus $\ln \|G(it)\|_{q_0}\le 0$. Dividing by $-\theta$ yields ii). The function $h(t)=-\ln \|G(1+it)\|_{q_1}$ is continuous, $\lim_{\theta\to 0} \frac{\sin(\pi \theta)}{\theta}$ converges to $1/\pi$ and the measures $\beta_{\theta}$ are uniformly bounded by $C e^{-|t|}$. Thus the dominated convergence theorem implies the assertion (see \cite{junge_universal_2018} for calculation of $\beta_0$).
\qd
Let us fix $0<q_1<q_0$ and
 \[ \frac{1}{q_{\theta}} \lel \frac{1-\theta}{q_0}+\frac{\theta}{q_1} \pl . \]
 We note that
  \[ \|G(it)\|_{L_{q_0}^1(\rho)}
  \lel \|\pi(g_{\hat{\rho},\hat{\eta}}^{it})e g_{\eta,\rho}^{it}\rho^{1/q_0}\|_{q_0}\kl 1 \pl .\]
and recall Lemma \ref{hir}. Hence
  \begin{align*}
   \int -\ln \|G(1+it)\|_{q_1} \beta_{\theta}(t) dt
  &\le \frac{\ln  \|G(\theta)\|_{q(\theta)}}{\theta}   \pl .
  \end{align*}
Our abstract recovery formula is summarized in the equation:
 \[ -\int \ln \|G(1+it)\|_{q_1} \beta_0(t) dt \kl \liminf_{\theta\to 0}
  \frac{-\ln  \|G(\theta)\|_{q(\theta)}}{\theta} \pl .\]
Before we launch into more fidelity estimates, we need a few $L_p$ norm inequalities. These will allow us to more formally state and prove the result.

\begin{rem}{\rm a)  For semifinite von Neumann algebras the $L_p$ continuity of
 \[ R_z^0(\hat{x}) \lel \eta^{z/2}\Phi^{\dag}(\hat{\eta}^{-z/2}\hat{x}    \hat{\eta}^{-z/2})\eta^{z/2} \]
 is an immediate application of Stein's analytic family interpolation theorem. However, for non-semifinite von Neumann algebras this map is not necessarily well-defined.

b) We have
 \[ R_z(\Phi(\eta)^{Re(z)})\lel \eta^{Re(z)} \]
for all $z$ in the  strip $\{z| 0\le Re(z)\le 1\}$.

c) For $z=\theta+it$ we see that
 \[ R_z \lel \si_{t/2}^{\eta}R_{\theta}\si_{-t/2}^{\hat{\eta}} \]
is indeed a rotated, generalized Petz recovery map.
}
\end{rem}

\begin{lemma}\label{fidid} Let $z=\theta+it$. Then the twirled Petz map (with respect to $\eta$) satisfies
  \[ \|G(z)\|_{L_{1/\theta}^1(\tilde{M},\rho)} \lel f_{1/\theta}(\rho^{\theta},R_z(\Phi(\rho)^{\theta})) \pl .\]
\end{lemma}

\begin{proof} Let $p=1/\theta$. Using the calculation in the Haagerup $L_p$ spaces we deduce from the definition of $R_z$ that

 \begin{align*}
  \|G(z)\|_{L_{p}^1(\tilde{M}, \rho)}^2 &= \|\pi(\hat{\rho}^{z/2}\hat{\eta}^{-z/2})e\eta^{z/2}\rho^{-z/2}\rho^{1/p}\|_{L_{p}(\tilde{M})}^2 = \|\rho^{1/p} G(z)^*G(z)\rho^{1/p}\|_{p/2} \\
  &= \|\rho^{1/2p-\theta}\rho^{-it/2} \eta^{-it/2}\eta^{+1/2p}   \Phi^{\dag}(\hat{\eta}^{it/2}\eta^{-\theta/2}\hat{\rho}^{\theta}\eta^{-\theta/2}\hat{\eta}^{-it/2})
  \eta^{+1/2p}\eta^{-it/2}\rho^{+it/2}\rho^{1/2p-\theta}\|_{p/2} \\
  &= \| \rho^{1/2p}\eta^{1/2p} \si_{t/2}^{\eta}  \Phi^{\dag}(\si_{-t/2}^{\hat{\eta}}( \hat{\eta}^{-\theta/2}\rho^{\theta}\hat{\eta}^{-\theta/2}))\eta^{1/2p} \rho^{1/2p}\|_{p/2} \\
  &= f_p(R_z(\hat{\rho}^{1/p}),\eta^{1/p})^2  \pl .  \qedhere
  \end{align*}
 \qd

\begin{cor} \label{fididcor} Let $z=\theta+it$. Then
 \[ f_{1/\theta}(\eta^{\theta},R_z(\Phi(\rho)^{\theta}))\le 1 \pl .\]
\end{cor}

\begin{proof} By H\"older's inequality,
 \begin{align*}
 \| \rho^{1/2p}\eta^{1/2p} \si_{t/2}^{\eta}  \Phi^{\dag}(\si_{-t/2}^{\hat{\eta}}( \hat{\eta}^{-\theta/2}\rho^{\theta}\hat{\eta}^{-\theta/2})\eta^{1/2p} \rho^{1/2p}\|_{p/2}
 &\le \|\rho^{1/2p}\|_{2p}^2
 \|\si_{t/2}R_p(\si_{-t/2}^{\hat{\eta}}(\hat{\rho}^{1/p}))\|_p \\
 &\le \|R_p(\si_{-t/2}^{\hat{\eta}}(\hat{\rho}^{1/p}))\|_p\
\\
&\le \|\si_{-t/2}^{\hat{\eta}}(\hat{\rho}^{1/p})\|_p \\
&\le \|\rho^{1/p} \|_p \pl .
\end{align*}
We use that $tr(\rho)=1$, the modular group extends to an isometry on $L_p$, and Proposition \ref{pest}. \qd

The analyticity of $G$ allows us to reformulate the interpolation formula for $G$ as an interpolation of complex families of fidelities.

\begin{rem} \label{inH} Theorem \ref{thm:fidhar} then follows from Lemma \ref{fidid} and Lemma \ref{norme}. We use Equation \eqref{eq:hir} as a reformulation of Lemma \ref{Hi} based on Lemma \ref{fidint}, after applying the re-iteration Theorem (see \cite{bergh_interpolation_1976} for more information), which allows us to replace the boundaries of the complex strip $i \RR$ and $1 + i \RR$ by $p_0 + i \RR$ and $p_1 + i \RR$.
\end{rem}

\begin{figure}[htb!] \scriptsize \centering
	\begin{subfigure}[b]{0.50\textwidth}
		\includegraphics[width=0.95\textwidth]{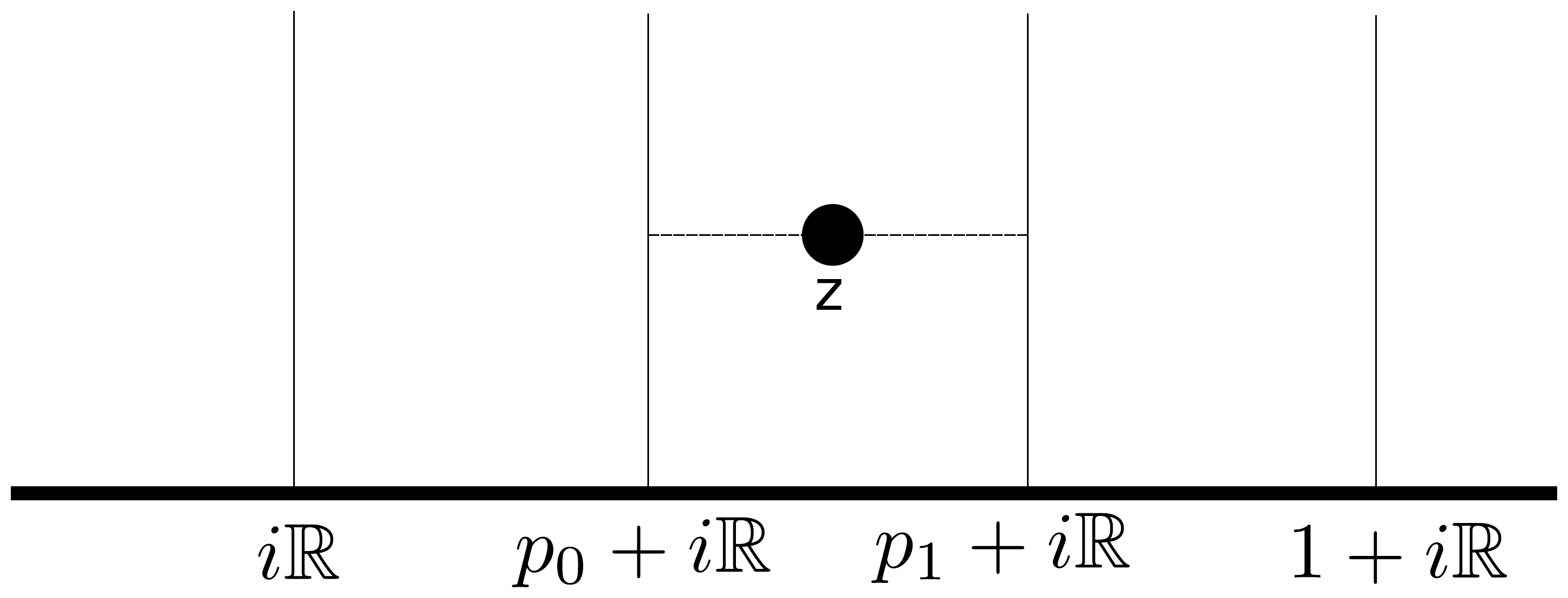}
	\end{subfigure}
	\hspace{10mm}
	%\begin{subfigure}[b]{0.35\textwidth}
	%	\includegraphics[width=0.95\textwidth]{interp2.pdf}
	%\end{subfigure}
	\caption{Using complex interpolation and the re-iteration theorem, we estimate the value of an analytic function at point $z \in \{0 \leq Re(z) \leq 1\}$ by the nearest points along the lines $p_0 + i \RR$ and $p_1 + i \RR$. \label{fig:interpolate}}
\end{figure}

\begin{rem} \label{rem:modular}
Within a finite-dimensional von Neumann algebra $M$, we may relate the Kosaki $p$-norm of $G(z)$ to a $p$-norm expression in terms of modular operators. For any $p$,
\begin{equation*}
\Delta_{\eta,\rho}^{z/2}(\rho^{1/p}) = \rho^{1/p - z/2} \eta^{z/2} = \rho^{-z/2} \rho^{1/p} \eta^{z/2} \pl,
\end{equation*}
and for any $\omega$ and $p$,
\begin{equation*}
\begin{split}
& \|(\hat{\rho}^{z/2}\hat{\eta}^{-z/2} \otimes \id^E) \omega \|_p 
 =  \|(\hat{\eta}^{-z/2} \otimes \id^E) \omega (\hat{\rho}^{z/2} \otimes \id^E)\|_p 
 =  \|(\Delta_{\hat{\rho},\hat{\eta}}^{z/2} \otimes \id^E) \omega\|_p \pl.
%& = \|(\hat{\eta}^{-z/2} \otimes \id^E)U \eta^{z/2}\rho^{-z/2} \sigma^{1/p} (\hat{\rho}^{z/2} \otimes \id^E)\|_p \\
%& = \|(\hat{\rho}^{z/2}\hat{\eta}^{-z/2} \otimes \id^E)U \eta^{z/2} \rho^{-z/2} \sigma^{2/p} \rho^{-z/2} \eta^{z/2} U^\dagger (\hat{\eta}^{-z/2} \hat{\rho}^{z/2} \otimes \id^E) \|_{p/2}
\end{split}
\end{equation*}
Hence
\[ \|G(z)\|_{L_p^1(M,\rho)} = \| (\Delta_{\hat{\rho},\hat{\eta}}^{z/2} \otimes \id^E) U \Delta_{\rho,\eta}^{-z/2} \|_{L_p^1(M,\rho)} \pl, \]
where $U$ is the finite-dimensional Stinespring isometry with environment $E$. This is not clear in type III, where we lack the tracial property. $G(z)$ is a more useful form in type III, due to results we leverage from operator algebras. In particular, we have
\[ G(z) = \pi(g_{\hat{\rho}, \hat{\eta}}^{z/2}) e g_{\eta, \rho}^{z/2} \pl, \]
and we use in proving Lemma \ref{fidint} that $g_{\eta, \rho}^{it}$ and $g_{\hat{\rho}, \hat{\eta}}^{it}$ are respectively in $M$ and $\hat{M}$. As noted in Remark \ref{cc}, $g_{\eta, \rho}$ has good analytic and algebraic properties that work well with the interpolation methods we require. The correspondence between $G(z)$ and its finite- equivalent in terms of modular operators may nonetheless merit future investigation.
\end{rem}

\subsection{Differentiation}

For the twirled recovery map we have to use a suitable differentiation result, first under the additional assumption of regularity  $\delta \eta \le \rho \le \delta^{-1}\eta$. More generally, we differentiate Kosaki norms for smooth functions with values in the underlying von Neumann algebra.

\begin{lemma} \label{diffn}Let $(M, \tau)$ be a finite von Neumann algebra with trace $\tau$. Let $h:I\to M$ be a differentiable function such that $h(0)=\id$. Let $\eta$ be a faithful state. Let $p$ be a differentiable function and $p(0)>1$. Then
 \begin{enumerate}
 \item[i)] $\frac{d}{d \theta} \|\eta^{1/2p(\theta)}h(\theta)\eta^{1/2p(\theta)}\|_{p(\theta)} \Big |_{\theta = 0} \lel  \lim_{\theta \to 0} \theta^{-1}( \|\eta^{1/2p(\theta)}h(\theta)\eta^{1/2p(\theta)}\|_{p(\theta)}-1) \lel -\frac{\eta(h'(0))}{p(0)}$;
 \item[ii)] $\lim_{\theta \to 0} \frac{-\ln \|\eta^{1/2p(\theta)}h(\theta)\eta^{1/2p(\theta)}\|_{p(\theta)}}{\theta}\lel \eta(h'(0))$.
\end{enumerate}
%If in addition $h(\theta)=v(\theta)^*v(\theta)$ and $v(\theta)$ is invertible, then then ii) also holds for $p(0)>1/2$.
\end{lemma}

\begin{proof} We consider $g(\theta)=\|\eta^{1/2p(\theta)}h(\theta)\eta^{1/2p(\theta)}\|_{p(\theta)}^{p(\theta)}$ and assume first that $p(\theta)>1$. We may assume by continuity that $h(\theta) > 0$ in a neighborhood of $\theta = 0$. Let $H(t)=\eta^{1/2p(\theta)}h(t\theta)\eta^{1/2p(\theta)}$. Using the differentiation formula for $p$-norms and convexity, we get for fixed $p=p(\theta)$ that
\begin{align*}
   & g(\theta)-1 = \|H(1)\|_{p}^{p}-\|H(0)\|_{p}^{p}
  \lel  p \int_0^1 \tau(H(t)^{p-1}H'(t)) dt \\
 &= p \theta \int_0^1 \tau(H(t)^{p-1}\eta^{1/2p}h'(t\theta)\eta^{1/2p}) dt \\
 &\lel p\theta \int_0^1 \tau((H(t)^{p-1}-H(0)^{p-1})\eta^{1/2p}h'(t\theta)\eta^{1/2p}) dt
 + p\theta  \int_0^1 \tau(\eta^{\frac{p-1}{p}}\eta^{1/2p}h'(t\theta) \eta^{1/2p}) dt \pl .
\end{align*}
For the second term we observe that
 \[ \tau(\eta^{\frac{p-1}{p}}\eta^{1/2p}h'(t\theta) \eta^{1/2p})\lel
 \tau(\eta h'(t\theta)) \]
and hence
 \[  p \theta \int_0^1 \tau(\eta^{\frac{p-1}{p}}\eta^{1/2p}h'(t\theta) \eta^{1/2p}) dt  \lel p \tau(\eta (h(\theta)-h(0))) \pl .\]
As for the error (first) term, we observe that
 \begin{align*}
 |\tau((H(t)^{p-1}-H(0)^{p-1}) \eta^{1/2p} h'(t\theta) \eta^{1/2p})|
 &\le \|(H(t)^{p-1}-H(0)^{p-1})\|_{p'} \| \eta^{1/2p} h'(t\theta)  \eta^{1/2p}\|_p
 \end{align*}
by H\"{o}lder's inequality. Now, we may use the continuity of the Mazur map, see \cite[Cor 2.3]{ricard_holder_2014} for $\al=p-1$, $p'=\frac{p}{p-1}$ and deduce that
\begin{equation*}
\begin{split}
& \|(H(t)^{p-1}-H(0)^{p-1})\|_{p'}\kl 3 (p-1) \|H(t)-H(0)\|_p \max \{ \| H(t) \|_{p}, \| H(0) \|_{p} \}^{p-2} \\
& \kl 3 (p-1) \|h(t\theta)-h(0)\|_{\infty} \max \{ \| H(t) \|_{p}, \| H(0) \|_{p} \}^{p-2} \\
& \kl 3 (p-1) \|h'\|_{\infty} t\theta \max \{ \| H(t) \|_{p}, \| H(0) \|_{p} \}^{p-2}
  \pl .
\end{split}
\end{equation*}\
We deduce that
 \begin{align*}
 & p\int_0^1 \tau((H(t)^{p-1}-H(0)^{p-1})  \eta^{1/2p} h'(t\theta)  \eta^{1/2p}) dt \\
 &\le \|h'\|_{\infty} 3 p (p-1) \int_0^1 \max \{ \| H(t) \|_{p}, \| H(0) \|_{p} \}^{p-2} \| \eta^{1/2p} h'(t\theta)  \eta^{1/2p}\|_p t\theta  dt\\
 &\le \|h'\|_{\infty} \|\eta^{1/2p} h'  \eta^{1/2p}\|_{\infty} 3 p (p-1) \theta \int_0^1 \max \{ \| H(t) \|_{p}, \| H(0) \|_{p} \}^{p-2} t dt \pl .
 \end{align*}
The faithfulness of $\eta$ and fact that $h(0) = 1$ imply that $\| H(0) \|_p > 0$ for all $p$, so the integral on the right hand side remains finite. As $\theta \rightarrow 0$, this term becomes 0. Thus for $p(0)>1$, we can find $\theta_0$  such that $p(\theta)-1>\delta$ for $\theta\le \theta_0$ and hence
 \[ \lim_{\theta\to 0} \frac{g(\theta)-1}{\theta} \lel p(0) \tau(\eta h'(0)) \pl .\]
Let us now define the function $F(\theta,p)=g(\theta)^{1/p}$ in two parameters. We find that $\frac{d}{d\theta}F=-\frac{1}{p^2}g(\theta)^{1/p-1}g'(\theta)$ and $\frac{dF}{dp}=-\frac{1}{p^2} g(\theta)^{1/p}\ln g(\theta)$. As $\eta$ is faithful, $g(\theta)$ is non-zero when $h(\theta)$ is always positive and not equal to zero. Hence $dF/dp$ is continuous and differentiable. To show that $dF(p, \theta(p))/d \theta$ is continuous and differentiable, we must also check the $dF/d\theta$ part, which involves $g'(\theta)$. We again apply separation of variables. First,
\begin{align*}
\frac{d}{d \theta} \|\eta^{1/2p}h(\theta)\eta^{1/2p}\|_{p}^{p} 
	= \|\eta^{1/2p}h(\theta)\eta^{1/2p}\|_{p}^{p} \bigg (
	\frac{d}{d \theta} \ln \|\eta^{1/2p}h(\theta)\eta^{1/2p}\|_{p} \bigg )
\end{align*}
The prefactor is continuous by the continuity of $g(\theta)$ for $p > 1$. We now use a fact of Banach spaces, that for any continuous, differentiable  function $H(\theta)$ and $p$ fixed,
\[ \frac{d}{d \theta} \|H(\theta)\|_p = \bigg \langle \Big ( \frac{H(\theta)}{\|H(\theta)\|_p} \Big )^{p/p'}, \frac{d}{d \theta} H(\theta) \bigg \rangle \pl. \]
Letting $H(\theta) = \|\eta^{1/2p(\theta)}h(\theta)\eta^{1/2p(\theta)}\|_{p(\theta)}$, left side of the braket is again the Mazur map and therefore continuous. For the right side,
\begin{align*}
 \frac{d}{d \theta}(\eta^{1/2p(\theta)}h(\theta)\eta^{1/2p(\theta)}) = \eta^{1/2p} h' (\theta) \eta^{1/2p} \pl.
\end{align*}
We again see continuity of this expression. Finally, positivity of $\theta$ and the chain rule for the natural logarithm give us continuity of the entire expression. We still however must contend with the $p$ derivative. Here we apply separation of variables yet another time, writing
\begin{align*}
& \frac{d}{d p} \|\eta^{1/2p}h(\theta)\eta^{1/2p}\|_{p}^{p} = \frac{d}{d p} \|\eta^{1/2q}h\eta^{1/2q}\|_{p}^{p} + \frac{d}{dq} \|\eta^{1/2q}h\eta^{1/2q}\|_{p}^{p} \Big |_{p=q} \pl.
\end{align*}
First, we deal with the $p$-derivative, noting that the quantity inside of the norm is assumed $p$-independent. We obtain
\begin{align*}
\frac{d}{d p} \|\eta^{1/2q}h(\theta)\eta^{1/2q}\|_{p}^{p}  = \frac{d}{d p} \tr((\eta^{1/2q}h\eta^{1/2q})^p)
 = \tr((\eta^{1/2q}h\eta^{1/2q})^p \ln(\eta^{1/2q}h\eta^{1/2q}) ) \pl.
\end{align*}
This is finite whenever $\eta^{1/2q}h\eta^{1/2q} > 0$, so this derivative is continuous. For the $q$ derivative,
\begin{align*}
\frac{d}{dq} \|\eta^{1/2q}h\eta^{1/2q}\|_{p}^{p} & = \frac{d}{d q} \tr((\eta^{1/2q}h\eta^{1/2q})^p) = p (\eta^{1/2q}h\eta^{1/2q})^{p-1} \frac{d}{d q} (\eta^{1/2q}h\eta^{1/2q}) \pl.
\end{align*}
Since we only care about continuity and will not rely here on explicitly evaluating this derivative, we merely note that the product rule allows us to differentiate the remaining factor, and that $\eta^{1/2q - 1}$ is finite by the positivity of $\eta$. This term is therefore continuous.

Hence $F$ is differentiable, and
  \[ \frac{d}{d\theta} F(\theta,p(\theta))
  \lel -\frac{1}{p(\theta)^2}g(\theta)^{1/p(\theta)-1}g'(\theta)-\frac{1}{p(\theta)} g(\theta)^{1/p(\theta)} \pl \ln g(\theta)\frac{d p(\theta)}{d\theta} \pl .\]
For $\theta=0$, we deduce from $g(0)=1$ that
 \[ \frac{d}{d\theta} F(\theta,p(\theta))|_{\theta=0} \lel -\frac{1}{p(0)}\eta(h'(0)) \pl .\]
This concludes the proof of i) in this case. For ii) we note that
 \[ \frac{\ln \|\eta^{1/2p(\theta)}h(\theta)\eta^{1/2p(\theta)}\|_{p(\theta)}}{\theta} \lel \frac{1}{p(\theta)} \frac{\ln g(\theta)}{\theta} \]
Using $\frac{d}{d\theta}\ln g(\theta)|_{\theta=0}=\frac{g'(0)}{g(0)}$ we deduce indeed ii).\qd

\begin{theorem}\label{rec1} Let $\delta \eta\le \rho\le \delta^{-1}\rho$ and $1\le p <\infty$. Then 
 \[ \int_{\rz} (-\ln f_{p}(\rho^{1/p},R_{\frac{1+it}{p}}(\Phi(\rho)^{1/p})))\pl  \pl \beta_0(t)dt  
 \kl \frac{D(\rho\|\eta)-D(\Phi(\rho)\|\Phi(\eta))}{2p}  \pl .\]
\end{theorem}

\begin{proof} Let $q\gl 1$ and $q_0>2$. We define $\frac{1}{q(\theta)}=\frac{1-\theta}{q_0}+\frac{\theta}{q}$. Then we may apply  Lemma \ref{fidint} for 
  \[ G_q(z) \lel G(z/q)\lel \pi(\hat{\rho}^{z/2q}\hat{\eta}^{-z/2q}\hat{f})ef\eta^{z/2q}\rho^{-z/2q} \]
which remains analytic as long as $q\gl 1$. Using $\|G_q(it)\|_{q_0}\le 1$, we deduce as in Lemma \ref{fidint} that 
 \[ \lim_{\theta \to 0} \frac{\ln \|G_q(\theta)\|_{L_1^{q(\theta)}}}{\theta}
 \kl \int \ln \|G_q(1+it)\|_{L_{q}^1} \pl \beta_{0}(t) dt \pl  .\] 
Let us recall that, according to Lemma \ref{fidid} we have 
   \[ \|G_q(1+it)\|_{L_{q}^1} \lel f_{q}(\rho^{1/q},R_{\frac{1+it}{q}}(\Phi(\rho)^{1/q})) \pl .\] 
However, we have used the dominated convergence theorem to interchange integral and limit, which is possible thanks to the continuity interpolated fidelity, proved in the next section. We are left to calculate the limit. We may introduce $p(\theta)=\frac{q(\theta)}{2}$ so that $p(0)>1$. Then we see that  
 \begin{align*}
  \|G_q(\theta)\|_{L_1^{q(\theta)}}^2
 &=  \|\rho^{1/q(\theta)} \rho^{-1/2q(\theta)}\eta^{1/2q(\theta)}
 \Phi^{\dag}(\hat{\eta}^{-1/2q(\theta)}\hat{\rho}^{1/q(\theta)}
 \hat{\eta}^{-1/2q(\theta)})\eta^{1/2q(\theta)}
 \rho^{-1/2q(\theta)}\rho^{1/q(\theta)}\|_{p(\theta)} \\
 &= \|\rho^{1/2p(\theta)}h_q(\theta)\rho^{1/2p(\theta)}\|_{p(\theta)} \end{align*}
holds for 
 \[ h_q(\theta) \lel  \rho^{-1/2q(\theta)}\eta^{1/2q(\theta)}
 \Phi^{\dag}(\hat{\eta}^{-1/2q(\theta)}\hat{\rho}^{1/q(\theta)}
 \hat{\eta}^{-1/2q(\theta)}) \eta^{1/2q(\theta)} \rho^{-1/2q(\theta)}
 \lel h \Big (\frac{\theta}{q} \Big ) \pl .\]
For $q=1$, our derivative of 
 \[ h(\theta) \lel \rho^{-\theta/2}\eta^{\theta/2}\Phi^{\dag}(\hat{\eta}^{-\theta/2}\hat{\rho}^{\theta/2}\hat{\eta}^{-\theta/2})
 \eta^{\theta/2}\rho^{-\theta/2} \]
satisfies
\[ h'(0) \lel -\ln \rho + \ln \eta + \Phi^{\dag}(\ln\hat{\rho})-\Phi^{\dag}(\ln \hat{\eta}) \pl.\]
This implies 
 \[ tr(\rho h'(0)) \lel -tr(\rho\ln \rho)+tr(\rho\ln(\eta))
  + tr(\Phi(\rho)\ln\Phi(\rho)-\ln\Phi(\eta))
  \lel -D(\rho|\eta)+D(\Phi(\rho)|\Phi(\eta)) \pl .\]
Using the chain rule, we get
 \[    -qtr(\rho h_q'(0)) \lel D(\rho\|\eta)-D(\Phi(\rho)\|\Phi(\eta)) \pl .\]
\qd
\begin{rem} In a type $III$ situation is is better to write
 \[ h(\theta) \lel \Delta_{\rho,\eta}^{\theta/2} \Phi^{\dag}((\Delta_{\hat{\rho},\hat{\eta}}^{\theta/2})^*\Delta_{\hat{\rho},\hat{\eta}}^{\theta/2})
 \Delta_{\rho,\eta}^{\theta/2}  \]
and hence
 \[ h'(0) \lel -\ln \Delta_{\rho,\eta}+ \Phi^{\dag}(\ln \Delta_{\hat{\rho},\hat{\eta}}) \pl .\]
This implies again 
 \begin{align*}
  tr(\rho h'(0)) &= -(\rho^{1/2},  \ln \Delta_{\rho,\eta}\rho^{1/2})
  + tr(\Phi(\rho)^{1/2},\Delta_{\Phi(\rho),\Phi(\eta)}\Phi(\rho)^{1/2}) \\
  &=-D(\rho\|\eta)+D(\Phi(\rho)\|\Phi(\eta)) \pl .
  \end{align*}
\end{rem} 

\section{Proofs of Recovery Results in Finite Algebras} \label{sec:finiterecov}
At this point, Theorem \ref{rec1} may appear to have nearly finished the proof of a universal recovery Theorem. The remaining technical step is to remove the condition that $\delta \eta \leq \rho \leq \delta^{-1}$, which absolves our analytic machinery from needing to handle infinite relative entropy. Within the finite-dimensional setting, this follows from a straightforward continuity argument. Infinite dimensions introduce additional subtleties with the continuity arguments, and it is not so simple to show that we can drop the restriction that $\delta \phi\le \rho\le \delta^{-1}\rho$. Section \ref{contsec} resolves these issues, extending recovery to type $II_1$. Since the finite-dimensional case is subsumed by these continuity results, we will not include another explicit proof of continuity for the finite case. Instead, we state the result:
 \begin{cor} \label{cor:univrecov}
 Let $1 \le p\le \infty$. Then
\[ \int_{\rz} (-\ln f_{p}(\rho^{1/p},R_{\frac{1+it}{p}}(\Phi(\rho)^{1/p})))\pl  \pl \beta_0(t)dt  
 \kl \frac{1}{2p}(D(\rho\|\eta)-D(\Phi(\rho)\|\Phi(\eta))) \pl .\]
Moreover, the (generally non-linear) universal recovery map
 \[ \tilde{R}_p(x) \lel \Big (\int R_{p,t}(x^{1/p})d\mu(t) \Big)^p \]
 satisfies
  \[ -\ln f_p(\rho, \tilde{R}_{p}(\Phi(\rho)))\kl \frac{1}{2p}(D(\rho\|\eta)-D(\Phi(\rho)\|\Phi(\eta))) \pl .\]
The same holds for the general von Neumann algebra version in Section \ref{sec:genrecov}.
\end{cor}

\begin{proof} We refer to Sections \ref{contsec} and \ref{sec:approx} for the discussion that assuming $\rho\le \la \eta$ is enough and to justify the differentiation Lemma. For the `moreover' part, we recall that $\ln$ is concave and $f_p$ is jointly concave, and hence
 \begin{align*}
& \int \ln f_p(\rho,R_{p,t}(\hat{\rho}^{1/p})^p)d\mu(t)\\
 &\le \ln \int f_p(\rho,R_{p,t}(\hat{\rho}^{1/p})^p)d\mu(t) \\
 &\le \ln f_p \Big (\int \rho d\mu(t),\int R_{p,t}(\hat{\rho}^{1/p})^p d\mu(t) \Big ) \\
 &= \ln f_p(\rho,\tilde{R}_p(\hat{\rho})) \pl .
 \end{align*}
\qd

\subsection{Measured Entropy Recovery}
Though Corollary \ref{cor:univrecov} generalizes \ref{jungerecov} to infinite dimensions, it does not immediately subsume the strengthened form of equation \eqref{eq:recmeas} from \cite{sutter_multivariate_2017}. As this entropy inequality from trace inequalities, we recall this original form of proof and port it to the general von Neumann algebra setting using Theorem \ref{gt}. In the infinite-dimensional setting, we define
 \[ D_M(\rho\|\eta) \lel \sup_{\Phi:L_1(M)\to \ell_1} D(\Phi(\rho)\|\Phi(\eta)) \pl , \]
replacing the POVM by an arbitrary channel from the Haagerup space $L_1(M)$ to the space $\ell_1$ of probability measures. In the finite case, this definition would coincide with that using arbitrary POVMs. We use the following variational forms of relative entropy (see \cite{petz_variational_1988, berta_variational_2017}):
\begin{equation}
D(\rho \| \eta) = \sup_{\omega > 0} tr(\rho \log \omega) + 1 - tr(\exp(\log \eta + \log \omega))
\end{equation}
and of the measured entropy,
\begin{equation} \label{eq:varment}
D_M(\rho \| \eta) = \sup_{\omega > 0} tr(\rho \log \omega) + 1 - tr(\eta \omega) \pl.
\end{equation}
Applying the Golden-Thompson inequality to the final term shows that the measured relative entropy is at most equal to the relative entropy, as does data processing. To justify that this form indeed equals the measured relative entropy as defined:
\begin{lemma}
For states $\rho, \eta$ on a von Neumann algebra $M$, where $M_{sa}$ denotes the subspace of self-adjoint operators in $M$,
\[D_M(\rho \| \eta)=\sup_{\om \in M_{sa}} \rho(\log(\om)) + 1-\eta(\om)\].
\end{lemma} 
\begin{proof} Let $\om$ be a self-adjoint element and $\pi:L_{\infty}(\si(\om),\mu)\to M$ be the normal $^*$-homomorphism. Let $\ez(\cdot)$ denote the expectation of the trace of an expression over values of $\omega$. Then $\pi_*:L_1(M)\to L_1(\mu)$ is a quantum-classical channel. We deduce that  
 \begin{align*}
 D(\Phi(\rho)\|\Phi(\eta))
 &=  \sup_{f} \ez( \Phi(\rho)\log f)+1-\ez ( \exp(\pi(\eta)+f))\\
 &= \sup_{f} \ez (\Phi(\rho)\log f)+1- \ez ( \Phi(\eta)f)\\
 &= \sup_f tr(\rho(\log f(\om))+1-tr(\eta f(\om)) \pl .
 \end{align*}
For $f(z)=z$, we deduce that $D_M$ is bigger than the right hand side, by approximation of $L_1$ by a finite $\si$-algebras. For the converse, we consider a channel $\Phi:L_1(M)\to \ell_1^m$ and $\Phi^*:\ell_{\infty}^m \to M$ which is unital and completely positive. Let $\Phi^*(e_j)=f_j$. Then we find that
  \begin{align*}
   D(\Phi(\rho)\|\Phi(\eta))
 &=  \sup_{\beta_j} \sum_j (e_j,\Phi(\rho))\log \beta_j)+1-\sum_j (e_j,\Phi(\eta))\beta_j \\
 &= tr(\rho \sum_j f_j\log \beta_j)+1-tr(\eta \sum_j f_j\beta_j) \\
 &\le tr(\rho\log(\sum_j f_j\beta_j))+1-gtr(\eta \sum_j f_i \beta_j)
 \pl,
 \end{align*}
thanks to the operator concavity (with respect to unital, completely positive maps) of the logarithm.\qd

Via Lance's Stinespring dilation (see Lemma \ref{adjoint} and \cite{lance_hilbert_1995}), a quantum channel $\Phi : L_1(M) \rightarrow L_1(N)$ has the adjoint form
\begin{equation}
\Phi^\dagger(x) = e \pi(x) e
\end{equation}
for some normal $*$-homomorphism $\pi : N \rightarrow \BB(l_2) \bar{\otimes} M$ and projection $e$, where $e = e_{1,1} \otimes \id_M$, and $\bar{\otimes}$ is the von Neumann algebra tensor product. It also holds for states $\rho, \eta$ when $M$ is finite that
\begin{equation} \label{eq:etrick}
\begin{split}
& tr(\Phi(\rho) \ln \Phi(\eta)) = tr(\rho \Phi^\dagger(\ln \Phi(\eta)))
\\ & = Tr(\rho e \pi(\ln(\Phi(\eta))) e ) = tr(\rho e \ln(\pi(\Phi(\eta))) e) \pl.
\end{split}
\end{equation}
When $M$ is not finite, $\BB(l_2) \bar{\otimes} M$ is not even semifinite, and the above equality may not have meaning. Here we show an entropy bound in the style of the desired recovery inequality (equation \eqref{eq:recmeas}), but where we perturb the quantum states to ensure faithfulness and set up for use in a crossed product $M \rtimes G$.
\begin{theorem} \label{entropybound}
Given $\rho, \eta \in M_*^+$ as states on semifinite von Neumann algebra $M$ and a channel $\Phi : L_1(M) \rightarrow L_1(N)$,
\[ D(\rho \| \eta) - D(\Phi(\rho) \| \Phi(\eta)) \geq
	\int_{\RR} \beta_0(t) D_M(\rho \| R^t_{\eta, \Phi} \circ \Phi(\eta)) dt \geq
	D_M(\rho \| \tilde{R}_{\eta, \Phi} \circ \Phi(\eta)) \pl, \]
where
\[ R^t_{\eta, \Phi}(\omega) = \eta^{(1+it)/2} \Phi^\dagger \big ( \Phi(\eta)^{(1+it)/2} 
	\Phi^\dagger(\omega)
  \Phi(\eta)^{(1-it)/2} \big ) \eta^{(1-it)/2} \pl, \]
and
\[ \tilde{R}_{\eta, \Phi}(\omega) = \int \beta_0(t) R^t_{\eta, \Phi}(\omega) dt \pl. \]
\end{theorem}
\noindent Here is the same rotated Petz map as in \cite{wilde_recoverability_2015}, and $\tilde{R}_{\eta, \Phi}$ is the integrated, rotated  Petz recovery map as in \cite{junge_universal_2018, sutter_multivariate_2017}.
\begin{proof}
Let $\gamma \in (1-e)(\BB(l_2) \bar{\otimes} M)(1-e)$ be a faithful state such that
\[ tr(\gamma (\ln(\pi(\Phi(\eta))) - \ln(\pi(\Phi(\rho))))) < \infty \pl, \]
$\epsilon > 0$, and
\begin{equation*}
\begin{split}
\rho_\epsilon = \rho + \epsilon \gamma = \begin{bmatrix} \rho & 0 \\ 0 & \epsilon \gamma \end{bmatrix}, 
\eta_\epsilon = \eta + \epsilon \gamma = \begin{bmatrix} \eta & 0 \\ 0 & \epsilon \gamma \end{bmatrix} \pl.
\end{split}
\end{equation*}
Let $\hat{\rho}_\epsilon = \rho_\epsilon / tr(\rho_\epsilon), \hat{\eta}_\epsilon = \eta\epsilon / tr(\eta\epsilon)$. We define
\[ c_{tr} := tr(\rho_\epsilon) = tr(\eta_\epsilon) = 1 + \epsilon tr((1-e) \gamma (1-e)) \pl. \]
 We then have that $c_{tr} D(\hat{\rho}_\epsilon \| \hat{\eta}_\epsilon) = D(\rho_\epsilon \| \eta_\epsilon)$. Via the block diagonal form, $D(\rho_\epsilon \| \eta_\epsilon) = D(\rho \| \eta)$.
 
We consider
\begin{equation*}
\begin{split}
I := \frac{1}{c_{tr}} D(\rho_\epsilon & \|
	\exp(\ln \rho_\epsilon - \ln \eta_\epsilon - \ln \pi(\Phi(\rho)) + \ln \pi(\Phi(\eta)))) \\ 
& = \frac{1}{c_{tr}} tr(\rho_\epsilon( \ln \rho_\epsilon - \ln \eta_\epsilon
		- \ln \pi(\Phi(\rho)) + \pi(\Phi(\eta)))) \pl .
\end{split}
\end{equation*}
 
We then use the variational form
\begin{equation} \label{eq:entdiff}
\begin{split}
c_{tr} I = \sup_{\omega \in \BB(l_2) \bar{\otimes} M : \omega > 0} tr(\rho_\epsilon \ln \omega) + 1
	- tr(\exp( \ln \omega + \ln(\pi(\Phi(\eta)))) - \ln(\pi(\Phi(\rho))) - \ln \eta_\epsilon)) .
\end{split}
\end{equation}
To use equation \eqref{eq:entdiff}, we apply the 4-term version of the generalized Golden-Thompson inequality for $p=2$, which states for real, faithful $\exp(H_0) \in \tilde{M}$ and Hermitian $H_1, H_2, H_3$ that
\begin{equation}
\begin{split}
& tr(\exp(H_0 / 2 + H_1 + H_2 + H_3)) \leq \\ &
	\int dt \beta_0(t) \ln Tr \Big ( \exp(H_0 / 2 ) \exp((1+it)H_1 / 2) \exp((1+it) H_2 / 2) \pl \times \\ & \pla
	 \exp(H_3) \exp((1-it) H_2 / 2) \exp((1-it)H_1 / 2) \exp(H_0 / 2 ) \Big )
\end{split}
\end{equation}
using equation \eqref{eq:kosdouble}. We identify
\begin{equation}
H_0 / 2 \leftarrow \ln \omega, \pl H_1 \leftarrow - \ln \eta_\epsilon , \pl 
H_2 \leftarrow  \ln \pi(\Phi(\eta)) , \pl H_3 \leftarrow  - \ln \pi(\Phi(\rho))  \pl.
\end{equation}
Via the supremum and positivity of $\omega$, we can replace $\omega$ by $\sqrt{\omega}$ or $\omega^2$ in equation \eqref{eq:entdiff} without changing the value. Hence
\begin{equation*}
\begin{split}
& c_{tr} I \geq \sup_{\omega > 0} tr(\rho_\epsilon \ln \omega) + 1
\\ & \pla - \int dt \beta_0(t) \ln tr \Big (\eta^{(1+it)/2}_\epsilon \pi(\Phi(\eta))^{(1+it)/2} 
	\pi(\Phi(\rho))
  \pi(\Phi(\eta))^{(1-it)/2} \eta^{(1-it)/2}_\epsilon \omega \Big ) \pl.
\end{split}
\end{equation*}
As $\pi$ is a homomorphism,
\begin{equation*}
\begin{split}
.. = \sup_{\omega > 0} tr(\rho_\epsilon \ln \omega) + 1 & -
	\int dt \beta_0(t) \ln tr \Big (\eta^{(1+it)/2}_\epsilon \pi \big ( \Phi(\eta)^{(1+it)/2} 
	\Phi(\rho)
  \Phi(\eta)^{(1-it)/2} \big ) \eta^{(1-it)/2}_\epsilon \omega \Big ) \pl.
\end{split}
\end{equation*}
Via the supremum over $\omega$, this expression only decreases if we assume that $\omega = e \tilde{\omega} e$ for some $\tilde{\omega}$, and observing that $[e, \eta] = 0$, we have
\begin{equation*}
... \geq \sup_{\tilde{\omega} > 0} tr(\rho \ln \tilde{\omega}) + 1 -
	\int dt \beta_0(t) \ln tr \Big (\eta^{(1+it)/2} \Phi^\dagger \big ( \Phi(\eta)^{(1+it)/2} 
	\Phi(\rho)
  \Phi(\eta)^{(1-it)/2} \big ) \eta^{(1-it)/2} \tilde{\omega} \Big ) \pl.
\end{equation*}
This step conveniently takes care of both eliminating the $\epsilon \gamma$ corrections and resulting in a recovery map form. We may compare directly to $R^t_{\eta, \Phi}$ and to equation \eqref{eq:varment} to see that
\[... = \int_{\RR} \beta_0(t) D_M(\rho \| R^t_{\eta, \Phi} \circ \Phi(\rho)) dt \]
as sought on the right hand side of the recovery inequality. We may also use the concavity of the logarithm to move the integral inside the logarithm, obtaining the sought form in terms of $\tilde{R}_{\eta, \Phi}$.
 
For the left hand side, Using equation \eqref{eq:etrick}, that $\rho = e \rho e$, and that $D(\rho \| \eta)  = tr(\rho(\ln \rho - \ln \eta))$,
\begin{equation} \label{ideff1}
\begin{split}
c_{tr} I & = tr(\rho_\epsilon(\ln \rho_\epsilon - \ln \eta_\epsilon
		- \ln(\pi(\Phi(\rho))) + \ln(\pi(\Phi(\eta)))))
 \\ & = D(\rho_\epsilon \| \eta_\epsilon) - D(\Phi(\rho) \| \Phi(\eta))
 	 + \epsilon tr(\gamma (\ln(\pi(\Phi(\eta))) - \ln(\pi(\Phi(\rho))))) \\
& = D(\rho \| \eta) - D(\Phi(\rho) \| \Phi(\eta)) + \epsilon tr(\gamma (\ln(\pi(\Phi(\eta))) - \ln(\pi(\Phi(\rho))))) \pl.
\end{split}
\end{equation}
Then we note that as $\epsilon \rightarrow 0$, the correction term that is linear in $\epsilon$ vanishes. This limit completes the Theorem. \qd

The obvious barrier in type III is the lack of a trace. Were this the only barrier, the Haagerup $L_p$ spaces and corresponding trace would suffice. The deeper problem is that the differentiability of $h(\theta)$ as used in Lemma \ref{diffn}, and the continuity of the trace of the operator logarithm are not clear without a finite trace. Hence we must approximate the crossed product by finite von Neumann algebras in Section \ref{sec:approx}, our main use of the techniques of \cite{haagerup_reduction_2008}.

\section{Continuity for fidelity of Recovery} \label{contsec}
In this section, we show some continuity results for the fidelity of recovery, which are not immediate in infinite dimension. We continue to use our standard assumptions on $\eta$, $\rho$ and $\Phi$.

\begin{lemma}\label{dconv} Let $A$ be an (possibly unbounded) positive operator on a Hilbert space $H$, $\xi$ in the domain of $A^{1/2}$ and $f_n : \RR \rightarrow \RR$ be sequence of functions such that
 \[ |f_n(x)|\kl C(1+|x|^{1/2}) \]
 and $\lim_n f_n(x)=f(x)$ for all $x$. Then
 \[ \lim_n  \|(f_n(A)-f(A))(\xi)\|_H \lel 0 \pl ,\]
where $f_n$ extends to operators by elementary functional calculus.
\end{lemma}

\begin{proof} Let $d\mu_{\xi}(x)$ be the spectral measure of $A$, i.e.
 \[ (\xi,f(A)\xi) \lel \int f(x)d\mu_{\xi}(x) \pl \]
for all measurable $f$. Then we observe by the triangle inequality that  $|f_n(x)-f(x)|^2 \kl 16 C^2(1+|x|)$ holds for all $\nen$ and moreover,
 \[ \|A^{1/2}\xi\|_H^2 \lel (A^{1/2}\xi,A^{1/2}\xi)
 \lel \int |x| d\mu_{\xi}(x) \]
Since $\xi$ has finite norm, we deduce that $x\mapsto (1+|x|)$ is in $L_1(\mu_{\xi})$. By the dominated convergence theorem, we deduce that
\begin{align*}
 \lim_n \|(f_n(A)-f(A))\xi\|_H^2
 &= \lim_n \int |f_n(x)-f(x)|^2 d\mu_{\xi}(x) \lel 0
 \pl . \qedhere
 \end{align*}
\qd

\begin{prop}\label{contin}  Let $\delta \eta\le \rho\le \delta^{-1}\rho$. Then the function
 \[ F(z) \lel f_{Re(z)}(\rho^{Re(z)},R_{z}(\Phi(\rho)^{Re(z)})) \]
is continuous in $z$ on $\{z| 0<Re(z)\le 1\}$.
\end{prop}
\begin{proof} Here we recall Lemma \ref{dconv} as a general fact.

Let $\rho$ and $\eta$ be states and $\psi(x)=\frac{\eta(x_{11})+\rho(x_{22})}{2}$ the corresponding positive functional on $M_2(M)$ considered by Connes \cite{pedersen_radon-nikodym_1973}. Then 
 \[ \eta^{1/2} \lel \Delta_{\eta,\rho}(\rho^{1/2}) \lel 
 \Delta_{\psi} \bigg ( \kla \begin{array}{cc} 0 &\rho^{1/2} \\  
                                      0&0 \end{array}\mer \bigg ) \]
belongs to the domain of $\Delta_{\psi}^{1/2}$. And hence 
 \[ \lim_{z\to w} \|\Delta_{\psi}^z-\Delta_{\psi}(|1\ran\lan 2|\ten \rho^{1/2})\| \lel 0 \]
as long as $\Re(z),\Re(w)\le 1/2$. Note that thanks to the calculation in the core  $M\rtimes\rz$ we know that 
 \[ \eta^{z/2}\rho^{-z/2}\rho^{1/2}\lel \eta^{z/2}\rho^{1/2}\rho^{-z/2}
 \lel 
\Delta_{\eta,\rho}^{z/2}(\rho^{1/2}) \cong \Delta_{\psi}(|1\ran \lan 2|\ten \rho^{1/2}) \pl .\] 
This means we have $L_2$ convergence in $z$ for $0\le \Re(z)\le 1$. Using Kosaki's interpolation result we deduce that 
 \[  \|(\eta^{z/2}\rho^{-z/2}- \eta^{w/2}\rho^{-w/2})\rho^{1/2p}\|_p
  \kl |(\eta^{z/2}\rho^{-z/2}- \eta^{w/2}\rho^{-w/2})\|_{\infty}^{1-1/p}
  \|(\eta^{z/2}\rho^{-z/2}- \eta^{w/2}\rho^{-w/2})\rho^{1/2}\|_2^{1/p} \pl .\] 
Therefore, we see deduce that $0\le \Re(z),\Re(w)\le 1$ we have 
 \[ \lim_{z\to w} \|(\eta^{z/2}\rho^{-z/2}- \eta^{w/2}\rho^{-w/2})\rho^{1/2p}\|_p \lel 0 \pl \] 
holds uniformly on compact sets. 

% \[ g(p) \lel  \|\rho^{1/2-1/2p}\eta^{1/2p}
% \Phi^{\dag}(\hat{\eta}^{-1/2p}}\hat{\rho}^{1/p)\hat{\eta}^{-\theta/2}}
% \eta^{\theta/2}\rho^{1/2-\theta/2}\|_{p}\] 

Now it is time we address the fidelity. We will use functional calculus and observe that
 \begin{align*} 
  \eta^{z/2}\rho^{-z/2}-\eta^{w/2}\rho^{-w/2}
  &= \eta^{z/2} (1-\eta^{w-z/2}\rho^{(z-w)/z}) \rho^{-z/2}  \pl .
  \end{align*}

Let us define the $^*$ homomorphism $\pi:C(\rz^2)\to B(L_2(M))$ given by $\pi(F_1\ten F_2)=L_{F_1(\rho)}R_{F_2(\eta)}$. Using
$|e^{a}-1|\kl a e^{|a|}$, we observe that

\begin{align*}
 |(x/y)^w-(x/y)^z| &= |(e^{\ln x-\ln y}(w-z)-1)(x/y)^z|
\kl  |w-z| |\ln(x/y)| |(x/y)^z| \pl .
 \end{align*}

Let $\delta\le D \le \delta^{-1}$ be a bounded operator. Using $|e^{x}-1|\kl x e^{|x|}$ and functional calculus we deduce that
 \[ \|D^w-D^z\| \lel \|(e^{(\ln D)w-z}-1)D^z\|
 \kl |w-z| |\ln \delta|  e^{|\ln \delta||w-z|}
\delta^{-|Re(z)|}  \lel |w-z| (\delta^{-1})^{|w-z|+|z|} \pl .\]
This allows us to estimate
 \begin{align*}
  \|G(z)-G(w)\| &= \|\pi(\Delta_{\hat{\rho},\hat{\eta}}^{z/2}) \Delta_{\eta,\rho}^{z/2}-
  \pi(\Delta_{\hat{\rho},\hat{\eta}}^{w/2})\Delta_{\eta,\rho}^{w/2}\|\\
  &\kl 2 (\delta^{-1})^{|w-z|+|z|} |w-z| \pl .
  \end{align*}
Let us know consider the case $p\le p_1$ where $\frac{1}{p}=Re(w)$, $Re(z)=\frac{1}{p_1}$. Then we find that
 \begin{align*}
  \|G(w)\|_{L_{p}(\L(\H_\M),\rho)}
  & \le \| G(w)-G(z)\|_{L_{p}(\L(\H_\M),\rho)} +
  \|G(z)\|_{L_{p}(\L(\H_\M),\rho)} \\
  &\le C(\delta,w,z) |w-z| +
  \|G(z)\|_{L_{p_1}(\L(\H_\M),\rho)} \pl .
    \end{align*}
Since $C(\delta,w,z)$ is bounded in bounded regions of $\cz$, we deduce continuity for $Re(w)\gl Re(z)$. More precisely, we have continuity for fixed $Re(z)$, and moreover,
 \begin{align}
 F(w) &\kl \liminf_{z\to w, Re(w)\gl Re(z) }
  F(z) \kl \limsup_{z\to w, Re(w)\gl Re(z)} F(z)
   \pl , \label{up1}\\
 \limsup_{z\to w, Re(z)\gl Re(w)} F(z) &\kl F(w)  \pl . \label{down1}
 \end{align}
To prove the missing inequality in \eqref{up1}, we may assume $Im(z)=Im(w)=0$. Let us now assume that   $Re(w)=\frac{1}{p}>Re(z)=\frac{1}{p_1}$, i.e. $p_1>p$ for fixed $p$.  Let $p_2\gl 1$,  Then we can find $\epsilon$ such that $\frac{1}{p_1} \lel \frac{1-\epsilon}{p}+\frac{\epsilon}{p_2}$.  We use that standard interpolation estimate and deduce from $\|G(1/p_2)\|\le 1$ that  
 \begin{align*}
  \|G(1/p_1)\|_{p_1}
  \kl 
 \Big (\int_{\rz} f_{p(\epsilon)}(\rho^{1/p},R_{\frac{1 + it}{p}}(\Phi(\rho)^{1/p_1(\epsilon)})) \beta_{\epsilon}(t) dt \Big )^{1-\epsilon} \pl .
  \end{align*}
Here $\frac{1}{q} =\frac{1}{p}-\frac{1}{p_2}$. We may now send $\epsilon \to 0$. Thanks to the continuity in the imaginary part and  the explicit form of the measure (see \cite[p=93]{bergh_interpolation_1976})
 \[ d\mu_{\epsilon}(t) \lel h_{\epsilon}(t)dt \pl ,\pl h_{\epsilon}(t)
  \lel  \frac{e^{-\pi t} \sin \pi \epsilon}{(1-\epsilon) (\sin^2\pi \epsilon+(\cos\pi \epsilon-e^{-\pi t})^2)} \pl \]
we deduce that
\begin{align*}
 \limsup_{\epsilon\to 0} \|G(1/p_1(\epsilon))\|_{p_1(\epsilon)} 
 & \kl \limsup_{\epsilon\to 0} 
 \Big (\int_{\rz} f_{p}(\rho^{1/p},R_{\frac{1 + it}{p}}(\Phi(\rho)^{1/p})) \beta_{\epsilon}(t) dt \Big )^{1-\epsilon} \\
 & \lel  f_{p}(\rho^{1/p},R_{\frac{1 + it}{p}}(\Phi(\rho)^{1/p})) \pl .
 \end{align*} 
This shows that
  \[ \limsup_{z\to w,Re(z)>Re(w)} F(z) \kl F(w) \pl.\]
Similarly, we prove the missing inequality 
  \[ F(w) \kl \liminf_{z\to w, Re(z)> Re(w)} F(z) \pl .\]
in \eqref{down1} using uniform continuity in the imaginary axes. All four inequalities together then yield continuity. \qd

\begin{lemma}\label{ct1} Let $2\le p<\infty$. The function
 \[ h(z) \lel g_{\eta,\rho}(z) \]
is continuous in $L_p^1(M,\rho)$.
\end{lemma}

\begin{proof} We will first prove the assertion for $p=2$. Following Connes we consider $M_2(N)$ and the state $\psi(x)=\frac{1}{2}(\eta(x_{11})+\rho(x_{22}))$. Let $e_{i,j}=|i\ran\lan j|$ be the matrix units in $M_2$. Then we see that
 \[ \Delta_{\psi}(e_{12}\ten \xi) \lel
 e_{12}\ten \eta \xi \rho^{-1}
 \lel e_{12}\ten \Delta_{\eta,\rho}(\xi) \pl .\]
Moreover, $\Delta^{1/2}(\rho^{1/2})\lel \eta^{1/2}$ shows that $e_{12}\ten \rho^{1/2}$ belongs to the domain. Note however, that, thanks to calculation in the core $M\rtimes \rz$ we have
 \[ g_{\eta,\rho}(z)\rho^{1/2}
 \lel \eta^{z/2}\rho^{-z/2}\rho^{1/2}
 \lel \Delta_{\eta,\rho}^{z/2}(\rho^{1/2}) \pl . \]
Let $\lim_n z_n=z$ such that $0\le \Re(z_n)\le 1$. Then $f_n(x)=x^{z_n/2}$ and $f(x)=x^{z/2}$ satisfy the assumption of Lemma \ref{dconv}, and hence we have convergence. For $2< p<\infty$ we deduce from Kosaki's interpolation theorem that also have
 \[ \|a\|_{L_p^1} \kl \|a\|_{L_2^1}^{1-\theta} \|a\|_{\infty}^{\theta} \]
provided $a$ is bounded and $\frac{1}{p}=\frac{1-\theta}{2}$. We apply this to $a=g_{\eta,\rho}(z_n)-g_{\eta,\rho}(z)$ which is uniformly bounded, see Remark \ref{cc}. Therefore, convergence in $L_2$ implies convergence for all $2\le p<\infty$.\qd

\begin{lemma}\label{normea} Let $a\in M$. Then
 \[ h(p) \lel \|a\|_{L_p^1(\rho)} \]
is continuous.
\end{lemma}

\begin{proof} Let $p\le q \le p_0$ and $\theta(q)$ such that
 \[ \frac{1}{q}=\frac{1-\theta}{p}+\frac{\theta}{p_0} \]
Then we deduce from Kosaki's interpolation theorem that
 \[ \|a\|_p \kl \|a\|_q \kl \|a\|_p^{1-\theta(q)}\|a\|_{p_0}^{\theta(q)} \pl .\]
Note that $q$ converges to $p$ iff $\theta(q)$ converges to $0$. This implies the assertion.
\qd

\begin{proof}(\ref{contin}) Let us consider $G_1(z) = \pi(g_{\hat{\rho},\hat{\eta}}(z/2))e$ and $G_2(z) = g_{\eta,\rho}(z/2)$ such that
 \[ G(z) \lel G_1(z)G_2(z)  \pl .\]
Let us the notation $\frac{1}{p(z)}=\Re(z)$.
From the triangle inequality we deduce that
 \begin{align*}
  |\|G(z)\|_{2p(z)}-\|G(w)\|_{2p(w)}|
  &\le |\|G(z)\|_{2p(z)}-\|G(w)\|_{2p(z)}|
  +|\|G(w)\|_{2p(z)}-\|G(w)\|_{2p(w)} | \\
  &\le \|G(z)-G(w)\|_{2p(z)} + |\|G(w)\|_{2p(z)}-\|G(w)\|_{2p(w)}|
  \end{align*}
A glance at (the proof of Lemma \eqref{normea}) show that because $\|G(w)\|\kl M$ uniformly for $Re(w)\le 1$ (see Remark \ref{cc}) we do have
 \[ \lim_{w\to z} |\|G(w)\|_{2p(z)}-\|G(w)\|_{2p(w)}| \lel 0 \pl .\]
For the first part we use Kosaki's interpolation result and get
 \[ \|G(z)-G(w)\|_{2p(z)}\kl \|G(z)-G(w)\|_2^{1-Re(z)} \pl .\]
Thus for $Re(z)>0$, it suffices to show that $L_2$ estimate. Then we observe that
 \begin{align*}
  \|G(z)-G(w)\|_2 &= \|G_1(z)G_2(z)-G_1(w)G_2(w)\|_2\\
  &\le \|G_1(z)(G_2(z)-G_2(w))\|_2+
  \|(G_1(z)-G_1(w))G_2(w)\|_2 \\
  &\le \|G_1(z)\|_{\infty} \|g_{\eta,\rho}(z/2)\rho^{1/2}-
  g_{\eta,\rho}(w/2)\rho^{1/2}\|_2 + \|(G_1(z)-G_1(w))G_2(w)\|_2
   \pl .
  \end{align*}
Thanks to Remark \ref{cc}, we deduce convergence for the first of the two terms from Lemma \ref{ct1}. Let us consider the remaining term and $w=1/q+it$. Then we deduce from H\"older's inequality and interpolation that
 \begin{align*}
  \|a G_2(w)\|_2
  &= \|a \eta^{w/2}\rho^{1/2-w/2}\|_2 \lel
   \|a \eta^{1/2q} \eta^{it/2}\rho^{-it/2}\rho^{1/2-1/q}\|_2\\
  &\le \|a\eta^{1/2q}\|_{2q} \kl   \|a\|_{\infty}^{1-1/q} \|a\eta^{1/2}\|_2^{1/q} \pl.
  \end{align*}
Therefore we are left with an $L_2$-norm estimate. In our case $a=\pi(G_1(z)-G_1(w))e$ and hence for $b=G_1(z)-G_1(w)$ we find that
 \begin{align*}
 \|a\eta^{1/2}\|_2^2
 &=Tr(\eta^{1/2}\Phi^{\dag}(b^*b)\eta^{1/2}) \lel
 Tr(\hat{\eta}(b^*b)) \\
 &= \|\hat{\rho}^{z/2}\hat{\eta}^{-z/2}\hat{\eta}^{1/2}-
 \hat{\rho}^{w/2}\hat{\eta}^{-w/2}\hat{\eta}^{1/2}\|_2^2 \pl .
 \end{align*}
Therefore Lemma  \ref{ct1} concludes the proof. \qd

\section{Approximation of relative entropy} \label{sec:approx}

In this section we will work with Lindblad's definition of relative entropy
 \[ D_{\Lin}(\rho\|\eta) \lel (\sqrt{\rho},\log \Delta_{\rho,\eta}(\sqrt{\rho}))+\eta(1)-\rho(1)\]
Indeed, $D_{\Lin}$ is the unique homogeneous joint extension of the relative $D$ entropy, i.e.
 \begin{enumerate}
 \item[i)] $D_{\Lin}(t\rho\|\eta)\lel t D_{\Lin}(\rho\|\eta)$;
 \item[ii)] $D_{\Lin}(\rho\|\eta)=D(\rho\|\eta)$ if $\rho(1)=\eta(1)=1$.
  \end{enumerate}

\subsection{Finite von Neumann algebras}

\begin{prop}\label{finite} Let $(N,\tau)$ be a finite von Neumann algebra and $a\le d_{\eta}\le a^{-1}$. Let $d_{\psi}$ be a density of a state $\psi$. Then
 \[ d_{M,\delta} \lel 1_{[0,M]}(d_{\psi})d_{\psi}+\delta d_{\eta} \]
satisfies
  \begin{enumerate}
   \item[0)] $\delta d_{\eta}\kl d_{M,\delta}\kl (a+\delta)d_{\eta}$;
  \item[i)] $\lim_{M\to\infty }\lim_{\delta\to 0} \|d_{M,\delta}-d_M\|_1\lel 0$;
   \item[ii)] $\lim_{M\to \infty}\lim_{\delta\to 0}
    D_{\Lin}(d_{M,\delta}\|d_{\eta}) \lel D(d\|d_{\eta})$.     \end{enumerate}
\end{prop}

\begin{proof} In the tracial setting, we have (see \cite{witten_aps_2018}) that
 \[ D(\psi|\eta) \lel D(d_{\psi}\|d_{\eta})
 \lel \tau(d_{\psi}\ln d_{\psi})-\tau(d_{\psi}\ln d_{\eta})
 \lel D_{\Lin}(d_{\psi}\|d_{\eta}) \pl .\]
For fixed $M$, we denote by $d_M=1_{[0,M]}(d_{\psi})d_{\psi}$ the  density obtained by functional calculus. Then $d_{M,\delta}=d_M+\delta d_{\eta}$ converges in operator norm, and $L_1$ norm to $d_M$. Therefore, the continuity of $f(x)=x\ln x$ implies that
 \[ \lim_{\delta\to 0} \tau(d_M+\delta d_{\eta}\ln d_M+\delta d_{\eta})-\tau((d_M+\delta d_{\eta})\ln d_{\eta})+\tau(d_{\eta})-\tau(d_M+\delta d_{\eta})
 \lel D_{\Lin}(d_M\|d_{\eta}) \pl . \]
Here we use that $d_{\eta}$ is bounded below and above and hence $\ln d_{\eta}$ is in $L_{\infty}(N)$. Using this fact again, we deduce from Fatou's lemma
 \begin{align*}
 \tau(d_{\psi}\ln d_{\psi})-\tau(d_{\psi}\ln d_{\eta})
 +\tau(d_{\eta})-\tau(d_{\psi})
 &=   \lim_{M\to \infty}  \tau(d_M\ln d_M)-\tau(d_M\ln d_{\eta})
 +\tau(d_{\eta})-\tau(d_M) \pl .
 \end{align*}
Note here that $D(d_{\psi}\|d_{\eta})$ is finite iff $\tau(d_{\psi}\ln d_{\psi})$ is finite. \qd

%This should be the same $\pi$ as in crossed product - just note where it's defined
For the convenience of the reader let us briefly review how to transition from trace free definition to the one using trace. Indeed, in $L_2(\N,\tau)$ the vector $\sqrt{d_{\eta}}$, the purification of the state $\eta$, implements the GNS representation with respect to the usual left-regular representation $\pi(x) \sqrt{d_{\eta}}=x \sqrt{d_{\eta}}$ for $x\in N$. We will use $\pi$ again in the Haagerup construction, section \ref{sec:haagerup} Moreover, using Connes' $2\times 2$ matrix trick, (see e.g. \cite{pedersen_radon-nikodym_1973}), we know for $\xi \in L_2(\N,\tau)$ that
 \[ \Delta_{\eta,\psi}(\xi) \lel d_{\eta}\xi d_{\psi}^{-1} \]
and hence
 \[ \Delta_{\psi,\eta}^{it}(x) \lel d_{\psi}^{it}xd_{\eta}^{-it} \pl .\]
This implies
 \[ \ln \Delta_{\psi,\eta}(d_{\psi}^{1/2})
 \lel \ln d_{\psi}d_{\psi}^{1/2}-d_{\psi}^{1/2}\ln d_{\eta} \pl .\]
Taking the inner product, we find
 \[ (d_{\psi}^{1/2}, \ln \Delta_{\psi,\eta}(d_{\psi}^{1/2}))
\lel \tau(d_{\psi}\ln d_{\psi})-\tau(d_{\psi}\ln d_{\eta})
  \lel D_{\tau}(d_{\psi}\|d_{\eta})
  \pl .\]

\subsection{Haagerup construction} \label{sec:haagerup}
Haagerup's construction for type $III$ algebras provides a convenient tool to deduce properties of type $III$ algebras from finite von Neumann algebras. 
\begin{rem}{\rm Let us recall two possible ways to represent the crossed product $M\rtimes G$ for an action $\al$ of a discrete group on Hilbert space. We may assume that $M\subset B(H)$ and consider $\ell_2(G,H)$. Then $M\rtimes G=\lan \la_H(G),\pi(M)\ran$ is generated by a copy  of $\la(G)$, the left regular representation of $G$, and $\pi(M)$. Here we may assume
\[ \pi(x) \lel \sum_{g} |g\ran\lan g| \ten \al_{g^{-1}}(x) \] 
is given by a twisted diagonal representation and $\la_H(g)=\la(g)\ten 1_H$. Alternatively, we may choose $\hat{\pi}(x)\lel 1\ten x$ and $\hat{\la}_H(g) \lel \la(g)\ten u_g$ such that $u_g^*xu_g=\al_{g^{-1}}(x)$. Both of these representations are used in the literature, and their equivalence is used in the proof of Takai's theorem. For the equivalence we note that
 \[ \la_H(g)^{-1}\pi(x)\la_H(g)
 \lel \pi(\al_{g^{-1}}(x)) \pl .\]
Similarly, $\la(g)^{-1}\ten u_g^{-1}(1\ten x)\la(g)\ten u_g=1\ten \al_{g}^{-1}(x)$. This shows that the algebraic  relations of these two representations coincide. Using a GNS construction this extends to the generated von Neumann algebras.}
\end{rem}

\begin{lemma}\label{up} Let $\rho, \eta$ be states on the von Neumann algebra $M$ with corresponding $\tilde{\rho}, \tilde{\eta}$ in $\tilde{M}_*$. Then $D(\tilde{\rho}\|\tilde{\varphi})=D(\rho\|\varphi)$.
\end{lemma}

\begin{proof}
We consider the Hilbert space $H=\ell_2(G,L_2(M))$ and still use the symbol $\la(g)$ instead of $\la_{L_2(M)}(g)$. Our first goal is to calculate the modular operator for an analytic state $\eta$ with density $d$ in $L_1(M)$, and $\tilde{\eta}=\eta\circ E$, $E:M\rtimes G\to M$ the canonical conditional expectation.
Then $\xi=|1\ran \ten d^{1/2}$ implements the state $\tilde{\eta}$ on the crossed product. In order to calculate the modular operator $\Delta=S^*S$, we recall that
 \[ (y\xi,\Delta(x\xi))\lel (x^*\xi,y^*\xi) \pl .\]
We start with finitely supported $y=\sum_g \la(g)\pi(y_g),z=\sum_g \la(g)\pi(z_g)$ and observe that 
 \[ (y\xi,z\xi) \lel (\sum_g |g\ran y_gd^{1/2},\sum_g |g\ran z_gd^{1/2})
 \lel \sum_g \eta(y_g^*x_g) \pl .\]
On the other hand, we find 
 \begin{align*}
 (x^*\xi,y^*\xi) &= (\sum_g |g^{-1}\ran \al_g(x_g^*)d^{1/2} , \sum_g |g^{-1}\ran \al_g(y_g^*)d^{1/2}) \lel  \sum_g \eta(\al_g(x_gy_g^*)) \pl .
\end{align*}
Let $d_{g^{-1}}=\al_{g}^{-1}(d)$. Then we see that 
  \begin{align*}
   \eta(\al_g(x_gy_g^*)) &= tr(d_{g^{-1}}x_gy_g^*) \lel tr(d^{1/2}y_g^*d_{g^{-1}}x_g d^{-1} d^{1/2})
   \lel (y_g d^{1/2},d_{g^{-1}} x_g d^{-1} d^{1/2}) \pl .
   \end{align*}
This means that the diagonal operator $\Delta_g(\xi_g)=\Delta_{d_{g^{-1}},d}$ is a good candidate for the modular operator, and is indeed well-defined for finitely supported  sequences of $\si^{t}_{\al_g^{-1}(\eta),\eta}$-analytic elements, which are dense. Now, it is easy to identify the polar composition using the isometry $J(\sum_g |g\ran \xi_g)=\sum_{g} |g^{-1}\ran \al_g(\xi_g^*)$ on $\ell_2(G,L_2(M))$, because $\al_g$ extends to an isometry on $L_2(M)$. This formula $S=J\Delta^{1/2}$ follows by calculation. Finally, we use Connes' $2\times 2$ matrix trick for  two states $\eta,\psi$ and the diagonal state $\hat{\eta}(x_{ab})=\eta(x_{11})+\psi(x_{22})$. Note  that $M_2(M)\rtimes G=M_2(M\rtimes G)$ and hence $\Delta_{\tilde{\eta},\tilde{\psi}}$ is the $1,2$ entry given by the $G$-diagonal operator $\Delta_{\al_{g}^{-1}(\eta),\psi}$. This implies
 \begin{align*}
 D(\tilde{\eta}\|\tilde{\psi})
 &= (\xi_{\psi},\log \Delta_{\tilde{\eta},\tilde{\psi}}(\xi_{\psi}))
  \lel (d_{\psi}^{1/2}, \Delta_{\al_{1}^{-1}(\eta),\psi}(d_{\psi}^{1/2})) \\
 &= (d_{\psi}^{1/2},\log\Delta_{\eta,\psi}(d_{\psi}^{1/2}))
 \lel D(\eta\|\psi) \pl .
 \end{align*}
Here we use that the relative entropy can be calculated on any representing Hilbert space. However, the representation of  $M\rtimes G$ is in standard form, which may be used as a definition of the relative entropy.
\qd 

A similar result holds for the fidelity.

\begin{theorem}\label{app} Let $\eta$ be a faithful state. Then there exists a sequence of states $\rho_{\al}$ such that
 \begin{enumerate}
 \item[i)] $\delta_{\al}\eta \le \rho_{\al} \kl \delta_{\al}^{-1}$ for some $\delta_{\al}>0$;
  \item[ii)] $\lim_{\al} \rho_{\al}=\rho$;
  \item[iii)] $D(\rho\|\eta) \lel \lim_{\al} D(\rho_{\al}\|\eta)$.
  \end{enumerate}
\end{theorem}

\begin{proof} Let us define $\psi_k=F_{k}(\tilde{\rho})$. Thanks to the Haagerup construction we know that $\lim_k \psi_k=\tilde{\rho}$. We may apply Proposition \ref{finite} and find $d_{k,m,\delta}=\al_{k,m,\delta} (1_{[0,m]}(d_{\psi_k})d_{\psi_k}+\delta d_{\eta_k})$, where $\al_{k,m,\delta}$ is chosen such that $d_{k,m,\delta}$ has trace $1$. Denote by $\psi_{k,m,\delta}^0$ the corresponding state on $\tilde{M}_k$ and $\psi_{k,m,\delta}=\psi_{k,m,\delta}^0\circ F_k$. Let $\rho_{k,m,\delta}$ be the restriction to $M$. Certainly, we find condition i). Moreover, by the data processing inequality (see Witten's notes \cite{witten_aps_2018})
 \[ D(\rho_{k,m,\delta}\|\eta)\kl D(\psi_{k,m,\delta}\|\eta) \]
and hence
 \begin{align*}
 \limsup_{k\to \infty,m\to \infty, \delta\to 0}
 D(\rho_{k,m,\delta}\|\eta) &\kl \limsup_k D(\psi_k\|\tilde{\eta}) \\
 &\le  D(\tilde{\rho}\|\tilde{\eta}) \lel D(\rho\|\eta) \pl .\end{align*}
However, we deduce from  $Hiii)$ and Proposition  \ref{finite} that
 \[ \lim_k \lim_m \lim_{\delta} \psi_{k,m,\delta} \lel \tilde{\rho} \pl .\]
Taking the conditional expectation $\E$ by restriction these state to $M$ preserves this property. Thus by the semicontinuity of $D_{\Lin}$, we deduce that
 \[ D(\rho\|\eta)\kl \liminf_{k,m,\delta} D(\rho_{k,m,\delta}\|\eta)
 \kl \limsup_{k,m,\delta} D(\psi_{k,m,\delta}\|\tilde{\eta})
\kl D(\rho\|\eta) \pl .\]
This allows us to find a suitable convergent subsequence. \qd

\section{Recovery Results} \label{sec:genrecov}
Finally, we are ready to show the general recovery results of this paper. In the following diagram, we illustrate the relationship of densities on the original algebra, crossed product, and approximating, finite algebras used to derive the final result:
\begin{equation}
\begin{array}{cccc}
	   \Phi: & L_1(M) & \rightarrow & L_1(N)
	\\ & \uparrow \pi_M^\dagger & & \downarrow \E_N^\dagger
	\\ & L_1(M \rtimes G) & & L_1(N \rtimes G)
	\\ & \uparrow \E_j^\dagger & & \downarrow \pi_k^\dagger
	\\ \Phi_{j,k}: & L_1(M_j) & \rightarrow & L_1(N_k)
\end{array}
\end{equation}
Here $\pi_m,\pi_k$ are inclusions maps in the Haagerup approximation, and $\E_N, \E_j$ are conditional expectations. We define an approximating sequence of quantum channels $\Phi_{j,k} : L_1(M_j) \rightarrow L_1(N_k)$ in the finite von Neumann algebras and apply Theorem \ref{entropybound}. Lemma \ref{hap} shows that the relative entropies in the crossed product converge to that of the original relative entropy in the von Neumann algebra with which we started. Theorem \ref{app} shows that we can construct an increasing sequence $L_1(M_j)$ and $L_1(N_k))$ in the finite algebras that converges to the relative entropy in the crossed product. We also may check that $\lim_{j,k} R_{\eta_j, \Phi_{j,k}} \rightarrow R_{\eta, \Phi}$. These steps follow those of \cite{junge_universal_2020}, introducing no new concepts, so we do not repeat them in detail here.

For the $p$-fidelities:
\begin{theorem}[technical version of \ref{thm:main}] Let $\eta$ and $\rho$ be states such that the corresponding support projections satisfy $e_{\rho}\le e_{\eta}$. Let $d_{\eta}, d_{\rho}$ their densities in $L_1(M)$. Let $\Phi:L_1(M)\to L_1(\hat{M})$ a complete positive trace preserving map with adjoint $\Phi^{\dag}$.  Then holds for $1\le p<\infty$.
 \[ -2 p \ln \|\rho^{1/2p}\eta^{1/2p}\Phi^{\dag}(\Phi(\eta)^{-1/2p}\Phi(\rho)^{1/p}
 \Phi(\eta)^{-1/2p})\eta^{1/2p}\rho^{1/2p}\|_{p/2}
+D(\Phi(\rho)\|\Phi(\eta))
\kl D(\rho\|\eta) \pl .
 \]
\end{theorem}
\begin{proof} Let $\rho_{\al}$ be as in Theorem \ref{app}.
We also need to fix a $k$ and consider $F_k(\tilde{\rho})$ together with  states $d_{k,m,\delta}$ and the density $\eta_k=F_k(\tilde{\eta})$ on the $\tilde{M}_k$. Then
$d_{k,m,\delta}$ and $\tilde{\eta}_k$ satisfy the assumptions and keep the notation of the proof of Theorem \ref{app}. Moreover, the map $\Phi_k=\Phi\circ E\circ F_k:L_1(\tilde{M}_k)\to L_1(\hat{M})$ is completely positive and trace preserving. This allows us to apply Theorem \ref{rec1} and deduce
 \begin{equation}\label{kmd} D(\Phi_k(d_{k,m,\delta})\|\Phi_{k}(\tilde{\eta_k}))-2 p \ln f_{p}(d_{k,m,\delta}^{1/p},R_{1/p}(\Phi_k(d_{k,m,\delta})^{1/p}))
 \kl D(d_{k,m,\delta}\|\eta_k) \pl .\end{equation}
Using lower semi-continuity we deduce that
  \[ D(\Phi(\rho)\|\Phi(\eta))\kl \liminf_{k,m,\delta}
  D(\Phi_k(d_{k,m,\delta})\|\Phi_{k}(\tilde{\eta_k})) \pl .\]
We also know that $\lim_{k,m,\delta}  D(d_{k,m,\delta}\|\eta_k)=D(\rho\|\eta)$. Note that
$\lim_{km,\delta} d_{k,m,\delta}=\tilde{\rho}$. Thus by norm continuity of the map $R_{1/p}$ and the Mazur map, we deduce that
 \[ \lim_{k,m,\delta}  f_{p}(d_{k,m,\delta}^{1/p},R_{1/p}(\Phi_k(d_{k,m,\delta})^{1/p}))
  \lel f_{p}(\tilde{\rho}^{1/p},R_{1/p}(\Phi(\tilde{\rho}^{1/p})) \pl .\]
By the definition of $R_{1/p}$ and Lemma \ref{fup}, we deduce that 
 \[ f_{p}(\rho^{1/p},\eta^{1/p})
 \lel \lim_{k,m,\delta} f_{p}(d_{k,m,\delta}^{1/p},R_{1/p}(\Phi_k(d_{k,m,\delta})^{1/p})) \pl .\] 
Thus taking  the limit in \eqref{kmd} implies the assertion.\qd  

Here we recall a shortened and slightly modified version of version of Lemma \ref{up}, which uses the Haagerup approximation method to relate the semifinite and type III relative entropies.
\begin{lemma} \label{hap} Let $G$ be a discrete group and $E:M\rtimes G\to M$ be a conditional expectation. Let $\tilde{\Phi}=w\circ E$ and $\tilde{\rho}=\rho\circ E$. Then
 \[ D_M(\tilde{\rho}\|\tilde{\phi})\lel D_M(\rho\|\phi) \pl .\] 
\end{lemma} 

\begin{proof} Since $M\subset M\rtimes G$, we deduce that
 \[ D_M(\rho\|\phi)\kl D_M(\tilde{\rho}\|\tilde{\phi}) \pl .\] 
For the converse consider $\Phi:L_1(M\rtimes G)\to \ell_1^m$ and the ucp-map $\Phi^*:\ell_{\infty}^m\to M\rtimes G$. The relative entropy is calculated with the help of the coefficients 
 \[ \al_j \lel \tilde{\rho}(\Phi^*(e_j))
 \lel \rho(E\Phi^*(e_j)) \] 
and $\beta_j=\phi(E(\Phi^*(e_j))$. Since $E\Phi^*:\ell_{\infty}^m\to M$ is a normal ucp map, we deduce the assertion.\qd
\noindent We also recall Theorem \ref{app}.

\begin{cor} \label{finalbound}
Let $\rho, \eta \in M_*^+$ be a pair of states on a von Neumann algebra $M$, and let $\Phi$ be quantum channel. Then
\begin{equation*}
D(\rho \| \eta) - D(\Phi(\rho) \| \Phi(\eta)) \geq
	\int_{\RR} \beta_0(t) D_M(\rho \| R^t_{\eta, \Phi} \circ \Phi(\eta)) dt \geq
	D_M(\rho \| \tilde{R}_{\eta, \Phi} \circ \Phi(\eta)) \pl.
\end{equation*}
\end{cor}
\noindent Corollary \ref{finalbound} is the technical version of Theorem \ref{entropybound}.

\section{Recovery of positive vectors} \label{sec:posvec}

In this section,  we explain how to recover certain vectors in a Hilbert space from a Petz recovery map. Our starting point is  representation of a von Neumann algebra $M\subset \BB(H)$ and a separating vector $h\in M$, i.e. the map $x\mapsto xh$ is injective. This implies that the corresponding normal state $\eta(x)=(h,xh)$ has full support in $M_*$.  Then we may apply the GNS construction and a partial isometry $U:\overline{Mh}\to L_2(M)$ via
 \[ U(xh) \lel x\eta^{1/2} \pl .\]
Indeed,
 \[ (U(xh),U(yh))\lel Tr(\eta^{1/2}x^*y \eta^{1/2}) \lel \eta(x^*y) \lel (xh,yh) \]
shows that $U$ extends to an isometry between $Mh$ and $L_2(M)$. Recall that the inclusion $M\subset \BB(L_2(M))$ is in standard position. This means there is a real subspace $L_2(M)_+\subset L_2(M)$ and partial isometry $J$ such that $J|_{L_2(M)_+}=id$. In fact, all these objects can be constructed by Tomita-Takesaki theory and $J_{\eta}=U^*JU$ is indeed the anti-linear part of $S=J\Delta^{1/2}$ in the polar decomposition of $S(xh)=x^*h$. Of particular importance here is the real subspace
 \[ H_+ \lel U^*(L_2(M)_+) \pl .\]
The space of positive vectors  is the range of Mazur map. Let us be more precise. For every norm one vector $k\in H$ we may consider the state
 \[ \om_k(x) \lel (k,xk) \]
which admits a density $d_k\in L_1(M)$ such that
 \[ \om_k(x) \lel Tr(d_kx) \pl .\]
Thanks to St{\o}rmer's inequality the map $d_k\mapsto d_k^{1/2}$ is continuous and hence
 \[ |k| \lel U^*d_k^{1/2} \in H_+  \pl .\]
This allows us to reformulate the usual polar decomposition theorem.
\begin{prop} Let $h$ be a separating vector and $H_h=Mh$. Then every element $k\in Mh$ admits a polar decomposition
 \[ k \lel v|k| \]
where $v\in M$ is a partial isometry, uniquely determined by $v^*v={\rm supp}(\om_k)$.
\end{prop}

\begin{rem}{\rm Since $U^*:L_2(M)\to \overline{Mh}$ we can also work with polar decomposition for the adjoint
 \[ U(k)\lel |U(k)^*|w \lel R_w(|U(k)^*)\]
where $w$ belongs to the $M$, $R_w$ is the right multiplication and hence
 \[ k \lel U^*R_wU U^*(|U(k)^*) \in M'H_+  \]
admits a polar decomposition with respect to the commutant. In this form the theorem extends to all of $H$. Indeed, let  \[ H \lel \sum_i \overline{Mh_i} \]
be a direct sum of irreducible subspaces with projections $e_iH=\overline{Mh_i}$ in $M'$. Then $Mh_i\cong L_2(M)f_i$ for some projection $f_i$ corresponding to the support of $h_i$. Using an isomorphism $V$ between $H$ and $\oplus_i L_2(M)f_i$ we see that $M'(Mh)=M'h$ is dense in $H$. Using this isomorphism, we now deduce that
 \[ k \lel wV^*(|V(k)^*|) \]
admits a polar decomposition with a partial isometry $w\in M'$ and $V^*(|V(k)^*|)\in H_+$.}
\end{rem}

For $1\le p \le \infty$ we may now consider the Kosaki interpolation space $L_p^1(M,\om_h)$ as embedded in $H$. Indeed, we have already the inclusion
 \[L_{\infty}(M,\om_h) \cong  Mh \subset H \cong L_2^1(M,\om_h) \]
and by interpolation we find an injective map
 \[ U^*_p:L_p^1(M,\om_h) \to H \pl . \]
This allows us to define the corresponding $p$-norm
 \[ \|k\|_p \lel \sup\{ |(ah,h)| \pl |\pl \|a\om_h^{1/p'}\|_p <\infty \}  \]
for $1\le p\le \infty$. For $1\le p\le 2$ the space
 \[ H^p \lel \{k \pl | \pl\|k\|_p <\infty\} \]
is dense in $H$ and isomorphic $L_p(M)$. Therefore we find natural cones
 \[ H_+^p \lel H^p\cap H_+ \]
as the range of $U^*(L_p(M)_+)$. Let us explain how these cones appear naturally in the context of Petz maps. We will assume that $\Phi:L_1(M)\to L_1(\hat{M})$ is a completely positive trace preserving map and, for simplicity, that $\eta$ and $\hat{\eta}=\Phi(\eta)$ have full support. Then the Petz map
 \[ R_{1/p}:L_p(\hat{M})\to L_p(M) \pl ,\pl R_{1/p}(\hat{\eta}^{1/2p}x\hat{\eta}^{1/2p})\lel \eta^{1/2p}\Phi^{\dag}(x)\eta^{1/2p} \]
is a contraction and sends $L_p(\hat{M})_+$ to $L_p(\hat{M})$. Therefore we also find a contraction
 \[ R_{1/p}: \hat{H}^p_+ \to H^p_+ \pl .\]
Let us describe this map more explicitly, by assuming that $\om_k\le C \om_h$ and hence, as above,
 \[ a(z) \lel \om_k^{z/2}\om_h^{-z/2}\quad, \quad
  \hat{a}(z) \lel \hat{\om}_k^{z/2}\hat{\om}_h^{-z/2} \]
are well defined. Then we find that
 \begin{align*}
  R_{1/p}(\hat{\om}_k^{1/p})&=
  \om_h^{1/2p}\Phi^{\dag}(\hat{a}(1/2p)^*\hat{a}(1/2p))\om_h^{1/2p}\\
  &\lel \Delta_{\om_h}^{1/2p}(\Phi^{\dag}(\hat{a}(1/2p)^*\hat{a}(1/2p))\om_h^{1/p}) \pl .
  \end{align*}
If we define  $b=\Phi^{\dag}(\hat{a}(1/2p)^*\hat{a}(1/2p))$ we see that
 \[
  R_{1/p}(\hat{\om}_k^{1/p})\lel \Delta_{\om_h}^{1/2p}(bh) \in H^p_+ \pl .\]
On the other hand we see that $k \in \hat{H}^p_+$ is represented $\hat{U}(k)\lel \hat{\om}_k^{1/p}\hat{\om}_h^{-1/p}\hat{\om}_h^{1/p}$. This implies
  \[  \hat{a}(1/2p)^*\hat{a}(1/2p)\lel \hat{\om}_h^{-1/2p}\hat{\om}_k^{1/p}\hat{\om}_h^{-1/2p}
 \lel \Delta_{\hat{\om}_h}^{-1/2p}(\hat{\om}_k^{1/p}\hat{\om}_h^{-1/p})  \pl .\]
Let us recall the map
 \[ \Phi^{\dag}_p(b\hat{\om}_h^{1/p}) \lel \Phi^{\dag}(b)\om_h^{1/p} \pl \]
which we extend to a densely  map on $H_p$ as follows
 \[ \Phi^{\dag}_p(b\hat{h}) \lel  \Phi^{\dag}(b)h \pl .\]
Then we can combine the calculations above and find that
 \begin{equation}\label{p22}
   R_{1/p} \lel \Delta_{\om_h}^{1/2p}\Phi_p^{\dag} \Delta_{\hat{\om}_h}^{-1/2p}  \pl .
 \end{equation} 
Our fidelity result can be formulated as follows:

\begin{cor}\label{Hver1} Let  $h$ be a separating vector for $M$ with associated vector state $\om_h$, and let   $\Phi^{\dag}:\hat{M}\to M$ be a normal, unital completely positive map and $\hat{\om}_h=\om_h \circ \Phi^{\dag}$ the associated vector state. Then map $R_{1/p}:\hat{H}_p\to H_p$
 \[ R_{1/p} \lel  \Delta_{\om_h}^{1/2p}\Phi_p^{\dag} \Delta_{\hat{\om}_h}^{-1/2p} \]
extends to a contraction and satisfies
 \begin{align*}
 -\ln f_p(k,R_{1/p}(\hat{k}))\kl \frac{1}{2p} ( D(\om_k\|\om_h)-D(\hat{\om}_k \| \hat{\om}_h)) \pl .
 \end{align*}
for every $k\in H^p_+$.
\end{cor}

Our next application tells us that if we use the standard form of representing a states on von Neumann algebras, then we may recover the implementing vector:
\begin{cor} Let $H=L_2(M)$. Then implementing vectors $\xi_{\rho}$ for $\rho$ and $\xi_{\hat{\rho}}$ satisfy  
 \[  \|\xi_{\rho}-R_{1/2}(\xi_{\hat{\rho}})\|_2^2
 \kl D(\rho|\eta)-D(\Phi(\rho)\|\Phi(\eta))\pl. \
 \pl .\]
%Moreover,  
\end{cor}

\begin{proof} Let us first consider $a,b\in L_2(M)_+$ of norm $1$ and $h=b-a$. Then
 \begin{align*}
  0 &= \|b\|^2-\|a\|^2 \lel \|a+h\|^2-\|a\|^2 \lel
   2(a,h)+\|h\|^2 \pl .
   \end{align*}
On the other hand 
 \begin{align*}
 1-f_2(a,b)^2 &= \|a\|^2-\|a^{1/2}b^{1/2}\|_2^2 
 \lel tr(a^2)-tr(ab) \lel tr(a(a-b)) \\
 &=-(a,h) \lel \frac{\|h\|^2}{2} \pl .
 \end{align*}
Then $\ln(1+x)\le x$ implies for $a=\rho^{1/2}$ and $b=R_{1/2}(\hat{\rho}^{1/2})$ that 
 \[ -\ln f_2(a,b)^2 \lel -\ln(1-(1-f_2(a,b)^2))\gl (1-f_2(a,b)^2)
 \gl \frac{\|a-b\|_2^2}{2} \pl .\] 
The assertion then follows from Theorem \ref{rec1}.\qd

\begin{rem} \label{rem:nonlin}
The proof of equations \eqref{eqnonlin1} and \eqref{eqnonlin2} in the introduction follows via the triangle and Cauchy-Schwarz inequalities.
\end{rem}

As an illustration we will now assume that $\hat{M}\subset M$ is a subalgebra and that there exists a normal conditional expectation $E:M\to \hat{M}$ such that
 \[ \om_h \lel \om_h|_{\hat{M}}\circ E \pl .\]
In this case $\Phi^{\dag}=\iota$ is just the inclusion map $\hat{M}\subset M$ and moreover, $\Phi^{\dag}$ commutes with the modular group (see \cite{junge_noncommutative_2003}). Then $E$ extends to map $E:L_2(M)_+\to L_2(\hat{M})_+$ via
 \[ E(x\om_h^{1/2})\lel E(x)\hat{\om}_h^{1/2} \pl .\]
Under these additional assumptions, we see that
$R_{1/p}:\hat{H}^p\to H^p$ is simply the inclusion map. In his particular case the fidelity can also be expressed easily. Indeed, according to the proof of Lemma \ref{fup} we know that
 \[ f_p(k',k) \lel \sup_{\|ak\|_{p'}\le 1} |(ak,\Delta_{k',k}^{1/2p}(k))| \pl .\]
The case $p=2$ is particularly interesting and gives the self-polar form
 \begin{align*}
 f_2(x,y)^2 &= \|x^{1/4}y^{1/4}\|_2^2 \lel Tr(x^{1/2}y^{1/2}) \pl .
 \end{align*}
For elements $k,k'\in H_+$ we may assume $k=a\om_h^{1/2}$ and $k'=b\om_{h}^{1/2}$,  and  $x^{1/2}=U(a\om_h^{1/2})$, $y^{1/2}=U(b\om_h^{1/2})$. This means
 \[ f_2(x,y) \lel Tr(\om_hb^*a)\lel (h,b^*ah)\lel (bh,ah)
 \lel (k',k) \pl .\]

\begin{cor} In addition to the assumption of \ref{Hver1} assume that $\om_h=\hat{\om}_h\circ E$ holds for a normal  conditional expectation. For $k\in H_+$
 \[ -\ln (k,E(k))\kl D(\om_h\|\om_k)-D(\hat{\om}_h\|\hat{\om}_k) \pl .\]
\end{cor}

\begin{rem}{\rm Without assuming the existence of $E$, we can still describe the Petz map for $L_2$ in this special case. Indeed, let us assume that $\hat{M}\subset M$ and denote by $\hat{\iota}:\hat{M}h\to Mh$ the canonical inclusion map. We will assume that $k\in H_+(\hat{M})$ and $\om_k\kl C \om_h$ (which implies $\hat{\om}_k\le C \hat{\om}_h$. Then 
 \[ \hat{\om}_k^{1/2} \lel \hat{\om}_k^{1/2}\hat{\om}_h^{-1/2}\hat{\om}_h^{1/2} \] 
 implies
  \[ k \lel \hat{\om}_k^{1/2}\hat{\om}_h^{-1/2}h \] 
 and 
  \[ \hat{\Delta}_{-1/4}(k) \lel  \hat{\Delta}_{-1/4}( \hat{\om}_k^{1/2}\hat{\om}_h^{-1/2})h \pl .\] 
Thanks to \eqref{p22} this implies
 \[ \xi \lel R_{1/2}(k) \lel \Delta_{1/4}( \hat{\iota}(
 (\hat{\Delta}_{-1/4}(k))) \pl .\] 
Let $P_{1/4}$ be the orthogonal projection onto the rotated space  $\hat{H}_{1/4}=\Delta_{1/4}(\hat{M}h)$. Then $\xi \in \hat{H}_{1/4}$ implies 
 \[ (|k|,\xi)\lel (P_{1/4}|k|,\xi) \lel \|P_{1/4}|k|\| \|\xi\|\kl \|P_{1/4}|k|\| \pl .\]
Therefore we deduce that 
 \[ -\ln \|P_{1/4}|k|\|\kl  D(\om_h\|\om_k)-D(\hat{\om}_h\|\hat{\om}_k) \pl .\]
In particular, if the relative entropy difference is small, then $P_{1/4}|k|\approx |k|$ implies that $U(|k|)$ almost commutes with $\om_h$. }\end{rem}

\section{Data processing inequality for $p$-fidelity} \label{pfiddp}

\begin{theorem}\label{pfid} Let $\Phi:L_1(M)\to L_1(\hat{M})$ be a channel. Then
 \[ f_p(\Phi(\rho),\Phi(\si))\gl f_p(\rho,\si) \pl .\]
\end{theorem}

We need the following $L_p$ norm inequality

\begin{prop}\label{kosdag} Let $\Phi^{\dag}$ be a normal, unital, completely, positive adjoint map of a channel $\Phi$, and $\eta$ be a normal state on $M$ such that $\Phi(\eta) = \hat{\eta}$. Then $\Phi_p:L_p(\hat{M})\to L_p(M)$ given by
\[ \Phi_p(x) \lel \eta^{1/2p}\Phi^{\dag}(\hat{\eta}^{1/2p}x\hat{\eta}^{1/2p})\eta^{1/2p} \]
is a completely positive contraction.
\end{prop}

\begin{proof} We may assume that the density $\eta$ of a given state has full support, let $\hat{e}$ be the support of $\hat{\eta}$, so that we may assume that $\Phi_p$ is defined on $\hat{e}L_p(\hat{M}))\hat{e}$. This allows us to use the Kosaki isomorphism $L_p(\hat{M})=L_p(\hat{M},\hat{\eta})$. With the help of this automorphism, we consider the densely defined  map
 \[ T(\hat{\eta}x\hat{\eta}) \lel \eta^{1/2}\Phi(x)\eta^{1/2} \pl .\]
Since $\Phi^{\dag}:\hat{M}\to M$ is contraction, we see that
 \[ \|T(x)\|_{\infty} \kl \|x\|_{\infty} \pl.\]  On the other hand let us assume that $x=ab$. Then we see deduce from the Cauchy-Schwarz inequality for completely positive maps that
 \begin{align*}
 \|T(\hat{\eta}^{1/2}ab\hat{\eta}^{1/2})\|
 &=\|\eta^{1/2}\Phi^{\dag}(ab)\eta^{1/2}\|_1
  \kl
 \|\eta \Phi^{\dag}(aa^*)\eta\|_1^{1/2}
 \|\eta \Phi^{\dag}(b^*b)\eta\|_1^{1/2}\\
 &\lel tr(\eta \Phi^{\dag}(aa^*))^{1/2}
tr(\Phi^{\dag}(b^*b)\eta)^{1/2} \\
 &= tr(\hat{\eta}(aa^*))^{1/2} tr(\hat{\eta}b^*b)^{1/2} \\
 & \lel \|\hat{\eta}a\|_2 \|b\hat{\eta}\|_2 \pl .
 \end{align*}
By density of $\hat{M}\hat{\eta}^{1/2}$ in $L_2(\hat{M})\hat{e}$, we deduce that
 \[ \|T(\xi \phi)\|_1\kl \|\xi\|_2 \|\phi\|_2 \]
for any $\xi$ and $\phi$. Thus $T$ extends to a completely positive contraction on $\hat{e}L_1(\hat{M})\hat{e}$. By the general Riesz-Thorin theorem (see \cite{bergh_interpolation_1976}), we deduce that $T:L_p(\hat{M},\hat{\eta})\to L_p(M,\eta)$ is a contraction. By Kosaki's theorem, this completes the proof.\qd

\begin{cor} Let $\eta,\rho$ be two densities of states. Then
 \[ T_p^{\eta,\rho}(x) \lel \eta^{1/2p}\Phi^{\dag}(\hat{\eta}^{-1/2p}x\hat{\rho}^{-1/2p})\rho^{1/2p} \]
extends to a contraction from $L_p(\hat{M})$ to $L_p(M)$.
\end{cor}
\begin{proof} We use Connes' matrix trick and consider $\si \lel \kla \begin{array}{cc}\rho&0\\\eta & 0\end{array}\mer$ on $M_2(M)$ for $\Phi_2=id_{M_2}\ten \Phi$. The assertion follows from applying Proposition \ref{kosdag} to $y=\kla \begin{array}{cc} 0 & x \\ 0 &  0\end{array}\mer$. \qd

\begin{proof}[Proof of \ref{pfid}] Let $x=\hat{\eta}^{1/2p}\hat{\rho}^{1/2p}$. Then we deduce that
 \[ T_p^{\eta,\rho}(\hat{\eta}^{1/2p}\hat{\rho}^{1/2p})
 \lel \eta^{1/2p}\rho^{1/2p} \pl .\]
Since $T_p^{\eta,\rho}$ is a contraction, we deduce that
  \begin{align*}
    f_p(\eta,\rho) &= \|\eta^{1/2p}\rho^{1/2p} \|_p \\
    &\le \|\hat{\eta}^{1/2p}\hat{\rho}^{1/2p}\|_p \lel   f_p(\hat{\eta},\hat{\rho})  \pl . \qedhere
    \end{align*}
\qd

\begin{cor} \label{equal1} If $D(\rho\|\eta)=D(\Phi(\rho)\|\Phi(\eta))$ for a channel $\Phi : L_1(M) \rightarrow L_1(\hat{M})$, then
 \[ T_{p}^{\eta,\eta}(\si_{s}^{\hat{\eta}})
 \lel \si_s^{\eta}(\rho^{1/p}) \]
holds for all $1\le p\le \infty$ and $s\in \rz$. Moreover, there exists a modular group intertwining channel $\Psi:L_1(M)\to L_1(\hat{M})$ such that $\hat{T}_p(x)\lel \si^{1/2p}\Psi^{\dag}(\hat{\si}^{-1/2p}x\hat{\si}^{-1/2p})\si^{1/2p}$ satisfies
 \[ \hat{T}_p(\hat{\rho}^{1/p})\lel \rho^{1/p} \]
and
 \[ \Psi(\rho) \lel \Phi(\rho) \pl .\] %This is indeed Phi, confirmed by Marius
\end{cor}

\begin{proof} In this case 
 \[ -\ln  f_p(\rho,R_{p,t}(\hat{\rho}^{1/p})^p) \lel 0 \]
holds $\mu$ almost everywhere. By continuity this holds for all $t$. In other words, thanks to the Mazur map, we get
 \[ \rho^{1/2p}\si^{1-it/2p}\Phi^{\dag}(\hat{\si}^{-(1-it/2p}\hat{\rho}^{1/p}\hat{\si}^{-(1+it)/2p})\si^{(1+it)/2p}\rho^{1/2p}
 \lel \rho^{2/p} \]
for all $t$. This implies
 \[ T_p^{\eta}(\si_{\hat{\eta}}(s) (\hat{\rho}^{1/p})) \lel \si_{\eta}(s) \rho^{1/p} \pl \]
for all $s$. For the moreover part we consider the family $R_p(x)=\hat{\si}^{-1/2p'}\Phi(\si^{1/2p'}x\si^{1/2p'})\hat{\si}^{1/2p}$. Thanks to data processing inequality for sandwiched relative entropy, this map is contraction, and hence
 \[  \Psi_2(x) \lel \lim_{T,\U} \int_{-T}^T  \si_{\hat{\eta}}(s) \Phi_2(\si_{\eta}(-s)(x)) \frac{ds}{2T}  \]
exists as a bounded operator on $L_2$. By density of $L_2$ in $L_1$ we deduce that
 \[   \Psi_1(\eta^{1/4}x\eta^{1/4}) \lel \hat{\eta}^{1/4} \Psi_2(x)\hat{\eta}^{1/4}
  \lel \lim_{T,\U}  \int_{-T}^{T} \si_{\hat{\eta}}(s) \Phi_2(\si_{\eta}(-s)(x)) \frac{ds}{2T}  \pl \]
is a completely positive map on $L_1(M)$. Its adjoint $\Psi_1^{\dag}$ is normal, unital completely positive map, defined as a point weak$^*$ limit of averages. Hence our assumption shows that $\hat{T}_p(x)=\eta^{1/2p}\Psi^{\dag}(\hat{\eta}^{-1/2p}x\hat{\eta}^{-1/2p})\eta^{1/2p}$ also satisfies
 \[  \hat{T}_{p}(\hat{\rho}^{1/p})\lel \rho^{1/p} \]
for $1\le p\le \infty$. For the final assertion, we have to establish a simple duality relation. Using Kosaki $L_p$ spaces, we see that the family of maps
 \[ \Phi_p \cong \Phi|_{\iota_p(L_p(M,\eta))} \]
is really the same map, via the topological embedding $\iota_p(x)=\eta^{1/2p}x\eta^{1/2p}$. Similarly,
 \[ \eta^{1/2p'}T_p(\hat{\eta}^{1/2p}x\eta^{1/2p})\eta^{1/2p'}
 \lel T_1(\hat{\eta}^{1/2}x\hat{\eta}^{1/2}) \] show that $T_p=T_1|_{\iota_p(L_p)}$ is also the same map. Moreover,
 \begin{align*}
 Tr(\Phi(\eta^{1/2}x\eta^{1/2}y))
 &= tr(\eta^{1/2}x\eta^{1/2}\Phi^{\dag}(y)) \lel tr(xT_1(\hat{\si}^{1/2}y\hat{\si}^{1/2})) \end{align*}
shows that $T_p=\Phi_{p'}^{\dag}$, by density. The same holds for $\hat{T}_p\lel \Psi_{p'}^{\dag}$. Now, it is easy to conclude. Our assumption implies
 \begin{align*}
  1 & \lel Tr(\rho^{1/p}\rho^{1/p'})
  \lel Tr(\rho^{1/p}\hat{T}_{p'}(\hat{\rho}^{1/p'})) \\
  &= (\iota_{p}(\rho^{1/p}),\hat{T}_1(\iota_{p'}(\hat{\rho}^{1/p'}))) \\
  &= (\Psi(\iota_p(\rho^{1/p}),\iota_{p'}(\hat{\rho}^{1/p'}))) \\
  &= Tr(\Psi_p(\rho^{1/p})\rho^{1/p'}) \pl .
  \end{align*}
By uniform convexity  of $L_p$ we deduce that
  \[ \Psi_p(\rho^{1/p}) = \hat{\rho}^{1/p} = \Phi(\rho)^{1/p}  \pl .\]
For $p\to 1$, we deduce  the assertion.
\qd

\section{$L_1$ isometries} \label{l1}

In the theory of von Neumann algebras completely isometric embeddings of $L_1(N)$ into $L_1(M)$ are completely characterized (see \cite{junge_classification_2005} for more information on the crucial work by Kirchberg). Indeed, a map $u:L_1(N)\to L_1(M)$ is complete isometry iff there exists a normal conditional expectation $E:M \to N \subset N_0$, a $^*$-homomorphism $\pi:M \to N_0$ and $J\in N_0'$ such that
\[ u(\eta^{1/2}x\eta^{1/2}) \lel \hat{\eta}\pi(x)J\hat{\eta} \pl .\]
Such a map is completely positive if $J$ is completely positive. Moreover, the inverse $u^{-1}$ extends to $L_1(M)$. Let us formulate a simple consequences of the the data processing inequalities.

\begin{lemma}\label{l1isolem} Let $u$ be a completely positive complete isometry $u:L_1(N)\to L_1(\tilde{N})$. Then
 \[ D(u(\eta)\|u(\rho))\lel D(\eta\|\rho) \pl \]
provided they are finite. Moreover,
 \[ f_p(u(\rho),u(\eta))\lel f_p(\rho,\eta) \pl .\]
 \end{lemma}

\begin{lemma}\label{sf1} Let $\hat{M}$ and $\hat{N}$ be semifinite and $\Phi:L_1(M)\to L_1(\hat{M})$, $\rho\le C \eta$  such that
 \[ D(\rho\|\eta)\lel D(\Phi(\rho)\|\Phi(\eta))\pl .\]
Then there exists completely positive $L_1$-isometry $u$ such that $\hat{\eta}=u(\eta)$ and $\hat{\rho}=u(\rho)$.
\end{lemma}

\begin{proof} Let $\Psi^{\dag}:\hat{M}\to \hat{N}$ the averaged map. Then we see that
 \[ \Psi^{\dag}(\hat{\rho}^{1/2})\lel \rho^{1/2} \]
and hence
 \[  \rho \lel \Psi^{\dag}(\hat{\rho}^{1/2})\Psi^{\dag}(\hat{\rho}^{1/2})  \kl \Psi^{\dag}(\hat{\rho})\lel \rho \pl .\]
Thus we equality in Kadison's inequality, and $\hat{\rho}$ belongs to the (extended) multiplicative domain $m\subset \hat{M}$. Since $\Psi$ is normal and invariant under $\si_{\hat{\eta}}$, we see that the multiplicative domain $m$ admits a $\eta$-invariant conditional expectation $E:\hat{M}\to m$ such that $\hat{\eta}E=\hat{\eta}$, see e.g. \cite{pedersen_radon-nikodym_1973} and also \cite{junge_noncommutative_2003}. In particular we have completely isometric, completely positive inclusion $\iota:L_1(m)\to L_1(\hat{M})$ such that
 \[ \iota(\hat{\eta}^{1/2}x\hat{\eta}^{1/2})
 \lel \hat{\eta}^{1/2}x\hat{\eta}^{1/2} \pl .\]
Let us denote by $\hat{M}(\hat{\rho},\hat{\eta})\subset  m$ be the smallest von Neumann algebra generated by $C^*(\hat{\rho})$ and $\si_t^{\hat{\eta}}$, which remains $\hat{\eta}$-complemented.
Let $f:\rz \to \rz$ be a bounded function.
Then we deduce that
\[  \Psi^{\dag}(f(\rho)) \lel f(\rho) \pl ,\pl \Psi^{\dag}(\si_{\hat{\eta}}(t)(f(\rho)))
\lel \si_{t}f(\rho) \pl .\]
This means that $\Psi^{\dag}$ extends to a natural isomorphism between $\hat{M}(\hat{\rho},\hat{\eta})$ and
 $M(\rho,\eta)$ such that
  \[ tr(\eta\Psi^{\dag}(x)) \lel tr(\Psi(\eta)x) \lel tr(\hat{\Phi}(x)) \pl .\]
The adjoint of $u=(\Psi^{\dag}|_{\hat{M}(\hat{\rho},\hat{\eta})})^{\dag}$ satisfies $u(\eta)\lel \hat{\psi}$ and
 \[ tr(u(\rho)x) \lel tr(\rho\Psi^{\dag}(x))\lel tr(\hat{\rho}(x)) \pl .\]
Since $M(\rho,\eta)$ is also $\eta$-conditioned, we deduce the assertion.\qd

\begin{rem}{\rm It follows easily that
 \[ u(\eta)^{1/p} \lel \hat{\eta}^{1/p} \]
and
 \[ u(\rho)^{1/p} \lel \hat{\rho}^{1/p} \]
holds for all $1\le p\le \infty$, under the assumptions above. }
\end{rem}

We want to extend this result to type $III$ von Neumann algebras. For this we need the notion of the multiplicative domain. For a completely positive unital map $\Phi:M\to N$ with Stinespring dilation $\Phi(x)=V^*\pi(x)V$, we recall that $x$ belongs to the right domain if
\begin{equation}\label{ridom}
  \Phi(x)^*\Phi(x) \lel \Phi(x^*x)
  \end{equation}
or equivalently $V^*\pi(x)(1-VV^*)\pi(x)V=0$. If $x$ and $x^*$ satisfy \eqref{ridom}, then  $[V,\pi(x)]=0$ holds for a minimal Stinespring dilation. The set
 \[  \mdom(\Phi)=\{x| [V,\pi(x)]=0\}
 \lel \{x | \Phi(x^*)\Phi(x)=\Phi(x^*x)\mbox{ and } \Phi(x)\Phi(x^*)=\Phi(xx^*) \}
 \]
is a sub-$C^*$-algebra of $M$ and for normal $\Phi$, hence normal $\pi$, see \cite{junge_decomposable_2004, junge_noncommutative_2005}, this is even a sub-von Neumann algebra.

\begin{lemma} Let $\Phi_n:\hat{M}\to M$ be a sequence of normal completely positive maps such that
 \begin{enumerate}
 \item[i)] The weak$^*$ limit
 \[ \Phi_{\infty}(x) \lel \lim_n \Phi_n(x) \pl;\]
 \item[ii)] $\Phi_n^{\dag}(\si)=\hat{\si}$ for normal faithful states $\si$ and $\hat{\si}$;
  \item[iii)] $(\si^{1/2}\Phi_n(x),\Phi_m(y)\si^{1/2})=
      (\si^{1/2}\Phi_{\min(n,m)}(x),\Phi_{\min(n,m)}(y)\si^{1/2})$.
 \end{enumerate}
 Let $(a_n)$ be a bounded sequence in the multiplicative domain of $\Phi_n$, converging strongly to $a$. Then $a$ belongs to the multiplicative domain of $\Phi$.
\end{lemma}

\begin{proof} We follow Kirchberg and use the $C^*$-algebra $C(\hat{\M})$ of all bounded sequences $(a_n)$ such that $a_n$ converges in the strong and strong $^*$-algebra. Similarly, we consider $C(\hat{M})$ and the corresponding quotient maps $\hat{q}$ and $q:C(M)\to M$ given by $q((a_n))\lel w^*\lim_n a_n$. We claim that $\Phi^{\bullet}C(\hat{M})\subset C(M)$. Indeed, assume that $\lim_n a_n-a$ converges to $0$ strongly. Then $a_n-a\hat{\si}$ converges to $0$ in $L_2(\hat{M})$.  Let us fix $n\le m m$. We find that
 \begin{align*}
  &\|(\Phi_n(a_n)-\Phi_m(a_m))\si^{1/2}\|_2 \lel  Tr(\si^{1/2}\Phi_n(a_n^*a_n)\hat{\si}^{1/2})
  +Tr(\si^{1/2}\Phi_m(a_m^*a_m)\hat{\si}^{1/2})\\
  &\quad
  - Tr(\si^{1/2}\Phi_n(a_n^*)\Phi_m(a_m)\si^{1/2})
  -Tr(\si^{1/2}\Phi_m(a_m)^*\Phi_n(a_n)\si^{1/2})\\
  &=Tr(\Phi_n^*(\si)(a_n^*a_n))+Tr(\Phi_m^*(\si)(a_m^*a_m))
 -Tr(\Phi_n^*(\si)(a_n^*a_m))-Tr(\Phi_n^*(\si)(a_m^*a_n))\\
 &= Tr(\hat{\si}(a_n^*a_n+a_m^*a_m-a_n^*a_m-a_m^*a_n))\\
 &= \|(a_n-a_m)\hat{\si}\|_2^2  \pl .
 \end{align*}
Since $\si$ is faithful and $(\Phi_n(a_n))$ bounded, we deduce that $\Phi_n(a_n)$ is also strongly convergent.

Let $\hat{M}_n\subset M$ the multiplicative domain of $\A=\{(x_n)|x_n \in \hat{M}_n\}$ the corresponding subalgebra of $\ell_{\infty}(\hat{M})$. Then $\Phi^{\bullet}:\A\to \ell_{\infty}(M)$ is a $^*$-homomorphism, and we may define $A=C(\hat{M})\cap \A$. Then
 \[ \Phi^{\infty}|_{A}:A\to C(M)  \]
is a $C^*$-homomorphism. Let $\hat{J}\subset C(\hat{M})$ be the kernel of the quotient  map $\hat{q}$. Since $\Phi^{\infty}$ preserves strong convergence, we deduce that $\Phi^{\infty}(\hat{J})\subset J$, $J$ the kernel $q$. We deduce that there exists a $^*$-homomorphism $\pi:\hat{q}(A)\subset C(\hat{M})/\hat{J}=\hat{M}$ to $M=C(M)/J$ such that
 \[ q\Phi^{\infty}(a_n) \lel \si(q(a_n)) \pl .\]
Note that $\si$ is the restriction of the completely positive map $\tilde{\Phi}:C(\hat{M})/\hat{J}\to C(M)/J$. By applying this map to the constant sequence $(b_n)=b$, we deduce that $\tilde{\Phi}=\Phi^{\infty}$. Thus for every strongly convergent sequence in $A$, we deduce that $a=\lim_n a_n$ belongs to the multiplicative domain of $\Phi^{\infty}$ because $\si(a^*a)=\si(a)^*\si(a)$ and $\si(a)^*\si(a)=\si(aa^*)$.\qd

\begin{theorem}[Technical version of Theorem \ref{l1intro}] \label{l1iso} Let $\rho\kl \la \eta$ for some $\lambda >0$, and $\Phi:L_1(M)\to L_1(\hat{M})$. Then the following are equivalent
\begin{enumerate}
\item[i)] $D(\Phi(\rho)\|\Phi(\eta))=D(\rho\|\eta)$;
 \item[ii)] There exists a $\eta$-conditioned subalgebra $M_0\subset M$ and an completely positive $L_1$-isometry $u$ such that
  \[ u(\eta)\lel \Phi(\eta)\pl,\pl u(\rho)\lel \Phi(\rho) \pl .\]
\end{enumerate}
\end{theorem}

\begin{proof} Thanks to Lemma \ref{l1isolem} we only have to prove $i)\Rightarrow ii)$.   In view of Corollary \ref{equal1}, we may assume that $\Phi=\Psi$ intertwines $\si_{\eta}$ and $\si_{\hat{\eta}}$. Let $G=\bigcup_{k} 2^{-k}\zz$. Since $\Psi$ is $\si$-invariant we know that $\Psi_G=\Psi\rtimes G$ extends to the cross product.  Recall that $\eta_G=\eta\circ E_G$, and $\rho_G=\rho\circ E_G$ naturally extend to the discrete crossed product. Let us recall that $\Psi^G$  extends to a map $T_1^G:L_1(\hat{M}_G)\to L_1(M_G)$ via
 \[ T_G(\hat{\eta}_G^{1/2}x\hat{\eta}_G^{1/2})
 \lel \eta_G^{1/2}\Phi_G^{\dag}\eta_G^{1/2} \pl .\]
Since $D(\rho_G\|\eta_G)=D(\rho\|\eta)$ and $D(\Psi_G(\rho)\|\Psi_G(\eta_G)) = D(\Psi(\rho)\|\Psi(\eta))$,
we deduce that
 \[ T_1^G(\hat{\rho}_G)\lel\rho_G \pl .\]
Let $\E_n$ be the conditional expectation given by the Haagerup construction. Note that $T_1^GE_n=E_nT_1^G$ follows from the fact  that $\Psi$ commutes with the modular group. Thus for every $\nen$, we may apply Lemma \ref{sf1} and find $A_n=\hat{M}_n(E_n(\rho_G)),E_n(\eta_G))$ in the multiplicative domain which is modular group invariant.

Let us now assume that $\rho=\eta^{1/2}h\eta^{1/2}$  for a bounded $h$ and hence (using the map $\Psi$ instead of $\Phi$) that
 \[   \hat{\rho}\lel
 \hat{\eta}^{1/2}\hat{h} \hat{\eta} \quad,\quad
  \hat{\rho}_{G}\lel \hat{\eta}_{G}\hat{h} \hat{\eta}_G \pl .\]
Let $d_n$ and $\hat{d}_n$ the densities of $\hat{\eta}_G|_{\hat{M}_n}$ and $\eta_G|_{M(n)}$, respectively. Recall that $\hat{d}_n$, and $d_n$ belong to the center of $\hat{M}(n)$ and $M(n)$. Then
 \[ E_n(\hat{\rho}_G) \lel \hat{d}_n^{1/2} E_n(\hat{h}) \hat{d}_n^{1/2} \]
implies that $\hat{h}_n=E_n(\hat{h})$ also belongs to the multiplicative domain of $\Psi_n^{\dag}\lel \Psi^{\dag}E_n$. In order to apply Lemma, we recall that $\eta_G$  and $\hat{\eta}_G$ are $E_n$ invariant. Since $\hat{M}_n$ are increasing, we deduce that  for $n\le m$
  \begin{align*}
  Tr(\eta_G^{1/2}E_n\Psi^{\dag}(a)E_m(b)\eta_G^{1/2})
  &=Tr(\eta_G^{1/2}\Psi^{\dag}(E_n(a)E_m(b))\eta_G^{1/2})\\
  &= Tr(\Psi(\eta_G)E_m(E_n(a)b))
  \lel Tr(\hat{\eta}_G(E_n(a)b)) \\
  &= Tr(\hat{\eta}_G(E_n(a)E_n(b)))
  \lel Tr(\eta_G^{1/2}E_n\Psi^{\dag}(a)E_n(b)\eta_G^{1/2}) \pl .
  \end{align*}
Note that for $\Phi_n=\Psi^{\dag}_GE_n$ we have $\Phi_{\infty}=\Psi^{\dag}$ and hence $\hat{k}=\lim_n \hat{k}_n$ belongs to the multiplicative domain of $\Psi^{\dag}_G$, and hence to multiplicative domain of $\Psi$. Indeed, we may consider $a_n\hat{k}_n^{1/2}$. Then $a_n^*a_n$ converges weakly to $\hat{k}$ if $a_n-\hat{k}^{1/2}$ converges strongly to $0$. Using
 \[ E_n(\hat{\rho}_G) \lel \hat{\eta}_G^{1/2}E_n(\hat{k})\hat{\eta}^{1/2}_G \]
we deduce that weak-convergence from the crucial inequality
 \[ \lim_n \|E_n(\hat{\rho}_G)-\hat{\rho}_G\|_1 \pl \]
in the Haagerup construction.  Note also that
 \[ \sqrt{\hat{d}_n^{1/2}\hat{k}_n\hat{d}_n^{1/2}}
 \lel \hat{d}_n^{1/4}\hat{k}_n^{1/2}\hat{d}_n^{1/4} \]
because $\hat{d}_n$ belongs to the center of $\hat{M}(n)$, which allows us to use St{\o}rmer's inequality. Since the multiplicative domain of $\Phi^{\dag}$ is invariant under the modular group of $\hat{\eta}$ and $\hat{k}$ belongs to the smallest modular group invariant von Neumann subalgebra $\hat{M}_0$ which is mapped to $M_0$ the smallest modular group invariant generated by $h$, we can now conclude as in Lemma \ref{sf1}.\qd

\section{Conclusions \& Outlook} \label{sec:conclusion}
The proofs in \cite{sutter_strengthened_2016} and more traditionally information-theoretic proofs of \cite{sutter_multivariate_2017} use an approach called the method of types \cite{csiszar_method_1998} (not to be confused with von Neumann algebra types). Classically, the key innovation of typicality in Shannon theory turns many copies of a complicated vector of different probabilities into a distribution that is nearly uniform on a set of typical outcomes and nearly unsupported elsewhere. The number of distinct eigenvalues of many copies of a density matrix grows only polynomially, while the dimension grows exponentially. The method of types is thereby powerful on quantities that scale linearly with tensored copies of a matrix.

A more mathematical approach to entropy bounds, used in \cite{wilde_recoverability_2015, gao_capacity_2018, junge_universal_2018} and in the second proof style of \cite{sutter_multivariate_2017}, uses complex interpolation to compare entropies as limits and logarithms of $p$-norms. These techniques are further from classical intuition, can lead to breakthroughs on problems that had resisted traditional information theoretic techniques, and often yield automatic $p$-R\'enyi generalizations. Furthermore, they naturally generalize to Kosaki spaces and do not rely on finite-dimensional assumptions.

Apparent in \cite{sutter_multivariate_2017} are direct correspondences between some information-theoretic methods and their interpolation analogs. Deeper work on this analogy may lead to a more intuitive understanding or mathematical duality. Renewed understanding of Shannon theory through analysis on operator algebras helps escape classical intuition and generalize beyond finite dimension, while the Shannon-theoretic analogy of results on operators may help clarify the physical justification of obtained inequalities. The Haagerup approximation method and Kosaki interpolation spaces add to understanding of this connection.

Holography in high energy physics proposes duality between entropy and geometry, suggesting spatial correspondences and intuition for famous entropy inequalities \cite{headrick_holographic_2007, headrick_lectures_2019} and operational techniques such as error correction \cite{almheiri_bulk_2015, pastawski_holographic_2015}. Many of these connections would manifest physically in field theories modeled as type III von Neumann algebras. The theory of entropy in holography will therefore benefit from an intuitive method of traceless entropy results.

\section{Acknowledgments}
We acknowledge correspondence with David Sutter as inspiring some inquiries within this project. We also acknowledge conversations with Li Gao, Nima Lashkari, and Mark Wilde.

{ \scriptsize

\bibliography{Merged}
\bibliographystyle{unsrt}

}

\end{document}